\documentclass[11pt,letterpaper]{article}

\pdfoutput=1
\usepackage{amsmath,amssymb,amsfonts}
\usepackage{mathrsfs}

\usepackage{tikz}
\usetikzlibrary{shapes.geometric}
\usetikzlibrary{calc}
\usepackage{float}

\usepackage{stackengine}

\newcommand{\sqrbullet}{%
  \scalebox{0.8}{%
    \stackinset{c}{0pt}{c}{-0.05ex}{$\bullet$}{$\square$}%
  }%
}

\newcommand{\bigcircbullet}{%
  \scalebox{0.8}{%
    \stackinset{c}{0pt}{c}{-0.05ex}{$\bullet$}{$\bigcirc$}%
  }%
}

\newcommand{\bulletdiagup}{%
  \mathrel{%
    \begin{tikzpicture}[baseline=(X.base),scale=0.7, transform shape]
      \node (X) {$\bullet$};
      \draw[thick] (X.center) -- ([xshift=2pt,yshift=2pt]X.north east);
    \end{tikzpicture}%
  }%
}

\newcommand{\bulletdiagdown}{%
  \mathrel{%
    \begin{tikzpicture}[baseline=(X.base),scale=0.7, transform shape]
      \node (X) {$\bullet$};
      \draw[thick] ([xshift=-2pt,yshift=2pt]X.north west) -- (X.center);
    \end{tikzpicture}%
  }%
}

\newcommand{\sqrdiagup}{%
  \mathrel{%
    \begin{tikzpicture}[baseline=(X.base), scale=0.7, transform shape]
      \node (X) {$\square$};
      \draw[thick] (X.center) -- ([xshift=2pt,yshift=2pt]X.north east);
    \end{tikzpicture}%
  }%
}

\usetikzlibrary{decorations.pathmorphing}

\newcommand{\sqrbulletdiagup}{%
  \mathrel{%
    \begin{tikzpicture}[baseline=(X.base),scale=0.7, transform shape]
      \node (X) {$\sqrbullet$};
      \draw[thick] (X.center) -- ([xshift=2pt,yshift=2pt]X.north east);
    \end{tikzpicture}%
  }%
}

\newcommand{\sqrbulletdiagdown}{%
  \mathrel{%
    \begin{tikzpicture}[baseline=(X.base),scale=0.7,transform shape]
      \node (X) {$\sqrbullet$};
      \draw[thick] ([xshift=-2pt,yshift=2pt]X.north west) -- (X.center);
    \end{tikzpicture}%
  }%
}

\newcommand{\sqrbulletV}{%
  \mathrel{%
    \begin{tikzpicture}[baseline=(X.base),scale=0.7,transform shape]
      \node (X) {$\sqrbullet$};
      \draw[thick] (X.center) -- ([xshift=2pt,yshift=2pt]X.north east);
      \draw[thick] ([xshift=-2pt,yshift=2pt]X.north west) -- (X.center);
    \end{tikzpicture}%
  }%
}

\newcommand{\os}{\;\hat{=}\;}

\newcommand{\op}[1]{\boldsymbol{#1}}

\usepackage{longtable}
\usepackage{array}

\newtheorem{theorem}{Theorem}[section]

\usepackage[dvipsnames]{xcolor}
\definecolor{airforceblue}{rgb}{0.36, 0.54, 0.66}
\definecolor{meb}{rgb}{0.01, 0.31, 0.59}
\usepackage[T1]{fontenc}
\def\title{Quantization of Gravity on Null Hypersurfaces} 
\usepackage{graphicx}
\usepackage[colorlinks=true, pdfstartview=FitV, linkcolor=blue, citecolor=magenta, urlcolor=blue]{hyperref}
\usepackage{cancel}
\usepackage{mdframed,framed}
\usepackage[dvipsnames]{xcolor}
\usepackage[normalem]{ulem}
\usepackage{layout}
\usepackage[nosort]{cite}

\usepackage{textgreek}

\newcommand{\beq}{\begin{eqnarray}}
\newcommand{\eeq}{\end{eqnarray}}
\newcommand{\beqn}{\begin{equation}
\begin{aligned}}
\newcommand{\eeqn}{\end{aligned}\end{equation}}
\newcommand{\bee}{\begin{equation} \begin{aligned}}
\newcommand{\eee}{ \end{aligned} \end{equation}}
\newcommand{\pa}{\partial}
\newcommand{\un}[1]{\underline{#1}}

\newcommand{\RR}{\mathbb{R}}

\newcommand{\cL}{{\cal L}}
\newcommand{\cO}{{\cal O}}

\newcommand{\fL}{\mathfrak{L}}

\newcommand{\osigma}{{\overline{\sigma}}}

\usepackage{tikz-cd}
\usepackage{pict2e}
\makeatletter
\newcommand{\variable@rule}[1]{%
  \fontdimen8  
  \ifx#1\displaystyle\textfont3\else
    \ifx#1\textstyle\textfont3\else
      \ifx#1\scriptstyle\scriptfont3\else
        \scriptscriptfont3\relax
  \fi\fi\fi
}
\usepackage{mathtools}

\usepackage{tocloft}
\newcommand{\ve}{\varepsilon}

\newcommand{\cN}{{\cal{N}}}

\newcommand{\cC}{{\cal{C}}}

\newcommand{\bq}{{\overline{q}}}
\newcommand{\bh}{{\overline{h}}}

\newcommand{\rd}{\text{d}}

\newcommand{\chkM}{{\color{red} \,\checkmark\kern-5pt{}_{M}}}

\newcommand{\be}{\begin{equation}}
\newcommand{\ee}{\end{equation}}
\newcommand{\bea}{\begin{eqnarray}}
\newcommand{\eea}{\end{eqnarray}}

\def\pa{\partial}

\newcommand{\ket}[1]{\vert #1 \rangle}

\newcommand{\bra}[1]{\langle #1 \vert}

\usepackage{ragged2e}
\usepackage{blindtext}

\newcommand{\perimeter}[1]{
	\centerline{
		\begin{minipage}[c]{0.7\textwidth}
			\begin{center}
			Perimeter Institute for Theoretical Physics,\\
			 31 Caroline St. N., Waterloo ON, Canada, N2L 2Y5\\
             \href{mailto:ciambelli.luca@gmail.com}{ciambelli.luca@gmail.com}
			\end{center}
		\end{minipage}
		}
	}

\newcommand{\caltech}[1]{
  \centerline{
    \begin{minipage}[c]{0.7\textwidth}
      \begin{center}
      ${}^{#1}$ Walter Burke Institute for Theoretical Physics \\
      and Leinweber Forum for Theoretical Physics,\\
      California Institute of Technology, Pasadena, CA 91125, U.S.A.\\
\href{mailto:klingerm@caltech.edu}{klingerm@caltech.edu}
      \end{center}
    \end{minipage}
  }
}

\begin{document}

\begin{center}
 \vspace*{1cm}
\textbf{\LARGE{\title{}}}
\end{center}
\vspace{0.5cm}
\begin{center}
Luca Ciambelli\\
\emph{\perimeter{}}
\vspace{0.5cm}
\end{center}
\begin{center}
Marc S. Klinger\\
\emph{\caltech{}}
\vspace{0.5cm}
\end{center}

\begin{center}
    \emph{This paper is dedicated to Robert G. Leigh.}
 \end{center}   

 \begin{center}
 \emph{
    Rob's fingerprints can be seen all over this work -- he laid its foundation over the course of his own remarkable research program. More importantly than that, he was an indelible mentor, collaborator, and dear friend to both of its authors. May he be somewhere beyond the horizon, still collecting stamps and insisting that the world is quantum -- and so, therefore, must be gravity.}
\end{center}

\begin{abstract}
\vspace{0.3cm}
The initial value problem for general relativity on spacelike hypersurfaces is famously captured by the ADM formalism. Less well known is the Cauchy problem for general relativity on \emph{null} hypersurfaces, which goes under the name of the Characteristic Initial Value Problem (CIVP). The CIVP is formulated on a pair of intersecting null hypersurfaces, encoding rich physics in the interplay between their respective initial data, and especially the gluing of these hypersurfaces at their shared codimension-two corner. In this work, we construct an operator-algebraic quantization of the CIVP. To do so, we first quantize each null hypersurface separately, using the corner symmetries as a guiding principle, and then glue them together at the joint initial cut. The resulting algebra admits an inner action of the full corner symmetry (semi)group consisting of superboosts, superrotations, and supertranslations. These are associated, respectively, with the quantization of the area, the Hájiček one-form, and the expansion along both null directions. Supertranslations do not act as algebra preserving maps, but instead are quantized to quantum channels and included into the algebra via a generalized form of the Stinespring dilation theorem. The inclusion of superboosts and superrotations implements the gravitational constraints arising from Einstein's equations at the level of the quantum algebra. By virtue of the CIVP, in its local caustic-free domain of validity, our construction yields a candidate for the on-shell algebra of a gravitational subregion subtended by a pair of null hypersurfaces.
\end{abstract}

\newpage

\tableofcontents

\newpage

\section{Introduction}

The formulation of general relativity as an initial value problem is one of the central achievements of classical gravitational physics \cite{Arnowitt:1962hi}; see also \cite{MisnerThorneWheeler1973,Wald1984}. In the spacelike setting, the ADM formalism \cite{Arnowitt:1962hi} recasts Einstein's equations as Hamiltonian evolution subject to constraint equations, thereby revealing gravity as a dynamical theory of geometry. This perspective is both conceptually powerful and technically indispensable: it identifies the physical degrees of freedom of the gravitational system, their gauge redundancies, and the equations that propagate them. Yet precisely these features become obstacles upon quantization
\cite{dewitt1967quantum,Isham:1992ms,Kuchar:1991qf}. In a closed gravitational
system, the Hamiltonian is constrained rather than absolute
\cite{Dirac:1958sc}, while diffeomorphism invariance obscures the localization
of observables and complicates the construction of a physical Hilbert space
\cite{Torre:1993fq}.\\

A natural complementary formulation is provided by the characteristic initial value problem (CIVP), originally introduced in \cite{Sachs:1962zzb}, rigorously studied in \cite{Friedrich:1981at,Rendall1990CharacteristicInitialValue,SAHayward_1993,PRBrady_1996,Choquet-Bruhat:2010vby}, and recast in modern terms in \cite{Reisenberger:2007ku,Reisenberger:2018xkn,Mars:2022gsa,Mars:2023hty,Chandrasekaran:2023vzb}. The CIVP analyzes general relativity as an initial value problem on null, rather than spacelike, hypersurfaces. This formulation is especially well adapted to causal propagation, radiation, horizons, and subregions bounded by lightlike surfaces. It therefore provides a causally adapted Lorentzian description of gravitational dynamics.

Null hypersurfaces have received a great deal of attention in recent years as a promising foundation for studying both classical and quantum features of gravity. Their classical geometry has been studied from both the embedded perspective \cite{Gourgoulhon:2005ng} and the intrinsic, or Carrollian, perspective
\cite{LevyLeblond1965,SenGupta:1966qer,Henneaux1979a,Duval:2014uva};
see also the review \cite{Ciambelli:2025unn}. Their dynamics \cite{Torre:1985rw,Hopfmuller:2016scf,Wieland:2019hkz}, radiative data \cite{Reisenberger:2007pq,Reisenberger:2012zq}, and boundary phase spaces \cite{Parattu:2015gga,Lehner:2016vdi,Chandrasekaran:2018aop,Adami:2021nnf,Ciambelli:2023mir,Odak:2023pga} have all been extensively developed. These structures are especially compelling because they place the gravitational constraints and radiative degrees of freedom in a form naturally suited to quantization \cite{Fuchs:2017jyk,Wieland:2021vef,Wieland:2024dop,Bak:2023wwo,
Ciambelli:2024swv,Wieland:2025qgx,Ciambelli:2025flo,Kowalski-Glikman:2026bwg}.\\

Toward this direction, symmetries provide one of the most robust anchors for quantization. This principle is particularly powerful in gravity, because apparent gauge symmetries are not merely redundancies: in the presence of boundaries, corners, or subregions they acquire non-trivial charges and organize the physical degrees of freedom \cite{Noether1918, Regge:1974zd, Wald:1999wa, Barnich:2001jy}. Null hypersurfaces are especially rich from this perspective. Their symmetry structure exhibits infinite-dimensional enhancements \cite{Donnay:2015abr, Donnay:2016ejv,Penna:2018gfx,Donnay:2019jiz, Ciambelli:2019lap}, closely related to the BMS symmetries of asymptotically flat spacetimes \cite{Bondi,Sachs,Sachs:1962wk,Barnich:2010eb}, but realized locally at finite null boundaries and corners. This viewpoint is central to the corner proposal, where gravitational subregions are organized by their corner symmetry algebras and associated edge modes \cite{Donnelly:2016auv,Speranza:2017gxd,Freidel:2020xyx,Freidel:2021cjp,Freidel:2021dxw,Varrin:2024sxe,Varrin:2025okc,Ciambelli:2025ztm,Neri:2025fsh}, see also the review \cite{Ciambelli:2022vot}. In the null setting, these symmetries are not only kinematical: their charges are directly tied to the geometric data of the CIVP. This makes the symmetry structure of null hypersurfaces a natural starting point toward constructing the quantum algebra for gravitational subregions.\\

In this work, we follow this principle and take a step toward the quantization of gravity on null hypersurfaces by using modern operator algebraic techniques to perturbatively quantize the CIVP data. This follows in the footsteps of recent work which has reformulated the null gravitational constraints in a form amenable to operator-algebraic quantization \cite{Ciambelli:2024swv,Klinger:2025hjp}. The algebraic methods we employ have found powerful applications
in gravitational settings, beginning with holographic systems
\cite{Leutheusser:2021frk,Witten:2021unn,Chandrasekaran:2022eqq,Gesteau:2023hbq}
and extending to de Sitter static patches \cite{Chandrasekaran:2022cip},
black-hole exterior algebras
\cite{Kudler-Flam:2023qfl,Faulkner:2024gst,Klinger:2026tws},
and generic gravitational subregions
\cite{Jensen:2023yxy,AliAhmad:2023etg,Klinger:2025tvg}. A central lesson of these developments is that imposing gravitational constraints in quantum systems can often be organized perturbatively through crossed-product techniques. This has sharpened long-held intuitions about the relationship between quantum gravity and quantum information, while also providing a framework for exploring possible observable signatures of quantum gravity in a rigorous language.\\

Because the main construction of this work is technical, we use this introduction to spell out the logical sequence of steps and to highlight the main conceptual points. To this end, we will now provide a skeleton of the analysis contained in this work, and in doing so provide an overview of its main insights. 

We begin in Section~\ref{sec: Classical} with a review of the classical CIVP. The main theorem, stated in Section~\ref{subsec:CIVPthm}, instructs us that the bulk metric in a spacetime region bounded by a pair of null hypersurfaces $\mathcal{N}_{(u)}$ and $\mathcal{N}_{(v)}$ intersecting at the initial joint codimension-2 corner $\mathcal{C}$ is uniquely characterized by the following set of data:\footnote{The various quantities introduced here will be carefully defined in the Section \ref{sec: Initial Data}.}
\begin{enumerate}
	\item \textbf{On the corner $\mathcal{C}$:}
	\begin{enumerate}
		\item A Riemannian metric $h_{ij}$ with unimodular part $\overline{h}_{ij}$ and volume element $\Omega\rvert_{\mathcal{C}}$,
		\item The value of the expansion scalars $\theta_{(u)}\rvert_{\mathcal{C}}$ and $\theta_{(v)}\rvert_{\mathcal{C}}$,
		\item The value of the Hájiček one-form $\pi_a\rvert_{\mathcal{C}}$,
		\item A scalar field $m\rvert_{\mathcal{C}}$.
	\end{enumerate}
	\item \textbf{In the bulk of each hypersurface $\mathcal{N}_{(i)}$:}
	\begin{enumerate}
		\item The inaffinity $\kappa_{(i)}$,
		\item The clock vector field $\un\ell_{(i)}$, 
		\item The shear $\sigma^{(i)}_{ab}$. 
	\end{enumerate}
\end{enumerate}
These initial data are depicted pictorially in Figure~\ref{fig:CIVPIntro}: 
\begin{figure}[H] 
\centering
\begin{tikzpicture}[
    scale=1.2,
    every node/.style={align=center},
    line/.style={thick},
    point/.style={circle, fill=black, inner sep=1.5pt}
]

% Intersection point
\node[point, label=below:$\mathcal{C}$] (C) at (0,0) {};

% Left and right lines (forming a V)
\draw[line] (C) -- (-4,3);
\draw[line] (C) -- (4,3);

% Labels for the null hypersurfaces
\node at (-3.2,2.8) {$\mathcal{N}_{(u)}$};
\node at (3.2,2.8) {$\mathcal{N}_{(v)}$};

% Data on left line
\node at (-3.5,1.2) {$\big(\kappa_{(u)},\, \ell_{(u)}^a,\, \sigma^{(u)}_{ab}\big)$};

% Data on right line
\node at (3.5,1.2) {$\big(\kappa_{(v)},\, \ell_{(v)}^a,\, \sigma^{(v)}_{ab}\big)$};
\node [draw, rectangle, minimum width=0.35cm, minimum height=0.35cm] at (C) {};
% Data at the intersection point
\node at (0,-0.8) {
$\big(
h_{ij},\theta_{(i)}\rvert_{\mathcal{C}}, \pi_a\rvert_{\mathcal{C}},m\rvert_{\mathcal{C}}
\big)$
};

\end{tikzpicture}
\caption{Pictorial representation of the CIVP. The indicated data determine the spacetime metric
in the local future development of the shared corner.}
\label{fig:CIVPIntro}
\end{figure}
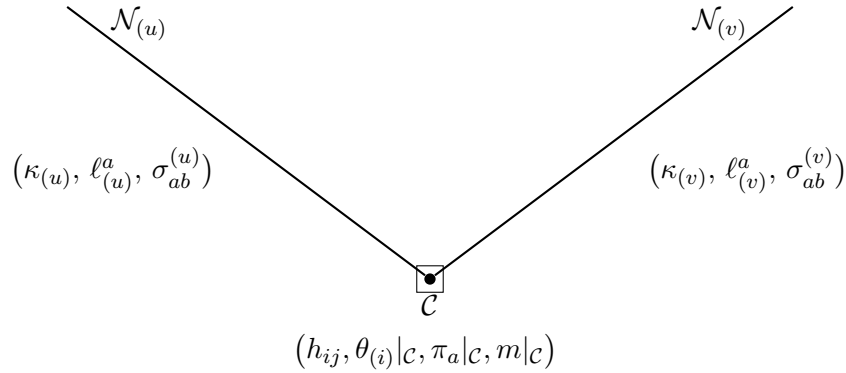

One of the major strategies we employ in this work is cutting and gluing together the CIVP data according to the relationships between its underlying degrees of freedom. This point of view serves to illustrate the symmetry structure underlying the CIVP, and ultimately prepares the ground for its quantization. To accommodate this cutting and gluing, we introduce a diagrammatic calculus that organizes the ingredients appearing in the CIVP and their interrelation. The CIVP data are split into four types -- pure hypersurface data labeled by a $\diagup$, pure corner data labeled by a $\bullet$, extended corner data labeled by a $\square$, and gluing data labeled by a $\bigcirc$. This division is visualized in Figure~\ref{fig:CIVPDeconstructedIntro}:
\begin{center}
\begin{figure}[H] 
\centering
\begin{tikzpicture}[
    scale=2.0,
    line/.style={thick},
    box/.style={draw, rectangle, minimum size=6pt, inner sep=0pt},
    dot/.style={circle, fill=black, inner sep=1.5pt},
    oval/.style={draw, ellipse, minimum width=1.2cm, minimum height=0.5cm},
    every node/.style={align=center, font=\small}
]

% --- Spacing parameters ---
\def\gapdotbox{0.5}
\def\gapboxline{0.5}
\def\linelength{2}
\def\ovalgap{0.7}

% --- RIGHT RAY (45°) ---
\coordinate (dotR) at (1,0);                     
\coordinate (boxR) at ($(dotR)+(45:\gapdotbox)$);
\coordinate (lineRstart) at ($(boxR)+(45:\gapboxline)$);
\coordinate (lineRend) at ($(lineRstart)+(45:\linelength)$);

\node[dot] at (dotR) {};
\node[box] at (boxR) {};
\draw[line] (lineRstart) -- (lineRend);

% Right labels (aligned with elements)
\node[right] at ($(dotR)+(0.15,-0.05)$) {$(\Omega_{(v)}\rvert_{\mathcal{C}},\, \pi^{(v)}_a\rvert_{\mathcal{C}})$};
\node[right] at ($(boxR)+(0.2,0)$) {$\theta_{(v)}\rvert_{\mathcal{C}}$};  % aligned with box
\node[right] at ($(lineRend)+(-0.7,-0.9)$) {$(\kappa_{(v)},\, \ell_{(v)}^a,\, \sigma^{(v)}_{ab})$};

% --- LEFT RAY (135°) ---
\coordinate (dotL) at (-1,0);                    
\coordinate (boxL) at ($(dotL)+(135:\gapdotbox)$);
\coordinate (lineLstart) at ($(boxL)+(135:\gapboxline)$);
\coordinate (lineLend) at ($(lineLstart)+(135:\linelength)$);

\node[dot] at (dotL) {};
\node[box] at (boxL) {};
\draw[line] (lineLstart) -- (lineLend);

% Left labels (aligned with elements)
\node[left] at ($(dotL)+(-0.15,-0.05)$) {$(\Omega_{(u)}\rvert_{\mathcal{C}},\, \pi^{(u)}_a\rvert_{\mathcal{C}})$};
\node[left] at ($(boxL)+(-0.2,0)$) {$\theta_{(u)}\rvert_{\mathcal{C}}$};  % aligned with box
\node[left] at ($(lineLend)+(0.7,-0.9)$) {$(\kappa_{(u)},\, \ell_{(u)}^a,\, \sigma^{(u)}_{ab})$};

% --- CENTRAL OVAL ---
\node[oval] at (0,-0.4) {$\overline{h}_{ij},\, m\rvert_{\mathcal{C}}$};

\end{tikzpicture}
\caption{A convenient way to organize the data of the CIVP.}
\label{fig:CIVPDeconstructedIntro}
\end{figure}
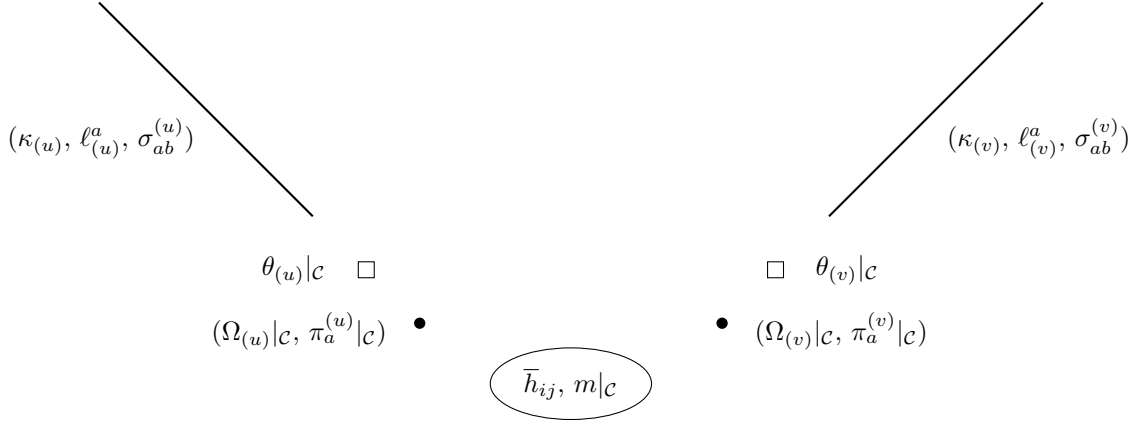
\end{center} 

Before gluing them together at the shared corner, the combinations $(\diagdown, \square_u, \bullet_u)$ and $(\diagup, \square_v, \bullet_v)$ can be regarded as defining two a priori separate null hypersurface theories. Indeed, each set of data is the natural physical input organizing the geometry, symplectic structure, and degrees of freedom of a single null hypersurface \cite{Torre:1985rw,Hopfmuller:2016scf,Reisenberger:2007pq}, and has recently served as a starting point for quantization
\cite{Wieland:2024dop,Ciambelli:2024swv,Klinger:2025tvg}. More precisely, the degrees of freedom associated with $\diagup$ define a phase space, $X_{\diagup}$, with kinematical symplectic form $\Omega_{\diagup}$. If the null hypersurface has an initial cut -- as in the case of the CIVP analysis -- this phase space admits an action by the corner symmetry semigroup:
\beq
	G_{\sqrbullet}^+ = \bigg( \text{Diff}(\mathcal{C}) \ltimes C^{\infty}(\mathcal{C})_B\bigg) \ltimes C^{\infty}(\mathcal{C})_T = G_{\bullet} \ltimes G_{\square}^+,
\eeq
in which $G_{\bullet}$ consists of superrotations and superboosts, while $G_{\square}^+$ consists of supertranslations.\footnote{This terminology is borrowed from the asymptotic symmetry literature on flat-space gravity -- the BMS group and its extensions originally introduced in \cite{Bondi,Sachs,Sachs:1962wk,Barnich:2010eb} -- although here it is used in the local corner/null-hypersurface sense.} We denote these actions by $a^{\bullet}: G_{\bullet} \times X_{\diagup} \rightarrow X_{\diagup}$ and $a^{\square}: G_{\square}^+ \times X_{\diagup} \rightarrow X_{\diagup}$, and emphasize that only $a^{\bullet}$ is phase space preserving. Supertranslations generally move the corner and therefore do not preserve the phase space. Moreover, only positive supertranslations are admitted, since in general a negative supertranslation could move the initial cut backward in time, and thus outside the causal domain of the CIVP. This is why we refer to the symmetry structure underlying the hypersurface theory as a \emph{semi}group. The various symmetry actions are collected in Figure \ref{Single Hyper Intro} below. It is important to appreciate the semidirect nature of the various symmetry actions, as this serves as a guiding principle for the perturbative quantization.\\

\begin{figure}[H]
\centering
\begin{tikzpicture}[
    scale=2.0,
    line/.style={thick},
    box/.style={draw, rectangle, minimum size=6pt, inner sep=0pt},
    dot/.style={circle, fill=black, inner sep=1.5pt},
    every node/.style={align=center, font=\small}
]

% --- Spacing parameters ---
\def\gapdotbox{0.5}
\def\gapboxline{0.5}
\def\linelength{2}

% --- RIGHT RAY (45°) ---
\coordinate (dotR) at (0,0);                     
\coordinate (boxR) at ($(dotR)+(45:\gapdotbox)$);
\coordinate (lineRstart) at ($(boxR)+(45:\gapboxline)$);
\coordinate (lineRend) at ($(lineRstart)+(45:\linelength)$);

\node[dot] at (dotR) {};
\node[box] at (boxR) {};
\draw[line] (lineRstart) -- (lineRend);

% Labels (named nodes)
\node[right] (labelDot) at ($(dotR)+(0.35,-0.1)$)
    {$G_{\bullet} = \mathrm{Diff}(\mathcal{C}) \ltimes C^{\infty}(\mathcal{C})_B$};

\node[right] (labelBox) at ($(boxR)+(0.35,0)$)
    {$G^+_{\square}\subset G_{\square} = C^{\infty}(\mathcal{C})_T$};

\node[right] (labelLine) at ($(lineRend)+(0.25,0.1)$)
    {$(X_{\diagup},\, \Omega_{\diagup})$};

% --- Curved arrows ---

% dot -> box (left side)
\draw[->, bend right=20]
    ([xshift=-1pt]labelDot.north west) to node[right] {$a^{\bullet,\square}$} ([xshift=-1pt]labelBox.south west);

% box -> line (center)
\draw[->, bend right=20]
    (labelBox.north) to node[right] {$a_+^{\square,\diagup}$} (labelLine.south);

% dot -> line (right side)
\draw[->, bend right=35]
    ([xshift=1pt]labelDot.north east) to node[right] {$a^{\bullet,\diagup}$} ([xshift=1pt]labelLine.south east);

\end{tikzpicture}
\caption{The classical data of a single hypersurface exhibit a nesting structure in which each piece of data acts on all of the data above it.}
\label{Single Hyper Intro}
\end{figure}
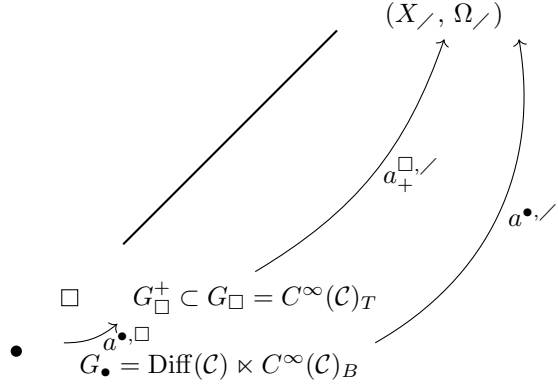

The Hamiltonians associated with the actions $a^{\bullet}$ and $a^{\square}$ decompose schematically as
\beq
	H_{(f_B,Y)} = \int_{\diagup} \bigg(f_B(\textrm{Ray}) + Y^a( \textrm{Dam}_a)\bigg) + \int_{\mathcal{C}} \bigg(f_B \Omega + Y^a \pi_a\bigg), \qquad H_{f_T} = \int_{\diagup} f_T (\textrm{Ray}) + \int_{\mathcal{C}} f_T \theta,
\eeq
where $\textrm{Ray}$ and $\textrm{Dam}_i$ are the projections of the Einstein equation to the null hypersurface -- the Raychaudhuri and Damour equations -- and $(f_B,Y)$ and $f_T$ parameterize elements of $G_{\bullet}$ and $G_{\square}$, respectively. On shell of the equations of motion, the Hamiltonians generating the actions $a^{\bullet}$ and $a^{\square}$ coincide with the corner supported Noether charges. The charge aspects of these generators coincide precisely with the CIVP data $\bullet$ and $\square$, respectively. This completes the classical analysis and recasts the CIVP in a form adapted to the quantum construction that follows.\\

In Sections~\ref{sec: QKin}, \ref{sec: QCon} and \ref{sec: gluing} we quantize the classical setup described above. To do so, we proceed in the following logical sequence of  steps:
\begin{enumerate}
	\item Kinematical Quantization (sec.~\ref{sec: QKin}),
	\item Extension by Supertranslations (sec.~\ref{sec: STCrossed}),
	\item Constraint Quantization (sec.~\ref{sec: Ray+Dam}),
	\item Corner Gluing (sec.~\ref{sec: gluing}). 
\end{enumerate}
In kinematical quantization, we promote the phase space data of each null hypersurface theory into an operator algebraic language. The phase space $X_{\diagup}$ is quantized to an operator algebra $\mathscr{A}_{\diagup}$. The symplectomorphic action $a^{\bullet}$ induces an action on $\mathscr{A}_{\diagup}$ by automorphisms (invertible, algebra preserving maps) $\alpha^{\bullet}: G_{\bullet} \rightarrow \text{Aut}(\mathscr{A}_{\diagup})$. Conversely, the non-symplectomorphic action $a^{\square}$ induces an action on $\mathscr{A}_{\diagup}$ by \emph{quantum channels}, i.e., completely positive and unital maps, $\alpha^{\square}: G_{\square}^+ \rightarrow \textrm{CP}(\mathscr{A}_{\diagup})$ which are generically neither invertible nor algebra preserving. The fact that supertranslations are represented by quantum channels rather than symplectomorphisms is a central result of our manuscript. At the same time, the Noether charges $Q_{(f_B,Y)}$ and $Q_{f_T}$ are quantized as generators of intrinsically defined operator algebras $\mathscr{A}_{\bullet}$ and $\mathscr{A}_{\square}$. The semidirect product structure of the semigroup $G^+_{\sqrbullet}$ implies that there is an automorphic action $\alpha^{\bullet,\square}: G_{\bullet} \rightarrow \text{Aut}(\mathscr{A}_{\square})$. We therefore obtain a straightforward quantization of Figure~\ref{Single Hyper Intro} into Figure~\ref{Single Quantum Hyper Intro}.

\begin{figure}[H]
\centering
\begin{tikzpicture}[
    scale=2.0,
    line/.style={thick},
    box/.style={draw, rectangle, minimum size=6pt, inner sep=0pt},
    dot/.style={circle, fill=black, inner sep=1.5pt},
    every node/.style={align=center, font=\small}
]

% --- Spacing parameters ---
\def\gapdotbox{0.5}
\def\gapboxline{0.5}
\def\linelength{2}

% --- RIGHT RAY (45°) ---
\coordinate (dotR) at (0,0);                     
\coordinate (boxR) at ($(dotR)+(45:\gapdotbox)$);
\coordinate (lineRstart) at ($(boxR)+(45:\gapboxline)$);
\coordinate (lineRend) at ($(lineRstart)+(45:\linelength)$);

\node[dot] at (dotR) {};
\node[box] at (boxR) {};
\draw[line] (lineRstart) -- (lineRend);

% Labels (named nodes)
\node[right] (labelDot) at ($(dotR)+(0.35,-0.1)$)
    {$\mathscr{A}_{\bullet}$};

\node[right] (labelBox) at ($(boxR)+(0.35,0)$)
    {$\mathscr{A}_{\square}$};

\node[right] (labelLine) at ($(lineRend)+(0.25,0.1)$)
    {$\mathscr{A}_{\diagup}$};

% --- Curved arrows ---

% dot -> box (left side)
\draw[->, bend right=20]
    ([xshift=-1pt]labelDot.north west) to node[right] {$\alpha^{\bullet,\square}$} ([xshift=-1pt]labelBox.south west);

% box -> line (center)
\draw[->, bend right=20]
    (labelBox.north) to node[right] {$\alpha^{\square,\diagup}$} (labelLine.south);

% dot -> line (right side)
\draw[->, bend right=35]
    ([xshift=1pt]labelDot.north east) to node[right] {$\alpha^{\bullet,\diagup}$} ([xshift=1pt]labelLine.south east);

\end{tikzpicture}
\caption{Quantum data of a single hypersurface.}
\label{Single Quantum Hyper Intro}
\end{figure}

We next form an extended algebra $\mathscr{A}_{\sqrdiagup}$ by combining together the algebras $\mathscr{A}_{\diagup}$ and $\mathscr{A}_{\square}$ through the action $\alpha^{\square}$. The resulting algebra resembles a crossed product, but is subtly different since $\alpha^{\square}$ is a non-automorphic action. Technically, the tool that makes this amalgamation possible is a generalized form of Stinespring's dilation theorem \cite{Longo:2017aet} which allows for the quantum channels implementing the supertranslation action to be related to an algebra preserving, but still non-invertible, action by endomorphisms. The extension of $\mathscr{A}_{\diagup}$ to $\mathscr{A}_{\sqrdiagup}$ can be interpreted as a quantum analog of the extended phase space formalism originally described in \cite{Donnelly:2016auv,Speranza:2017gxd}, and further developed in \cite{Ciambelli:2021vnn,Ciambelli:2021nmv,Carrozza:2022xut, Klinger:2023qna}, see also the review \cite{Ciambelli:2022vot}. \\

The tensor product algebra $\mathscr{A}_{\textrm{kin.}} = \mathscr{A}_{\sqrdiagup} \otimes \mathscr{A}_{\bullet}$ describes the kinematical degrees of freedom of the hypersurface theory prior to the imposition of the gravitational constraints. As we have alluded to above, in the classical theory the constraints can be formulated as the on-shell equality
\beq
	H_{(f_B,Y)} \os Q_{(f_B,Y)}.
\eeq
In the quantum theory, $H_{(f_B,Y)}$ and $Q_{(f_B,Y)}$ are the generators of two different representations of the group $G_{\bullet}$. The Hamiltonian generates the representation of $G_{\bullet}$ on the full kinematical algebra $\mathscr{A}_{\textrm{kin.}}$, while the Noether charge generates the action of $G_{\bullet}$ on the group algebra $\mathscr{A}_{\bullet}$. The constraint can therefore be reinterpreted as the equality of these two representations for physical operators. That is, the set of physical operators is given by the subalgebra
\beq
	\mathscr{A}_{\sqrbulletdiagup} = \{\op{\mathcal{O}} \in \mathscr{A}_{\textrm{kin.}} \; | \; e^{i \op{H}_{(f_B,Y)}} \op{\mathcal{O}} e^{-i\op{H}_{(f_B,Y)}} = e^{i \op{Q}_{(f_B,Y)}} \op{\mathcal{O}} e^{-i\op{Q}_{(f_B,Y)}}, \quad \forall \ \ (f_B,Y) \in G_{\bullet} \}. 
\eeq
This is the standard crossed-product implementation of a group action \cite{takesaki2003toa2}: the
physical algebra is equivalently obtained by adjoining to
$\mathscr{A}_{\sqrdiagup}$ the unitaries implementing the action of
$G_{\bullet}$,
\beq
    \mathscr{A}_{\sqrbulletdiagup} 
    \simeq \mathscr{A}_{\sqrdiagup} \rtimes G_{\bullet}.
\eeq
As we discussed above, crossed-products have played a central role in the construction of semiclassical gravitational algebras and the formalization of the generalized entropy \cite{Witten:2021unn,Chandrasekaran:2022cip}. They were further developed as a mechanism for constraint quantization in \cite{Klinger:2023auu,AliAhmad:2024wja}, which is the perspective adopted here.

The on-shell algebra satisfying all of the hypersurface constraints can be understood as consisting of pure hypersurface degrees of freedom dressed to the full corner symmetry $G_{\sqrbullet}^+$, expansion operators dressed to $G_{\bullet}$, and area and Hájiček operators that live on the immediate corner, as pictorially represented in the figure below.\footnote{We denote operators in bold, albeit at this stage one should interpret this as heuristic. More precisely, these objects are quantized by appealing to a proper smearing, as described in Sec.~\ref{sec: QKin}.}

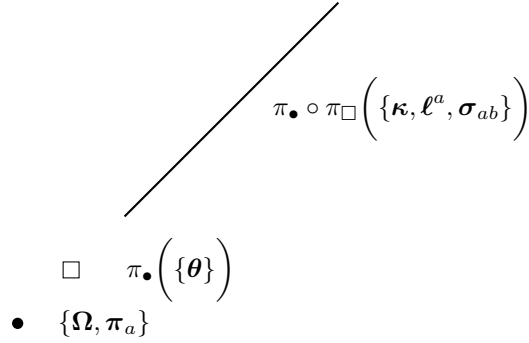
\begin{figure}[H]
\centering
\begin{tikzpicture}[
    scale=2.0,
    line/.style={thick},
    box/.style={draw, rectangle, minimum size=6pt, inner sep=0pt},
    dot/.style={circle, fill=black, inner sep=1.5pt},
    every node/.style={align=center, font=\small}
]

% --- Spacing parameters ---
\def\gapdotbox{0.5}
\def\gapboxline{0.5}
\def\linelength{2}

% --- RIGHT RAY (45°) ---
\coordinate (dotR) at (0,0);                     
\coordinate (boxR) at ($(dotR)+(45:\gapdotbox)$);
\coordinate (lineRstart) at ($(boxR)+(45:\gapboxline)$);
\coordinate (lineRend) at ($(lineRstart)+(45:\linelength)$);

\node[dot] at (dotR) {};
\node[box] at (boxR) {};
\draw[line] (lineRstart) -- (lineRend);

% Labels (named nodes)
\node[right] (labelDot) at ($(dotR)+(0.2,0)$)
    {$\{\op{\Omega},\op{\pi}_a\}$};

\node[right] (labelBox) at ($(boxR)+(0.3,0)$)
    {$\pi_{\bullet}\bigg(\{\op{\theta}\}\bigg)$};

\node[right] (labelLine) at ($(lineRend)+(-0.5,-0.7)$)
    {$\pi_{\bullet} \circ \pi_{\square}\bigg(\{\op{\kappa}, \op{\ell}^a, \op{\sigma}_{ab}\}\bigg)$};

\end{tikzpicture}
\caption{Constraint quantization of a single hypersurface. Here, $\pi_{\bullet}$ and $\pi_{\square}$ are dressing representations associated with $G_{\bullet}$ and $G_{\square}^+$, respectively. The hierarchy of actions is translated into a hierarchy of dressings; every piece of data is dressed to all data below it.}
\label{Single Hyper Quant Const Intro}
\end{figure}

\begin{figure}[ht] 
\centering
\begin{tikzpicture}[
    scale=2.0,
    line/.style={thick},
    sqdot/.style={draw, rectangle, minimum size=6pt, inner sep=0pt},
    dot/.style={circle, fill=black, inner sep=1.2pt},
    oval/.style={draw, ellipse, minimum width=1.2cm, minimum height=0.5cm},
    every node/.style={align=center, font=\small}
]

% --- Spacing parameters ---
\def\linelength{2}
\def\ovalgap{0.7}

% --- RIGHT RAY (45°) ---
\coordinate (lineRstart) at (1,0);
\coordinate (lineRend) at ($(lineRstart)+(45:\linelength)$);

\draw[line] (lineRstart) -- (lineRend);

% square-with-dot at bottom
\node[sqdot] (sqR) at (lineRstart) {};
\node[dot] at (sqR.center) {};

% Right labels
\node[right] at ($(lineRstart)+(0.6,0.5)$) {$\mathscr{A}_{\sqrbulletdiagup}$};
\node[right] at ($(lineRend)+(0.25,0.1)$) {};

% --- LEFT RAY (135°) ---
\coordinate (lineLstart) at (-1,0);
\coordinate (lineLend) at ($(lineLstart)+(135:\linelength)$);

\draw[line] (lineLstart) -- (lineLend);

% square-with-dot at bottom
\node[sqdot] (sqL) at (lineLstart) {};
\node[dot] at (sqL.center) {};

% Left labels
\node[left] at ($(lineLstart)+(-0.6,0.5)$) {$\mathscr{A}_{\sqrbulletdiagdown}$};
\node[left] at ($(lineLend)+(-0.25,0.1)$) {};

% --- CENTRAL OVAL ---
\node[oval] at (0,-0.3) {$(\bar{q}_{AB},\, m\rvert_{\mathcal{C}})$};

\end{tikzpicture}
\caption{The CIVP algebra prior to gluing.}
\label{fig:CIVPDeconstructed2intro}
\end{figure}
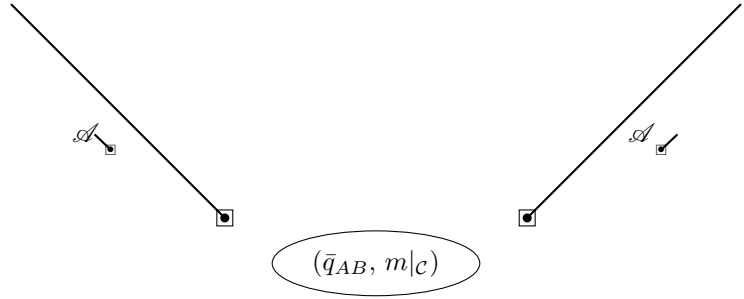

Applying the same procedure to both hypersurface theories, we arrive at the situation described by Figure~\ref{fig:CIVPDeconstructed2intro}. We have two on-shell algebras $\mathscr{A}_{\sqrbulletdiagup}$ and $\mathscr{A}_{\sqrbulletdiagdown}$ satisfying the full set of Einstein equations on either hypersurface. However, this set up is not yet a complete quantization of the CIVP. Comparing with Figure~\ref{fig:CIVPIntro}, we see that this set up overcounts the number of unique degrees of freedom and fails to account for the non-trivial interaction between the two hypersurfaces as encoded in the remaining CIVP data $\bigcirc$. 

We address these two features by gluing together the hypersurface algebras as
\beq
	\mathscr{A}_{\sqrbulletV} = \bigg(\mathscr{A}_{\sqrbulletdiagdown} \otimes_{\bigcirc} \mathscr{A}_{\sqrbulletdiagup}\bigg)/\sim_{\bullet},
\eeq
where $\sim_{\bullet}$ is an equivalence relation. This equivalence relation can be seen to implement two quotients. It enforces at the corner
\beq
	\op{\Omega}_{(u)} \sim_{\bullet} \op{\Omega}_{(v)}, \qquad \op{\pi}^{(u)}_a \sim_{\bullet} \op{\pi}^{(v)}_a,
\eeq
so that we have only one set of pure corner degrees of freedom in the full CIVP algebra. At the same time, the relative tensor product $\otimes_{\bigcirc}$ encodes the non-trivial action of a $u$-supertranslation on a $v$-supertranslation (and vice versa)
\beq
	[\op{\theta}_{(u)}, \op{\theta}_{(v)}] = \mathcal{F}_{\bigcirc}(\op{\theta}_{(u)}, \op{\theta}_{(v)}, \op{1}),
\eeq
As the notation indicates, this commutation relation is determined by the data $\bigcirc$ in the CIVP.\\

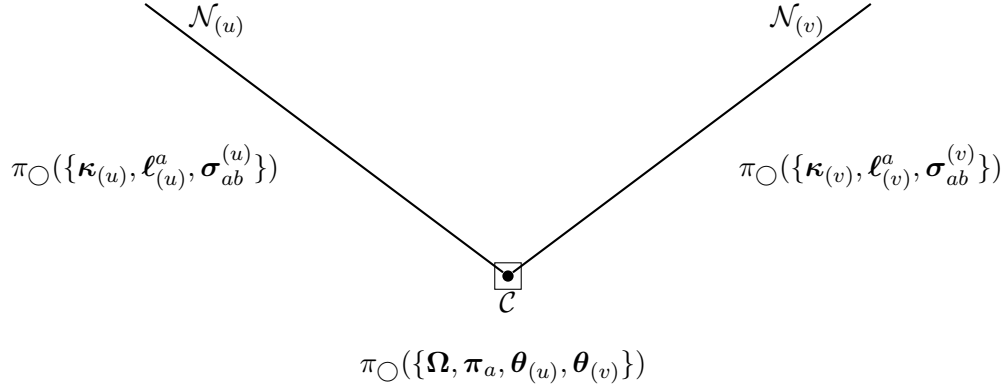
\begin{figure}[ht] 
\centering
\begin{tikzpicture}[
    scale=1.2,
    every node/.style={align=center},
    line/.style={thick},
    point/.style={circle, fill=black, inner sep=1.5pt}
]

% Intersection point
\node[point, label=below:$\mathcal{C}$] (C) at (0,0) {};

% Left and right lines (forming a V)
\draw[line] (C) -- (-4,3);
\draw[line] (C) -- (4,3);

% Labels for the null hypersurfaces
\node at (-3.2,2.8) {$\mathcal{N}_{(u)}$};
\node at (3.2,2.8) {$\mathcal{N}_{(v)}$};

\node [draw, rectangle, minimum width=0.35cm, minimum height=0.35cm] at (C) {};

% Data on left line
\node at (-4,1.2) {$\pi_{\bigcirc}(\{\op{\kappa}_{(u)}, \op{\ell}_{(u)}^a, \op{\sigma}^{(u)}_{ab}\})$};

% Data on right line
\node at (4,1.2) {$\pi_{\bigcirc}(\{\op{\kappa}_{(v)}, \op{\ell}_{(v)}^a, \op{\sigma}^{(v)}_{ab}\})$};

% Data at the intersection point
\node at (0,-1) {
$\pi_{\bigcirc}(\{\op{\Omega},\op{\pi}_a, \op{\theta}_{(u)}, \op{\theta}_{(v)}\})$
};
\end{tikzpicture}
\caption{Operator algebra of the CIVP.}
\label{fig:OA_CIVPIntro}
\end{figure}

We therefore arrive at our proposed operator-algebraic quantization of the CIVP which takes the form as indicated in Figure~\ref{fig:OA_CIVPIntro}. 

As we heuristically derived in this Introduction, the operator algebra $\mathscr{A}_{\sqrbulletV}$ is obtained by gluing the quantum data of two null hypersurfaces. Each hypersurface carries various pieces appropriately joined together. We recollect them in the table below.
\begin{table}[H]
\centering
\begin{tabular}{llll}
\hline
\textbf{Classical data} & \textbf{Symmetry/charge} & \textbf{Quantum object} & \textbf{Role} \\
\hline
Data on $\cN$ 
& Radiative/shear data 
& $\mathscr{A}_{\diagup}$ 
& Kinematical null algebra \\

$\Omega,\pi_a$ 
& Superboosts/superrotations 
& $\mathscr{A}_{\bullet}$ 
& Corner algebra / constraints \\

$\theta$ 
& Supertranslations 
& $\mathscr{A}_{\square}$ 
& CP-map extension \\

$h_{ij},m$ 
& Gluing data 
& $\otimes_{\bigcirc}$ 
& Double-null coupling \\
\hline
\end{tabular}
\caption{Classical and quantum organization of the CIVP data.}
\label{tab:civp-quantization-dictionary}
\end{table}
The outcome of our construction is thus the operator algebra
$\mathscr{A}_{\sqrbulletV}$, which we propose as the quantum algebra associated with the double-null characteristic data. In this sense, the paper brings
together several structures that have so far mostly been treated separately: the quantization of null hypersurface degrees of freedom, the algebraic implementation of gravitational constraints, the organizing structure of the corner symmetry, the non-automorphic action of
supertranslations, and the gluing of two null branches across a shared corner. The novelty of the present work is to organize these ingredients into a single operator-algebraic construction adapted to the full double-null CIVP: the two null hypersurface algebras are first constructed separately, the supertranslation sector is incorporated through completely positive maps and their Stinespring dilation, the gravitational constraints are
implemented through crossed products, and the two branches are finally glued at the corner. The operator algebra $\mathscr{A}_{\sqrbulletV}$ is therefore the
central contribution of the paper.

\section{The Classical CIVP} \label{sec: Classical}

In this section, we review and reformulate the characteristic initial value problem of gravity. Our viewpoint is both intrinsic to the initial hypersurfaces and embedded in the bulk spacetime. We begin by introducing the relevant fields and geometric ingredients on the null hypersurfaces, and by stating the CIVP theorem. We then develop a diagrammatic language that cleanly separates the different components of the CIVP, each of which will later play a distinct role. With this structure in place, we construct the kinematical phase space and identify the associated off-shell symplectic data. We next review the construction of symmetries and charges in the covariant phase space formalism, with particular emphasis on the contrast between the symplectomorphic action of superboosts and superrotations and the non-symplectomorphic action of supertranslations. We conclude by summarizing the full symmetry structure of a single null hypersurface and the main lessons that will be carried forward.

\subsection{Geometry, Equations of Motion, and Initial Data} \label{sec: Initial Data}

In this subsection, we review the salient features of null hypersurfaces in the language of Carrollian geometry, and then discuss the gravitational equations of motion which are realized in terms of the Raychaudhuri and Damour constraints. This leads naturally to the identification of the main ingredients of the CIVP, which we then formulate and analyze. A central outcome of this subsection is the introduction of a precise and useful diagrammatic language for the CIVP, which will play a pivotal role in the quantum section that follows.

\subsubsection{Geometry of a Null Hypersurface}

As in the spacelike initial value problem, the CIVP identifies the data on an initial hypersurface
that determine the corresponding spacetime development. However, the initial data of the CIVP consists of a pair of null hypersurfaces intersecting at a common initial codimension-$2$ surface, henceforth called the initial corner -- or cut -- $\cC$. The first step is thus to consider the intrinsic geometry of a null hypersurface. The double-null initial surface is then obtained by taking two such hypersurfaces and imposing the
appropriate gluing conditions at their common corner. \\

We first review the intrinsic definition of a null hypersurface using the Carrollian framework. We defer the reader to \cite{Ciambelli:2025unn} for notation and more references on the subject. A  Carroll structure $(\mathcal{N}, q, \un{\ell})$ is a $(d+1)$-dimensional\footnote{We use $d$ to denote the dimension of corners of $d+1$-dimensional hypersurfaces in a $d+2$-dimensional ambient spacetime, the bulk.} manifold, $\mathcal{N}$, equipped with a degenerate corank-$1$ metric $q$ and a nowhere-vanishing null vector field $\un{\ell}$ satisfying
\beq
	i_{\un{\ell}} q = 0\,.
\eeq
Therefore, the metric-dual one-form of $\un\ell$ is zero. It is convenient to introduce a tangent-bundle dual one-form $k$ satisfying $i_{\un{\ell}} k = 1$. This one form plays the role of an Ehresmann connection on $\mathcal{N}$, thereby providing a notion of horizontal subbundle. A vector $\un{\xi} \in T\mathcal{N}$ can thus be decomposed in its vertical and horizontal parts as $\un{\xi} = f \un{\ell} + \un{Y}$ with $i_{\un{Y}}k = 0$ and $f \in C^{\infty}(\mathcal{N})$. The Carrollian manifold $\mathcal{N}$ can be elegantly described as fibered over a $d$-dimensional base manifold, see \cite{Ciambelli:2019lap}. The exterior derivative of $k$ defines the Carrollian acceleration $\varphi$ and vorticity $\varpi$,
\beq\label{dk}
\rd k=-(k\wedge \varphi+\varpi)\,,
\eeq
which are both horizontal. In particular, $\varphi=-\cL_{\un\ell}k$. In the following, we will identify $k$ with the one-form normal to the foliation of cuts $\cC_{u(x)}$, such that $k= \rd (u-u(x))$. This conveniently sets $\varphi=0=\varpi$.\\

The degenerate metric $q$ can be restricted to cuts $\cC$ of $\cN$, where it gives rise to a standard Riemannian metric $h$. To be concrete, using coordinates $x^a=(u,x^i)$ on $\cN$, the degenerate metric
\beq
\rd s^2=q_{ab}(u,x) \rd x^a \rd x^b\,,
\eeq
can be evaluated on a generic cut $\cC_{u(x)}$ and gives
\begin{equation}
\begin{aligned}
\rd s^2\vert_{\cC_{u(x)}}&=q_{uu}(u(x),x) \pa_i u(x) \rd x^i \pa_j u(x) \rd x^j+2q_{uj}(u(x),x) \pa_i u(x) \rd x^i  \rd x^j+q_{ij}(u(x),x) \rd x^i \rd x^j\\
&=h_{ij}(u(x),x)\rd x^i \rd x^j\,,
\end{aligned}
\end{equation}
such that $h_{ij}$ is the non-degenerate Riemannian metric induced on cuts. Denoting by $\ve_{\mathcal{C}}$ the volume form induced by the metric $q$ on $\mathcal{C}$, we can define a volume form $\ve_{\mathcal{N}} = k \wedge \ve_{\mathcal{C}}$ on $\mathcal{N}$. Given any null hypersurface embedded in a spacetime, it is always possible to define a Carroll structure by appropriate projection \cite{Freidel:2022vjq}. This is conveniently done using the Rigging formalism \cite{Mars:1993mj}.\\

The second fundamental tensor on $\cN$ is
\beq
	\theta_{ab} = \frac{1}{2} \mathcal{L}_{\un{\ell}} q_{ab}\,,
\eeq
which is symmetric and horizontal, since $\ell^a\theta_{ab}=0$.
To extract its trace and traceless parts, we must construct the mixed-indices tensor $\theta_a{}^b$. Given a rank-$n$ covariant tensor $T_{a_1 ... a_n}$, we can construct mixed-indices tensors by enforcing the properties
\beq
	T_{a_1 ... a_n}{}^b q_{bc} = T_{a_1 ... a_n c}\,, \qquad T_{a_1 ... a_n}{}^b k_b = 0\,,
\eeq	
which amount to first projecting to the horizontal and then raising indices there. This, importantly, does not provide inverses, as there is a projection involved. For instance, one can construct 
\beq\label{proj}
	q_a{}^b = \delta^b_a - k_a \ell^b\,, \qquad q_a{}^bq_{bc}=q_{ac}\,,\qquad q_a{}^bk_b=0\,,
\eeq
which, by construction, is a projector ($q_a{}^c q_c{}^b=q_a{}^b$), and satisfies $q_a{}^a=d$.
One can similarly construct the tensor $\theta_a{}^b$, which admits a decomposition as
\beq
	\theta_a{}^b = \sigma_a{}^b + \frac{\theta}{d} q_a{}^b\, , 
\eeq 
where $\theta=\theta_a{}^a$ is the expansion and $\sigma_a{}^b$ is the traceless shear. These quantities will play an important role in the gravitational constraints on $\cN$; the topic which we turn to now. 

\subsubsection{The Gravitational Constraints}

The projection of the Einstein equations to $\cN$ gives rise to the Raychaudhuri and Damour constraints. To understand them intrinsically,\footnote{More precisely, the Raychaudhuri equation is independent of the connection while the explicit form of the Damour equation, since it is a horizontal constraint, depends on the horizontal covariant derivative chosen.} we need to specify the covariant derivative chosen on $\cN$. On a null manifold, there is no analogue of the Levi-Civita theorem, and thus no preferred connection. Since we are interested in embedding this manifold as a null hypersurface in an ambient pseudo-Riemannian bulk, one can choose to work with the induced connection from the bulk Levi-Civita connection. This is the Rigging connection \cite{Mars:1993mj} which, from a purely intrinsic standpoint, is the standard Carrollian connection thoroughly discussed in \cite{Ciambelli:2025unn}. This connection is torsionless but not metric compatible. Moreover, it is not entirely dictated by intrinsic geometric data, and thus one must supplement extra information to characterize it. Its under-determinacy is controlled by a scalar $\kappa$ and a horizontal one form $\pi_a$.\footnote{The action of the connection on $k$ also involves the symmetric horizontal tensor $\overline{\theta}_{ab}$, see \cite{Ciambelli:2025unn}. This tensor parametrizes part of the freedom in the choice of rigging, not an independent datum of the CIVP. In the present discussion we fix this freedom by choosing $\overline{\theta}_{ab}=0$.} Once the null hypersurface is embedded in the bulk, $\kappa$ is the inaffinity of $\un\ell$ and $\pi_a$ is the Hájiček one form \cite{Chandrasekaran:2018aop}. The standard Carrollian connection features these tensors in its action on the geometric data
\begin{equation}
\begin{aligned}
	D_aq_{bc} &=-k_b \theta_{ac}-k_c\theta_{ab}\\
    D_a\ell^b&= \theta_a{}^b+\omega_a\ell^b\,,
\end{aligned}
\end{equation}
where we introduced the one-form $\omega_a=\kappa k_a+\pi_a$. \\

The gravitational constraints on a null hypersurface are obtained by suitably projecting Einstein equations on it. The vertical and horizontal projections give the Raychaudhuri and Damour equations, respectively. In the vacuum, they can be written using the data above as
\beqn\label{RayDam}
	(\mathcal{L}_{\un{\ell}} + \theta) \theta &= \mu \theta - \sigma_a{}^b \sigma_b{}^a\,,  \\
	(\mathcal{L}_{\un{\ell}} + \theta) \pi_a &=q_a{}^d  q_e{}^c D_c(\mu q_d{}^e- \sigma_d{}^e)\,, 
\eeqn
where we have defined $\mu = \kappa + \frac{d-1}{d} \theta$. We now wish to understand which data one must specify in order to solve these constraints on $\cN$.\\

To do so, we need to isolate the unimodular part of the degenerate metric $q$. There is another reason for doing so. The shear $\sigma_a{}^b$ is a symmetric, traceless, horizontal tensor. It thus has $\frac{d^2+d-2}{2}$ independent components. The degenerate metric is a symmetric and horizontal tensor, thus containing $\frac{(d+1)d}{2}$ components. This means that $q$ contains an extra independent component compared to $\sigma$, which indeed is its conformal factor. Since we will later show that the unimodular part of $q$ and $\sigma$ are symplectic partners, it is important to isolate the unimodular part of $q$. This is achieved by decomposing $q$ as
\beq\label{uni}
	q_{ab} = \Omega^{2/d} \bq_{ab}\,.
\eeq
That is, $\Omega = \sqrt{|q|}$ is the infinitesimal area element, with $|\bq| = 1$.\footnote{Since $q_{ab}$ has zero determinant, by $|q|$ we mean the determinant of the first minor, which is by construction non-vanishing as $q$ has rank $d$.} Note that, on a cut $\cC$, by design one has $\Omega\vert_\cC=\sqrt{|h|}$, with $h_{ij}=\Omega^{2/d}\vert_\cC\overline{h}_{ij}$. 

It is then straightforward to show that\footnote{Since $\Omega$ is a density, $\mathcal{L}_\ell\Omega=\partial_a(\ell^a\Omega)$, so that
$\mathcal{L}_\ell\log\Omega=\Omega^{-1}\partial_a(\ell^a\Omega)$.}
\beq\label{exp}
	\theta = \mathcal{L}_{\un{\ell}} \log \Omega\,, 
\eeq
and
\beq\label{sigma}
	\sigma_{ab} = \frac{\Omega^{2/d}}{2} \mathcal{L}_{\un{\ell}} \bq_{ab}\,. 
\eeq
Thus, we may regard $\theta$ and $\sigma_{ab}$ as the modular and unimodular contributions to the derivative of $q_{ab}$ which are conjugate to $\Omega$ and $\bq_{ab}$, respectively. \\

With these geometric ingredients, we can now discuss the data required to solve \eqref{RayDam}.
Consider a null hypersurface $\cN$ with an initial cut $\cC$. Given \eqref{exp}, the Raychaudhuri equation -- the first equation in \eqref{RayDam} -- is a second-order differential equation for $\Omega$. A key ingredient is the choice of null time $\un\ell$, which must be specified on $\cN$ as part of the initial setup. Once $\un\ell$ is fixed, the equation is solved by prescribing the extrinsic datum $\kappa$ and the shear $\sigma_a{}^b$ on $\cN$, together with the two integration constants $\theta|_\cC$ and $\Omega|_\cC$ at the initial cut.

Next consider the Damour equation -- the second equation in \eqref{RayDam}. This is a first-order differential equation for $\pi_a$. Its coefficients depend on $\theta$, now determined from the Raychaudhuri equation, and on the prescribed data $\sigma_a{}^b$ and $\kappa$ on $\cN$. The remaining integration constant is the value of $\pi_a$ at the initial cut, $\pi_a|_\cC$. 

Thus, \eqref{RayDam} are solved specifying $(\kappa,\un\ell,\sigma_{ab})$ on $\cN$, and $(\theta,\Omega,\pi_a)$ at the initial cut $\cC$. As we will shortly see, this discussion provides already the main ingredients of the CIVP.

\subsubsection{The Characteristic Initial Value Problem Theorem} \label{subsec:CIVPthm}

We are now fully equipped to state the CIVP theorem. Let us first review the geometric setup. The CIVP requires two intersecting null hypersurfaces, which means that the above data must be specified twice, together with the conditions that enforce a smooth gluing at the corner $\cC$, which is the initial cut of both hypersurfaces. It is important to be precise about the future domain of validity of this construction. In a generic null congruence, once the expansion becomes negative, the Raychaudhuri equation \eqref{RayDam} -- in vacuum or assuming NEC -- implies focusing of the null generators as one evolves along the null direction. This leads to the formation of caustics where neighboring generators intersect and the foliation description breaks down, with $\Omega\to 0$.\footnote{As one approaches a caustic, the induced volume element on the cuts tends to zero. Therefore, provided a $d$-form ${\cal P}$ remains finite there, the future boundary contribution vanishes and Stokes' theorem on $\cN$ reduces to
\beq\label{Stokes}
\int_{\cN}\rd {\cal P}=-\int_{\cC} {\cal P}\vert_{\cC}\,.
\eeq} From the intrinsic viewpoint, the Carrollian structure defined above only covers the portion of the null hypersurface prior to the first caustic. Accordingly, from the CIVP perspective, the theorem below applies only from the initial cut up to the formation of the first caustic. The physics and geometry of caustics are subtle and important, see e.g. \cite{Gadioux:2023pmw}. Extending both the CIVP and our intrinsic geometric description beyond caustics remains an active area of investigation.\\

Given these remarks, we can now define the data of the CIVP as follows. First, denoting the null times on each null hypersurface as $u$ and $v$, we introduce two null hypersurfaces $\mathcal{N}_{(i)}$ with $i = u, v$, which share a common initial cut $\mathcal{C}$, the corner. Then, we endow this geometric setup with the following set of data:
\begin{enumerate}
	\item \textbf{On the corner $\mathcal{C}$:}
	\begin{enumerate}
		\item A Riemannian metric $h_{ij}$ with unimodular part $\overline{h}_{ij}$ and volume element $\Omega\rvert_{\mathcal{C}}$,
		\item The value of the expansion scalars $\theta_{(u)}\rvert_{\mathcal{C}}$ and $\theta_{(v)}\rvert_{\mathcal{C}}$,
		\item The value of the Hájiček one-form $\pi_a\rvert_{\mathcal{C}}$,
		\item A scalar field $m\rvert_{\mathcal{C}}$.
	\end{enumerate}
	\item \textbf{In the bulk of each hypersurface $\mathcal{N}_{(i)}$:}
	\begin{enumerate}
		\item The inaffinity $\kappa_{(i)}$,
		\item The clock vector field $\un\ell_{(i)}$, 
		\item The shear $\sigma^{(i)}_{ab}$. 
	\end{enumerate}
\end{enumerate}
Note that in the CIVP the shear $\sigma^{(i)}_{ab}$ is often replaced by the unimodular metric $\bq_{ab}$. This is because, given \eqref{sigma}, once the initial data are specified, the data on each $\cN$ can be thought of as $(\kappa_{(i)},\un\ell_{(i)},\sigma^{(i)}_{ab})$ or, equivalently, $(\kappa_{(i)},\un\ell_{(i)},\bq^{(i)}_{ab})$. We prefer the formulation in terms of $\sigma_{ab}$ as this relates more directly to the discussion on the resolution of the equations of motion in the previous subsection.\\

We shall denote the collection of data featuring the CIVP by 
\beq
\mathfrak{i}=\{h_{ij},\theta_{(i)}\rvert_{\mathcal{C}}, \pi_a\rvert_{\mathcal{C}},m\rvert_{\mathcal{C}}\ , \ \kappa_{(i)},\un\ell_{(i)},\sigma^{(i)}_{ab}\}.
\eeq
In the spacetime picture, this would correspond to a pair of embedded null hypersurfaces intersecting at a corner. Let us denote by $\mathfrak{i}[M,g]$ the structure induced by such an embedding. Note that the only data in $\mathfrak{i}$ that we did not involve in the resolution of the constraints on each single null hypersurface are $\overline{h}_{ij}$ and $m\rvert_{\mathcal{C}}$. These data are needed to specify the gluing of the two null hypersurfaces at the corner. In the bulk spacetime picture, $e^{-m}$ determines the angle between the two clock vector fields
\beq
	-e^{-m} = \un{\ell}_{(u)} \cdot \un{\ell}_{(v)}\,.
\eeq
Thus, $\overline{h}_{ij}$ instructs us about the local geometry on $\cC$, whereas $m\vert_\cC$ tells us the angle between the two null hypersurfaces. While they are not dynamical data on the null hypersurfaces, they are required to determine how these two are joined at the initial corner.\\

With all this machinery in place, the content of the CIVP can be simply stated as follows:
\begin{theorem}[CIVP] Given sufficiently regular characteristic initial data $\mathfrak{i}$ satisfying the appropriate
corner compatibility conditions, there exists a unique spacetime development $(M,g)$, up to
gauge equivalence, admitting a pair of intersecting embedded null hypersurfaces such that
$\mathfrak{i}=\mathfrak{i}[M,g]$ and $(M,g)$ solves the vacuum Einstein equations in the local
future domain of dependence of the initial null hypersurfaces, prior to the formation of caustics.
\end{theorem}

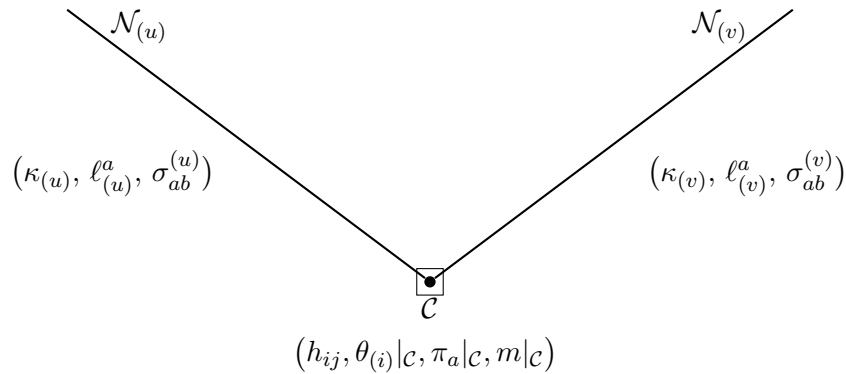
\begin{figure}[H] \label{fig:CIVP}
\centering
\begin{tikzpicture}[
    scale=1.2,
    every node/.style={align=center},
    line/.style={thick},
    point/.style={circle, fill=black, inner sep=1.5pt}
]

% Intersection point
\node[point, label=below:$\mathcal{C}$] (C) at (0,0) {};

% Left and right lines (forming a V)
\draw[line] (C) -- (-4,3);
\draw[line] (C) -- (4,3);
\node [draw, rectangle, minimum width=0.35cm, minimum height=0.35cm] at (C) {};

% Labels for the null hypersurfaces
\node at (-3.2,2.8) {$\mathcal{N}_{(u)}$};
\node at (3.2,2.8) {$\mathcal{N}_{(v)}$};

% Data on left line
\node at (-3.5,1.2) {$\big(\kappa_{(u)},\, \ell_{(u)}^a,\, \sigma^{(u)}_{ab}\big)$};

% Data on right line
\node at (3.5,1.2) {$\big(\kappa_{(v)},\, \ell_{(v)}^a,\, \sigma^{(v)}_{ab}\big)$};

% Data at the intersection point
\node at (0,-0.8) {
$\big(
h_{ij},\theta_{(i)}\rvert_{\mathcal{C}}, \pi_a\rvert_{\mathcal{C}},m\rvert_{\mathcal{C}}
\big)$
};

\end{tikzpicture}
\caption{Pictorial representation of the CIVP. The indicated data determine the spacetime metric in the local future development of the shared corner,
within the domain where the null foliation remains smooth.}
\end{figure}

A convenient way to read the CIVP is depicted in Figure \ref{fig:CIVPDeconstructed}. To begin, we imagine that we have two null hypersurface theories specified by a priori independent Carroll structures. We then specify the data required to make Raychaudhuri and Damour well posed for these two `theories' individually. The sources, which are supported in the full null hypersurfaces, are represented by the outstretched lines $\diagdown$ and $\diagup$. The expansion, which is supported on the corner but depends on the clock vector field, is represented by a box $\square$. The volume element and Hájiček are represented by a dot $\bullet$. Since at this stage we are treating the two hypersurfaces as independent we have included two copies of this data. We know from the CIVP, however, that this is an overcounting: eventually the two dots will have to be identified as the shared initial cut. The remaining data, $(\overline{h}_{ij}, m\rvert_{\mathcal{C}})$, are represented by a single oval $\bigcirc$ between the two hypersurfaces and specify the local geometry of the corner, and the clocks' scalar product. These data are used to link together the two a priori independent hypersurface theories. Eventually, the gluing procedure will be interpreted as an operation which both sets the two dots equal (by quotient) to each other and determines the necessary algebraic relations between the two independent expansions. 

\begin{center}
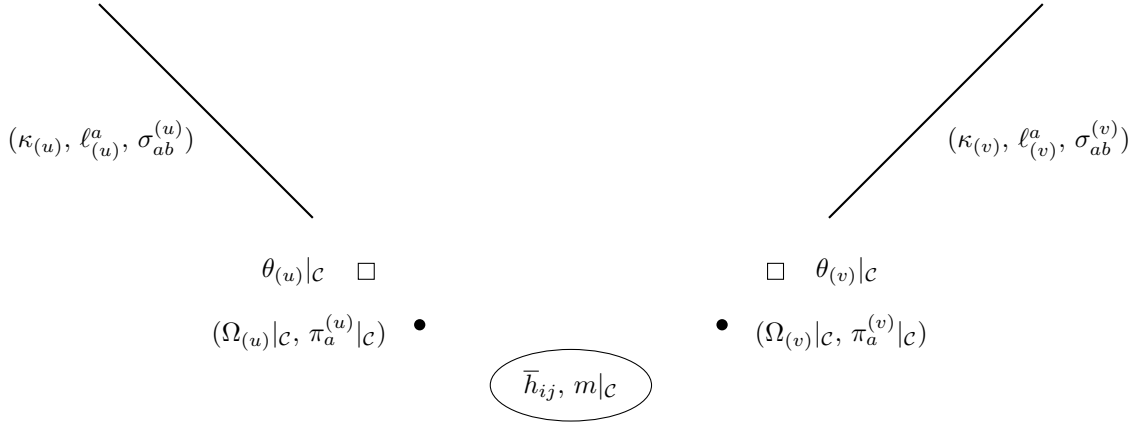
\begin{figure}[H] 
\centering
\begin{tikzpicture}[
    scale=2.0,
    line/.style={thick},
    box/.style={draw, rectangle, minimum size=6pt, inner sep=0pt},
    dot/.style={circle, fill=black, inner sep=1.5pt},
    oval/.style={draw, ellipse, minimum width=1.2cm, minimum height=0.5cm},
    every node/.style={align=center, font=\small}
]

% --- Spacing parameters ---
\def\gapdotbox{0.5}
\def\gapboxline{0.5}
\def\linelength{2}
\def\ovalgap{0.7}

% --- RIGHT RAY (45°) ---
\coordinate (dotR) at (1,0);                     
\coordinate (boxR) at ($(dotR)+(45:\gapdotbox)$);
\coordinate (lineRstart) at ($(boxR)+(45:\gapboxline)$);
\coordinate (lineRend) at ($(lineRstart)+(45:\linelength)$);

\node[dot] at (dotR) {};
\node[box] at (boxR) {};
\draw[line] (lineRstart) -- (lineRend);

% Right labels (aligned with elements)
\node[right] at ($(dotR)+(0.15,-0.05)$) {$(\Omega_{(v)}\rvert_{\mathcal{C}},\, \pi^{(v)}_a\rvert_{\mathcal{C}})$};
\node[right] at ($(boxR)+(0.2,0)$) {$\theta_{(v)}\rvert_{\mathcal{C}}$};  % aligned with box
\node[right] at ($(lineRend)+(-0.7,-0.9)$) {$(\kappa_{(v)},\, \ell_{(v)}^a,\, \sigma^{(v)}_{ab})$};

% --- LEFT RAY (135°) ---
\coordinate (dotL) at (-1,0);                    
\coordinate (boxL) at ($(dotL)+(135:\gapdotbox)$);
\coordinate (lineLstart) at ($(boxL)+(135:\gapboxline)$);
\coordinate (lineLend) at ($(lineLstart)+(135:\linelength)$);

\node[dot] at (dotL) {};
\node[box] at (boxL) {};
\draw[line] (lineLstart) -- (lineLend);

% Left labels (aligned with elements)
\node[left] at ($(dotL)+(-0.15,-0.05)$) {$(\Omega_{(u)}\rvert_{\mathcal{C}},\, \pi^{(u)}_a\rvert_{\mathcal{C}})$};
\node[left] at ($(boxL)+(-0.2,0)$) {$\theta_{(u)}\rvert_{\mathcal{C}}$};  % aligned with box
\node[left] at ($(lineLend)+(0.7,-0.9)$) {$(\kappa_{(u)},\, \ell_{(u)}^a,\, \sigma^{(u)}_{ab})$};

% --- CENTRAL OVAL ---
\node[oval] at (0,-0.4) {$\bh_{ij},\, m\rvert_{\mathcal{C}}$};

\end{tikzpicture}
\caption{A convenient way to organize the data of the CIVP.}
\label{fig:CIVPDeconstructed}
\end{figure}
\end{center}

In other words, from this perspective we can view the entire CIVP as the product of two initial data structures for a priori independent null hypersurface theories up to a quotienting operation that governs the relation between data at the shared corner. Pictorially:

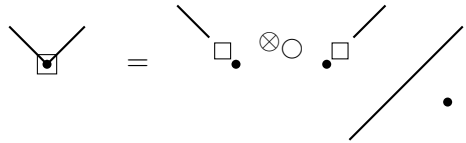
\begin{figure}[H] 
\centering
\begin{tikzpicture}[
    line/.style={thick},
    box/.style={draw, rectangle, minimum size=6pt, inner sep=0pt},
    dot/.style={circle, fill=black, inner sep=1.2pt},
    oval/.style={draw, ellipse, minimum width=0.6cm, minimum height=0.25cm}
]

% --- LHS: V diagram ---
\coordinate (Vdot) at (0,0);
\draw[line] (-0.5,0.5) -- (Vdot);
\draw[line] (0.5,0.5) -- (Vdot);
\node[dot] at (Vdot) {};

\node [draw, rectangle, minimum width=0.15cm, minimum height=0.15cm] at (Vdot) {};

% --- Equal sign ---
\node at (1.2,0) {$=$};

% --- RHS: tensor product numerator "A" ---
% Left sequence (reflected)
\coordinate (dotL) at (2.5,0);
\coordinate (boxL) at ($(dotL)+(135:0.25)$);
\coordinate (lineLstart) at ($(boxL)+(135:0.25)$);
\coordinate (lineLend) at ($(lineLstart)+(135:0.6)$);
\node[dot] at (dotL) {};
\node[box] at (boxL) {};
\draw[line] (lineLstart) -- (lineLend);

% Right sequence
\coordinate (dotR) at (3.7,0);
\coordinate (boxR) at ($(dotR)+(45:0.25)$);
\coordinate (lineRstart) at ($(boxR)+(45:0.25)$);
\coordinate (lineRend) at ($(lineRstart)+(45:0.6)$);
\node[dot] at (dotR) {};

\node[box] at (boxR) {};
\draw[line] (lineRstart) -- (lineRend);

% Tensor product symbol
\node at (3.1,0.25) {$\otimes_{\bigcirc}$};

% Forward-slash for quotient (longer, 45°)
\coordinate (slashStart) at (4,-1);
\coordinate (slashEnd) at (5.5,0.5); % longer, 45° angle
\draw[line] (slashStart) -- (slashEnd);

% Oval for equivalence relation (~), lowered
\node[dot] at (5.3,-0.5) {};

\end{tikzpicture}
\caption{CIVP as quotient.}
\label{fig:CIVP = CIVPDecon}
\end{figure}

\noindent The mathematical structure described in figure \ref{fig:CIVP = CIVPDecon} will be our guidepost for the remainder of the manuscript. For the rest of this section, it will allow us to identify the classical features associated with the different parts of the data $\mathfrak{i}$. Then, it will instruct us on how to quantize the full CIVP by first quantizing the separate pieces kinematically, and subsequently putting them back together by implementing the gravitational constraints and our gluing procedure. 

\subsection{Phase Space on a Null Hypersurface}\label{2.2}

In this subsection, we introduce the classical structure associated with a single line in Fig.~\ref{fig:CIVPDeconstructed}. This is the structure carried by an individual null hypersurface embedded in spacetime. More precisely, we construct the corresponding kinematical phase space and identify its off-shell symplectic data. This phase space admits two complementary interpretations: from the ambient spacetime perspective, the null hypersurface is an embedded geometric object; from the Carrollian perspective, it is treated intrinsically. Our goal is to describe this phase space in a clear and precise way, isolating the natural symplectic pairs that organize its degrees of freedom. \\

Following \cite{Ciambelli:2023mir,Odak:2023pga,Chandrasekaran:2021vyu},\footnote{See \cite{Iyer:1994ys} for the original reference on the Noether formalism employed here.} the canonical presymplectic potential associated with the Carroll structure on $\mathcal{N}$ can be written in the form
\beq\label{th}
	\Theta_{\mathcal{N}} = \int_{\mathcal{N}} \ve_{\mathcal{N}} \bigg(\frac{1}{2} \tau^{ab} \delta q_{ab} - \tau_a \delta \ell^a\bigg)\,.
\eeq
Here, 
\beq
	\tau_{ab} = \frac{1}{8 \pi G} (\sigma_{ab} - \frac{1}{2} \mu q_{ab})\,, \qquad \tau_a = \frac{1}{8 \pi G} (-\theta k_a + \pi_a)\,,
\eeq
can be interpreted as stresses which are conjugate to the geometric variables $(q_{ab}, \ell^a)$, a covariantization of the Carrollian momenta introduced in \cite{Ciambelli:2018ojf}. The Einstein-Hilbert potential pulled back to the null hypersurface gives rise to the presymplectic potential \eqref{th}, plus $\delta$- and $\rd$-exact terms. The covariance of the symplectic structure on null hypersurfaces has been studied at length in \cite{Odak:2023pga}.\\

The symmetries of this phase space consist of diffeomorphisms on the null hypersurface Diff$\cN$, rescaling of the null clock $\un\ell$, and shifts of the Ehresmann connection $k$ \cite{Ciambelli:2025unn}.\footnote{Here, the terminology symmetry is used in a loose sense, see \cite{Ciambelli:2022vot}.} The shift symmetry of the Ehresmann connection is pure gauge in the present setup;
we therefore fix this freedom by taking $\delta k_a=0$, without loss of generality. A detailed discussion of this gauge fixing in the Carrollian phase-space language
will appear elsewhere, \cite{Ciambelli:ToAppear}. With this choice, and using the unimodular decomposition \eqref{uni}, the presymplectic potential can be recast in the form\footnote{In particular, we used that the shear is traceless, and that $\frac12 q^{ab}\delta q_{ab}=\delta\ln\Omega$.}
\beq\label{presymp}
	\Theta_{\mathcal{N}} = \frac{1}{8 \pi G} \int_{\mathcal{N}} \ve_{\mathcal{N}} \bigg(-\mu \delta \ln \Omega - \pi_a \delta \ell^a + \frac{1}{2} \sigma^{ab} \Omega^{2/d}\delta \bq_{ab}\bigg). 
\eeq
The volume form $\ve_{\cN}=k\wedge \ve_{\cC}$ is not a phase space constant, since $\delta \ve_{\cC}\neq 0$. Given the discussion below \eqref{uni}, we can isolate its field dependency $\ve_{\cN}=\Omega \ve^{(0)}_{\cN}$, where $\ve^{(0)}_{\cN}=\rd^{d+1}x$ is the bare measure, such that $\delta \ve^{(0)}_{\cN}=0$. Moreover, we can define the unimodular shear $\osigma^{ab}$ starting from \eqref{sigma} in the following way. The unimodular tensor $\bq^{ab}$ is given by $q^{ab}=\Omega^{d/2}\bq^{ab}$, since it must satisfy \eqref{proj}. Then, we have\footnote{Since $\sigma_{ab}$ is horizontal, raising the indices using our procedure outlined above is the same as using $q^{ab}$.}
\beq
\sigma^{ab}=q^{ac}\sigma_{cd}q^{bd}=\frac{\Omega^{2/d}}{2}q^{ac}q^{bd}\cL_{\un\ell}\bq_{cd}=\frac{\Omega^{d/2}}{2}\bq^{ac}\bq^{bd}\cL_{\un\ell}\bq_{cd}=\Omega^{d/2}\osigma^{ab}\,,
\eeq
where, by design, the unimodular shear $\osigma^{ab}=\frac{1}{2}\bq^{ac}\bq^{bd}\cL_{\un\ell}\bq_{cd}$ is independent of $\Omega$, and thus it is built using exclusively $\un\ell$ and $\bq_{ab}$.\\

Utilizing this, and taking a further phase space variation, we arrive at the presymplectic $2$-form
\beq \label{Hypersurface Symp}
	\Omega_{\mathcal{N}} = \frac{1}{8 \pi G} \int_{\mathcal{N}} \ve_{\mathcal{N}}^{(0)} \bigg(-\delta \mu \wedge \delta \Omega - \delta(\Omega \pi_a) \wedge \delta \ell^a + \frac{1}{2} \delta(\Omega \osigma^{ab}) \wedge \delta \bq_{ab}\bigg)\,. 
\eeq
The redefinition leading to $\osigma^{ab}$ is essential, since $\osigma^{ab}$ is the natural symplectic partner of $\bq_{ab}$. If we now treat both $\mu$\footnote{Note that $\kappa$ in $\mu$ is already an independent datum, so we could have understood $\mu$ as a field redefinition.} and $\osigma^{ab}$ as independent of $\Omega$, $\un\ell$, and $\bq_{ab}$, then \eqref{Hypersurface Symp} is in Darboux form, with three sets of conjugate pairs:
\begin{table}[H]
\centering
\begin{tabular}{c|c}
 & Symplectic Partners \\
\hline
Spin 0 & $(\Omega,\mu)$ \\
Spin 1 & $(\ell^a,\Omega\pi_a)$ \\
Spin 2 & $(\bq_{ab},\Omega\osigma^{ab})$ \\
\end{tabular}
\end{table}
As we have indicated, these organize according to their tensor structure into pairs of spin-0, spin-1, and spin-2 degrees of freedom. We also note the specific appearance of $\Omega$ in the spin-1 and spin-2 momenta $\Omega\pi_a$ and $\Omega\osigma^{ab}$. The spin-0 propagates into these symplectic data, making the construction of the Poisson brackets a non-trivial endeavor \cite{Reisenberger:2007ku,Ciambelli:2023mir, Ciambelli:ToAppear,Wieland:2017zkf, Adami:2023wbe}.\\

These symplectic data can be understood through the lens of the CIVP as follows. Once the relation between $\osigma^{ab}$ and $\bq_{ab}$ is imposed, the specification of $\kappa$, $\un\ell$, and $\bq_{ab}$ on $\cN$ determines all symplectic data except for $\Omega$, $\pi_a$, and $\theta$, the latter being contained in $\mu$. These are precisely the remaining data that the CIVP provides at the initial corner and subsequently evolves through the Einstein equations. This explains why, at the level of the kinematical phase space, $\Omega$ and $\pi_a$ must be supplemented on the entirety of $\cN$: unlike in the CIVP, the gravitational constraints have not yet been implemented. We therefore work first with the kinematical phase space, and impose the constraints only at the quantum level. This also allows us to properly account for the possible emergence of quantum anomalies \cite{Ciambelli:2024swv}.

To organize this structure, we rely on the aforementioned diagrammatic calculus that will be used extensively in sections devoted to quantization. The hypersurface data are summarized as follows: to each $\diagup$ in the CIVP we assign a null hypersurface $\mathcal{N}{\diagup}$ equipped with a kinematical phase space $(X{\diagup}, \Omega_{\diagup})$, defined by the presymplectic form \eqref{Hypersurface Symp}. Since the hypersurface data of the CIVP are doubled, the full kinematical contribution consists of two such phase spaces, $(X_{\diagdown}, \Omega_{\diagdown})$ and $(X_{\diagup},\Omega_{\diagup})$.

\subsection{Symmetries and Charges}

Having dealt with $\diagdown$ and $\diagup$, we now turn to the physics of the corner. Following the organization of Fig.~\ref{fig:CIVPDeconstructed}, we treat the corners of the two hypersurface theories as distinct, since we are working first at the kinematical level, and we want to construct two independent null hypersurfaces prior to applying the gluing procedure. As indicated in the figure, it is useful to split the degrees of freedom supported at the corner into two classes: the box $\square$, which identifies the expansion, and the point $\bullet$, which identifies the infinitesimal area element and the Hájiček one-form.

This distinction is already natural from the viewpoint of the CIVP. The expansion extends out into the hypersurface, at least infinitesimally, through its dependence on the clock vector field. As a result, even after imposing the gravitational constraints, the two expansion scalars remain distinct, whereas $\Omega$ and $\pi_a$ must be identified at the corner. We will now see that the same distinction is reflected in the symmetry structure. The degrees of freedom associated with $\bullet$ generate invertible symmetry transformations that preserve each hypersurface and act symplectomorphically on $\cN$. By contrast, the degree of freedom associated with $\square$ generates transformations that move the corner along the null direction, failing to act symplectomorphically on the hypersurface phase space.

\subsubsection{Superboosts and Superrotations}\label{subsub}

The phase space $(X_{\diagup}, \Omega_{\diagup})$ admits a symplectomorphic action of the group
\beq
	G_{\bullet} = \text{Diff}(\mathcal{C}) \ltimes C^{\infty}(\mathcal{C})_B \subset \text{Diff}(\mathcal{N}_{\diagup})\,. 
\eeq
That is, there exists a map
\beq
	a^{\bullet}: G_{\bullet} \times X_{\diagup} \rightarrow X_{\diagup}\,,
\eeq
such that $a^{\bullet}_{g}{}^* \Omega_{\diagup} = \Omega_{\diagup}$ for each $g \in G_{\bullet}$.\footnote{Infinitesimally, this corresponds to $\fL_{\hat{\un\xi}}\Omega_{\diagup}=0$, where $\hat{\un\xi}$ is the Hamiltonian vector field associated with the vector field $\un\xi$ in the algebra ${\mathfrak g}_{\bullet}$ of the group $G_{\bullet}$, and $\fL_{\hat{\un\xi}}=I_{\hat{\un\xi}}\delta+\delta I_{\hat{\un\xi}}$ is the phase space Lie derivative along $\hat{\un\xi}$, see \cite{Ciambelli:2022vot} for a review of the phase space calculus utilized here.} We recall that the notation $\mathcal{C}$ refers to the initial corner of the hypersurface $\mathcal{N}_{\diagup}$. The infinitesimal generators of the group $G_{\bullet}$ are vector fields $\un\xi$ living in the algebra ${\mathfrak g}_{\bullet}$ of the form
\beq
	\un{\xi}_{(f_B,Y)} = f_B \un{\ell} + \un{Y}\,,
\eeq
where $\un{Y} \in T \mathcal{C}$ and $f_{B} \in C^{\infty}(\mathcal{N}_{\diagup})$ satisfying
\beq\label{boo}
	f_B\rvert_{\mathcal{C}} = 0, \qquad \underbrace{\cL_{\un{\ell}} \cdots\cL_{\un{\ell}}}_{{n \ \text{times}}} f_B \rvert_{\mathcal{C}} = 0 \quad \forall \  n > 1\,. 
\eeq
We refer to diffeomorphisms generated by $f_B \un{\ell}$ as superboosts and diffeomorphisms generated by $\un{Y}$ as superrotations. Physically, the former corresponds to an angle-dependent boost, while the latter is a generic diffeomorphism of the corner. \\

To be explicit, suppose e.g. $\un\ell=\un\pa_u$ and the initial corner is located at $u=0$. Then, this vector field can be written as
\beq\label{symm}
\un\xi=\underbrace{b(x)}_{C^{\infty}(\mathcal{C})_B} u \ \un\pa_u +\underbrace{Y^i(x)}_{\text{Diff}(\mathcal{C})}\un\pa_i\,.
\eeq
Thus, we see that the data of $f_B$ is encoded in a function $b(x)$ on the corner, and so we will denote the collection of such functions by $C^{\infty}(\mathcal{C})_B$. One can directly inspect that these vector fields generate the algebra $\mathfrak{g}_{\bullet}$ via Lie bracket
\beq
[\un\xi_1,\un\xi_2]=[b_1(x)u\un\pa_u+Y^i_1(x)\un\pa_i,b_2(x)u\un\pa_u+Y^j_2(x)\un\pa_j]=\left(\un Y_1(b_2)-\un Y_2(b_1)\right)u\un\pa_u+[\un{Y}_1,\un{Y}_2]_{\cC}\,,
\eeq
where $\un Y(b)=Y^i(x)\pa_i b$, and $[\un{Y}_1,\un{Y}_2]_{\cC}=(Y^i_1\pa_i Y^j_2-Y^i_2\pa_i Y^j_1)\un\pa_j$ is the corner Lie bracket.\\

Returning to the general construction, we can employ the covariant phase space formalism to derive the charges and fluxes associated with these symmetries. Let $\hat{\un\xi}_{(f_B,Y)}$ denote the vector field on phase space which generates a superboost parameterized by $f_B$ (and thus determined by some $b\in C^{\infty}(\mathcal{C})_B$ as above) and a superrotation parameterized by $\un{Y} \in T\mathcal{C}$. Using the notation $\Theta_{\diagup}$ for \eqref{presymp}, and defining the local presymplectic potential as $\Theta_{\diagup}=\int_{\cN_\diagup}\theta_{\diagup}$, the associated local Noether current is given by\footnote{The Noether current of a spacetime vector field $\un{\xi}$ is given by
\beq\label{totj}
	j_{\un{\xi}} = I_{\hat{\un{\xi}}} \theta - i_{\un{\xi}} \mathscr{L}\,,
\eeq
where $\mathscr{L}$ is the Lagrangian of the theory. Since all of the vector fields we are considering are tangent to the null hypersurface the latter term always vanishes. We will return to this point in and around eq. \eqref{intch}.}
\beq\label{jxi}
	j_{(f_B,Y)} = I_{\hat{\un\xi}_{(f_B,Y)}} \theta_{\diagup}\,,
\eeq  
where we denoted $j_{(f_B,Y)}=j_{\un\xi_{(f_B,Y)}}$ to lighten the notation. The integrated Noether currents define functions on phase space:
\beq\label{ham}
	H_{(f_B,Y)} = \int_{\mathcal{N}_{\diagup}} j_{(f_B,Y)}\,. 
\eeq
For a symplectomorphism, $H_{(f_B,Y)}$ are precisely the Hamiltonian functions associated to the Hamiltonian vector fields $\hat{\un\xi}_{(f_B,Y)}$, that is,
\beq\label{symp}
	I_{\hat{\un\xi}_{(f_B,Y)}} \Omega_{\diagup}=- \delta H_{(f_B,Y)} \,. 
\eeq
This allows us to conclude that the collection $H_{(f_B,Y)}$ form a subset of observables in the Poisson algebra of $X_{\diagup}$ which are homomorphic to the Lie algebra of the group $G_{\bullet}$,
\beq
	\{H_{(f_B,Y)}, H_{(f'_B,Y')}\} = -H_{[(f_B,Y),(f'_B,Y')]}\,,
\eeq
where $[(f_B,Y),(f'_B,Y')]$ denotes the Lie bracket between the two vector fields: $[(f_B,Y),(f'_B,Y')]=[\un\xi_{(f_B,Y)},\un\xi_{(f'_B,Y')}]$, and $\{\cdot,\cdot\}$ are the Poisson brackets derived inverting \eqref{Hypersurface Symp}.\\

Noether's second theorem enforces that the Noether current \eqref{jxi} splits into two terms 
\beq\label{Noe}
	j_{(f_B,Y)} = c_{(f_B,Y)} + \rd q_{(f_B,Y)}\,.
\eeq  
The first term $c_{(f_B,Y)}$ is a $(d+1)$-form  which vanishes on-shell of the equations of motion. This term encodes the constraints of the theory.\footnote{The constraints are given in terms of the equations of motion $1$-form $E$ as $I_{\hat{\un\xi}}E=\rd c_{\un\xi}$.} The second term gives the surface (aka corner) charges of the theory, which are $d$-forms. In \cite{Ciambelli:2023mir}, it is shown that the resulting charges are given by\footnote{The minus sign compared to \cite{Ciambelli:2023mir} comes from the fact that our corner $\cC$ is an initial corner, see \eqref{Stokes}. Note also that $Y^a$ is the lift of $Y^i$ into the horizontal subbundle of $T\cN_{\diagup}$.}
\beqn\label{chbullet}
	Q_{(f_B,0)} &= \int_{\mathcal{N}_{\diagup}} \rd q_{(f_B,0)} = -\frac{1}{8\pi G} \int_{\mathcal{C}} \ve_{\mathcal{C}}\,\mathcal{L}_{\un{\ell}} f_B= -\frac{1}{8\pi G} \int_{\mathcal{C}} \ve^{(0)}_{\mathcal{C}} \,\mathcal{L}_{\un{\ell}} f_B \, \Omega \\
	Q_{(0,Y)} &= \int_{\mathcal{N}_{\diagup}} \rd q_{(0,Y)} = -\frac{1}{8\pi G} \int_{\mathcal{C}} \ve_{\mathcal{C}} \,\,Y^a \,\pi_a= -\frac{1}{8\pi G} \int_{\mathcal{C}} \ve^{(0)}_{\mathcal{C}} \,\,Y^a \,\Omega \pi_a\,,
\eeqn
with $\ve_{\cC}=\Omega \ve^{(0)}_{\cC}$, such that $\ve^{(0)}_{\cC}=\rd^d x$ contains no fields. Thus, the Noether charge aspects associated with superboosts and superrotations are, respectively, the volume element $\Omega$ and the spin-1 symplectic datum $\Omega\pi_a$ restricted to the corner. These are precisely the degrees of freedom which appear in the component $\bullet$ of the CIVP and the spin-0 and spin-1 symplectic data in \eqref{Hypersurface Symp}. Consequently, to each $\bullet$ in the CIVP we assign a corner $\mathcal{C}$ supporting the infinitesimal generators of a group $G_{\bullet}$, which acts symplectomorphically on the null hypersurface phase space via the map $a^{\bullet}: G_{\bullet} \times X_{\diagup} \rightarrow X_{\diagup}$. \\

To conclude, we emphasize an important kinematical distinction. The Hamiltonian functions in \eqref{ham} and the surface charges in \eqref{chbullet} should not be identified off shell. The former generate transformations on the kinematical phase space, whereas the latter arise as the corner terms in the Noether decomposition of the corresponding currents. These two notions coincide only after imposing the equations of motion, when the constraint contribution to the Noether current in \eqref{Noe} vanishes. Since our quantization is performed before imposing the gravitational constraints, we keep $H_{(f_B,Y)}$ and $Q_{(f_B,Y)}$ as distinct kinematical objects. Their on-shell equality will later be enforced as the quantum constraint that selects the physical algebra.

\subsubsection{Supertranslations}

Another class of symmetries that we can consider are diffeomorphisms generated by vector fields of the form $\un{\xi}_{f_T} = f_T \un{\ell}$, where $f_T$ is simply a function supported on the corner. These are called supertranslations, and form a group which we denote by $G_{\square} = C^{\infty}(\mathcal{C})_T$. Like the group $G_{\bullet}$, $G_{\square}$ acts on the null hypersurface phase space
\beq\label{asq}
	a^{\square}: G_{\square} \times X_{\diagup} \rightarrow X_{\diagup}\,.
\eeq
However, unlike $a^{\bullet}$, the action $a^{\square}$ is not symplectomorphic, because $a^{\square}_{g}{}^* \Omega_{\diagup} \neq \Omega_{\diagup}$ for $g \in G_{\square}$. 

This point is clearly seen by computing the symplectic flux of the supertranslation. Let $\hat{\un{\xi}}_{f_T}$ denote the vector field on phase space that generates a supertranslation parameterized by $f_T$, and define by 
\beq
j_{f_T} = I_{\hat{\un{\xi}}_{f_T}} \theta_{\diagup}
\eeq
the associated local Noether charge. The associated Hamiltonian is
\beq
	H_{f_T} = \int_{\mathcal{N}_{\diagup}} j_{f_T}\,. 
\eeq
For superboosts and superrotations, we claimed in \eqref{symp} that the functions $H_{(f_B,Y)}$ were Hamiltonian with respect to the generating vector fields $\hat{\un{\xi}}_{(f_B,Y)}$. It is instructive to derive this result, as it will help us demonstrate how this ceases to be true in the supertranslation case.\\

Consider a generic vector field $\hat{\un\xi}$ in field space. Using $\Omega_{\diagup}=\delta \Theta_{\diagup}$, we have
\beq
	 I_{\hat{\un{\xi}}} \Omega_{\diagup} =- \delta I_{\hat{\un\xi}}\Theta_{\diagup}+\fL_{\hat{\un{\xi}}} \Theta_{\diagup}\,. 
\eeq
If the presymplectic potential is generally covariant, one has $\fL_{\hat{\un{\xi}}} \Theta_{\diagup}=\int_{\cN_{\diagup}}\cL_{{\un{\xi}}} \theta_{\diagup}$.\footnote{This is true in the unextended phase space \cite{Ciambelli:2021nmv}. The presymplectic potential may be non-covariant, such that $\Delta_{\hat{\un\xi}}\theta_\diagup=(\fL_{\hat{\un\xi}}-\cL_{{\un\xi}})\theta_\diagup\neq 0$. In this case, one has to utilize the phase space anomaly formalism introduced in \cite{Hopfmuller:2018fni}.} Using the covariant phase space identity $\rd \theta_\diagup=\delta \mathscr{L}+E$, where the phase space $1$-form $E$ are the equations of motion, we have
\beq\label{intch}
I_{\hat{\un{\xi}}} \Omega_{\diagup}=- \delta \int_{\cN_\diagup}\left(I_{\hat{\un\xi}}\Theta_{\diagup}-i_{\un\xi}\mathscr{L}\right)+\int_{\cN_\diagup}\left(i_{\un\xi}E+\rd i_{\un\xi}\theta_\diagup\right)\,.
\eeq
The first parenthesis is exactly the local Noether current \eqref{totj} and, as we explained, the term $i_{\un\xi}\mathscr{L}$ vanishes for diffeomorphisms tangent to $\cN_\diagup$. This equally applies to $i_{\un\xi}E$, such that, for a generic Diff$\cN_\diagup$, we can write
\beq\label{N2t}
I_{\hat{\un{\xi}}} \Omega_{\diagup}=- \delta H_{\un\xi}+\int_{\cN_\diagup}\rd i_{\un\xi}\theta_\diagup=- \delta H_{\un\xi}-\int_{\cC}i_{\un\xi}\theta_\diagup\,.
\eeq
The last term, often called symplectic flux, is the obstruction of $\hat{\un\xi}$ to be represented canonically by $H_{\un\xi}$, \cite{Wald:1999wa}. \\

In our specific setup, using \eqref{th} and a generic vector field $\un{\xi} = f \un{\ell} + \un{Y}$ in Diff$\cN_\diagup$, the symplectic flux at the initial corner reads
\beq
\int_{\cC}i_{\un\xi}\theta_\diagup=\int_{\mathcal{C}} i_{\un{\xi}} \ve_{\mathcal{N}_{\diagup}} \Big(\frac{1}{2} \tau^{ab} \delta q_{ab} - \tau_a \delta \ell^a\Big) = \int_{\mathcal{C}} f\rvert_{\mathcal{C}} \, \ve_{\mathcal{C}} \Big( \frac{1}{2} \tau^{ab} \delta q_{ab} - \tau_a \delta \ell^a\Big)\,. 
\eeq
Therefore, the symplectic flux will vanish for any vector field $\un{\xi}$ in which $f\rvert_{\mathcal{C}} = 0$. This in particular comprises the collection of superboosts (thanks to \eqref{boo}) and superrotations (since they have no vertical component, $f=0$) and justifies that they give rise to Hamiltonian vector fields in \eqref{symp}. On the other hand, it is also clear that the symplectic flux does not vanish for supertranslations, and thus they satisfy
\beq\label{flux}
I_{\hat{\un{\xi}}_{f_T}} \Omega_{\diagup}=- \delta H_{f_T}-\int_{\cC}i_{{\un\xi}_{f_T}}\theta_\diagup\,.
\eeq
Physically, this expresses a simple but crucial distinction: boosts and rotations preserve the position of the corner, whereas translations move it along the null direction. Equation \eqref{flux} therefore isolates the obstruction that makes supertranslations qualitatively different from the other corner symmetries. Unlike superboosts and superrotations, they do not act symplectomorphically on the hypersurface phase space, and must therefore be treated separately. \\

The local Noether current for supertranslations also splits into the constraints and the corner charges
\beq
	j_{f_T} = c_{f_T} + \rd q_{f_T}\,,
\eeq
where the latter is given by
\beq
	Q_{f_T} = \int_{\mathcal{N}_{\diagup}} \rd q_{f_T} = \frac{1}{8\pi G} \int_{\mathcal{C}} \ve_{\mathcal{C}} \, f_T\, \theta\,. 
\eeq
Thus, the Noether charge aspect associated with supertranslations is the expansion at the corner. This is precisely the degree of freedom represented by the component $\square$ in the CIVP. Accordingly, to each $\square$ we associate an extended corner $\mathcal{C}$, which supports the infinitesimal generators of the group $G_{\square}$. As shown above, $G_{\square}$ acts on the hypersurface phase space, but its action is not symplectomorphic. This failure of symplectomorphicity is the classical feature that will require supertranslations to be quantized differently from the symmetries generated by $G_{\bullet}$.\\

A final feature that will play an important role in the quantization is the fact that, in the CIVP, only supertranslations that move the corner to the future should be allowed. Although supertranslations do not act symplectomorphically on the hypersurface phase space, a positive displacement of the initial corner along each null hypersurface $\cN_{(i)}$ is nevertheless a well-defined operation: it replaces the initial cut $\cC$ by a later cut contained in the domain already determined by the CIVP. By contrast, translating the corner to the past would require access to a larger spacetime region, lying before the original initial data surface, for which the CIVP does not provide a well-posed evolution problem. Thus, once an initial corner has been chosen, only future-directed supertranslations are admissible.

To make this statement explicit, let $u$ be a null parameter on $\cN$, with the initial corner located at $u=0$. A supertranslation acts locally as
\beq\label{st}
u'(u,x)=u+f_T(x)\,,
\eeq
where $f_T\in C^\infty(\cC)_T$. The transformation preserves the future domain of the CIVP only if
\beq
f_T(x)\geq 0
\qquad \forall \, x\in \cC\,.
\eeq
Thus, at the infinitesimal level, the allowed generators do not span the full Lie algebra $\mathfrak{g}_{\square}$, but only its future-directed cone
\beq
\mathfrak{g}^+_\square
=
\left\{\un\xi=
f_T(x)\un\ell\in \mathfrak{g}_\square \; \big| \; f_T(x)\geq 0
\quad \forall \, x\in \cC
\right\}.
\eeq
Exponentiating this cone gives the positive supertranslation semigroup
\beq
G^+_\square
=
\left\{
e^{f_T(x)\un\ell}\in G_\square \; \big| \; f_T(x)\geq 0
\quad \forall \, x\in \cC
\right\}\,.
\eeq
This is a semigroup rather than a group: the composition of two future-directed supertranslations is again future-directed,
\beq
e^{f^1_T(x)\un\ell}\circ e^{f^2_T(x)\un\ell}
=
e^{(f^1_T(x)+f^2_T(x))\un\ell}\,,
\qquad
f^1_T(x),f^2_T(x)\geq 0\,,
\eeq
whereas the inverse would be generated by $-f_T(x)\un\ell$ and is therefore not admissible unless $f_T(x)=0$. In the following, the map
\beq
	a_+^{\square}: G^+_{\square} \times X_{\diagup} \rightarrow X_{\diagup}\,,
\eeq
is the restriction of \eqref{asq} to this subsemigroup.\\

\begin{figure}[H]
\centering
\begin{tikzpicture}[scale=1.0]

% Full hypersurface strip
\coordinate (A) at (0,0);
\coordinate (B) at (6,2.2);
\coordinate (C) at (0,2.4);
\coordinate (D) at (6,4.6);

% Angle-dependent shifted cut
\coordinate (E0) at (2.00,0.73);
\coordinate (E1) at (2.28,1.55);
\coordinate (E2) at (1.95,2.35);
\coordinate (E3) at (2.32,3.25);

% Shade the full hypersurface
\fill[gray!6] (A) -- (B) -- (D) -- (C) -- cycle;

% Shade the nested future domain
\fill[gray!20]
(E0)
  .. controls (2.25,1.15) and (2.35,1.35) .. (E1)
  .. controls (2.05,1.95) and (1.85,2.10) .. (E2)
  .. controls (2.20,2.70) and (2.35,2.95) .. (E3)
-- (D) -- (B) -- cycle;

% Boundary of the full null hypersurface
\draw[thick] (A) -- (B);
\draw[thick] (C) -- (D);
\draw[thick] (A) -- (C);
\draw[thick] (B) -- (D);

% Initial cut
\draw[very thick] (A) -- (C);
\node[left] at (0,1.2) {$\cC$};

% Later angle-dependent cut
\draw[very thick]
(E0)
  .. controls (2.25,1.15) and (2.35,1.35) .. (E1)
  .. controls (2.05,1.95) and (1.85,2.10) .. (E2)
  .. controls (2.20,2.70) and (2.35,2.95) .. (E3);

\node[above] at (E3) {$\cC_{f_T}$};

% Supertranslation arrow
\draw[->,thick] (0.55,1.25) -- (1.65,1.63);
\node[below] at (1.12,1.22) {$f_T(x)\geq 0$};

% Labels for domains
\node at (1.05,2.3) {$\cN$};
\node at (4.35,2.7) {$\cN_{f_T}\subset \cN$};

% Indicate future direction
\draw[->] (0.35,-0.35) -- (2.0,0.25);
\node[below] at (1.25,-0.15) {$u$};

\end{tikzpicture}
\caption{A positive supertranslation moves the initial cut $\cC$ to a later, angle-dependent cut $\cC_{f_T}$. The full null hypersurface is denoted by $\cN$, while the darker region $\cN_{f_T}$ is the portion of $\cN$ lying to the future of $\cC_{f_T}$. Thus $\cN_{f_T}\subset\cN$, and positive supertranslations act as inclusion maps between nested future domains, rather than as invertible automorphisms.}
\label{fig:PositiveSupertranslation}
\end{figure}
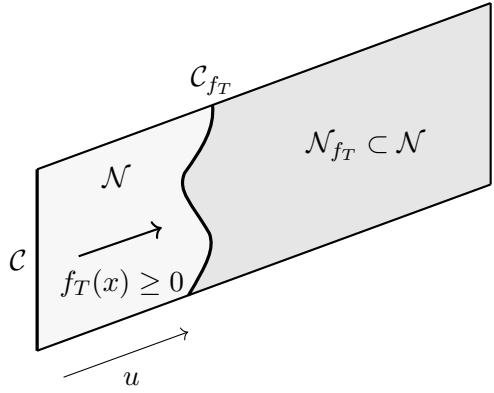

The loss of invertibility has a simple geometric meaning. A future-directed supertranslation does not define a reversible transformation on $\cN$. Rather, it defines a nesting map: it sends the original initial cut to a later cut, so that the corresponding future domain is included in the original one,
\beq
\cN_{f_T}\subset \cN,
\qquad
\cN_{f_T}:=\{(u,x)\in \cN \mid u\geq f_T(x)\}.
\eeq
Thus positive supertranslations are naturally interpreted as inclusion maps between nested future domains, rather than as automorphisms of the full CIVP domain. Their inverse would move the corner to the past and would require data outside the region determined by the original initial cut. This nesting is illustrated in Fig.~\ref{fig:PositiveSupertranslation}. The restriction to $G^+_{\square}$, rather than the full group $G_{\square}$, is therefore not merely a technical choice: it encodes the causal orientation and nesting structure of the characteristic initial value problem. This is the semigroup that will be represented quantum mechanically.

\subsection{The Final Symmetry Structure and Classical Setup}

The kinematical data of a single null hypersurface can therefore be collected as follows. The bulk of the hypersurface, $\diagup$, corresponds to a presymplectic manifold $(X_{\diagup},\Omega_{\diagup})$, encoding the hypersurface source data of the CIVP. The corner data split into two pieces. The point $\bullet$ corresponds to the group
\beq
G_\bullet=\mathrm{Diff}(\cC)\ltimes C^\infty(\cC)_B\,,
\eeq
which acts symplectomorphically on $X_{\diagup}$ and whose Noether charge aspects are the volume element and the Hájiček one-form. The box $\square$ corresponds to supertranslations. At the level of the ambient symmetry algebra these form the abelian group
\beq
G_\square= C^\infty(\cC)_T\,,
\eeq
whose Noether charge aspect is the expansion. However, once an initial corner has been chosen, the CIVP only admits the future-directed subsemigroup
\beq
G^+_\square\subset G_\square \,.
\eeq
This is the physically relevant structure for the characteristic evolution.\\

The group $G_\bullet$ acts naturally on supertranslations and preserves the future-directed cone. Indeed, superrotations act by pullback on the function $f_T$, while superboosts rescale it by a positive factor. Schematically, 
\beq \label{Dot action on Square}
f_T(x)\mapsto e^{b(x)}\, f_T(Y^{-1}(x)) \,.
\eeq
Hence a non-negative supertranslation parameter is mapped to another non-negative supertranslation parameter. The positive cone $\mathfrak g^+_\square$ is therefore stable under the action of $G_\bullet$.

We thus obtain the admissible corner symmetry semigroup
\beq
G^+_{\sqrbullet}
=
G_\bullet\ltimes G^+_\square \,.
\eeq
Equivalently, one should view $G_{\sqrbullet} = G_{\bullet} \ltimes G_{\square}$ as the ambient group from which the CIVP-admissible semigroup is obtained by restricting the supertranslation factor to $G^+_\square$. The corresponding ambient group is\footnote{In the explicit coordinates used in \eqref{symm}, supertranslations are simply given by $\un\xi_{f_T}=f_T(x)\un\pa_u$, and the algebra $\mathfrak{g}_{\sqrbullet}$ associated with $G_{\sqrbullet}$ can be easily verified.}
\beq\label{Gsqb}
	G_{\sqrbullet} = (\text{Diff}(\mathcal{C}) \ltimes C^{\infty}(\mathcal{C})_B) \ltimes C^{\infty}(\mathcal{C})_T\,. 
\eeq
Here, we have introduced the new symbol $\sqrbullet$ to indicate the amalgamation of the two pieces of corner data into one. The distinction between $G_{\sqrbullet}$ and $G^+_{\sqrbullet}$ will be useful in the quantum theory: the former provides a convenient representation-theoretic setting, while the physical supertranslation sector is represented by the latter.\\

Given this semidirect structure, $G_\bullet$ acts on both the supertranslation sector and the hypersurface phase space, while $G^+_\square$ acts only on $X_{\diagup}$, as schematically reported in the figure below.
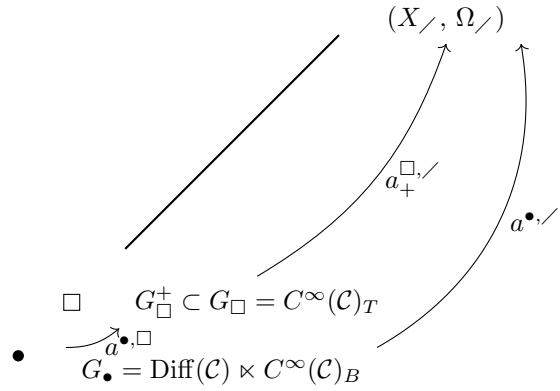
\begin{figure}[H]
\centering
\begin{tikzpicture}[
    scale=2.0,
    line/.style={thick},
    box/.style={draw, rectangle, minimum size=6pt, inner sep=0pt},
    dot/.style={circle, fill=black, inner sep=1.5pt},
    every node/.style={align=center, font=\small}
]

% --- Spacing parameters ---
\def\gapdotbox{0.5}
\def\gapboxline{0.5}
\def\linelength{2}

% --- RIGHT RAY (45°) ---
\coordinate (dotR) at (0,0);                     
\coordinate (boxR) at ($(dotR)+(45:\gapdotbox)$);
\coordinate (lineRstart) at ($(boxR)+(45:\gapboxline)$);
\coordinate (lineRend) at ($(lineRstart)+(45:\linelength)$);

\node[dot] at (dotR) {};
\node[box] at (boxR) {};
\draw[line] (lineRstart) -- (lineRend);

% Labels (named nodes)
\node[right] (labelDot) at ($(dotR)+(0.35,-0.1)$)
    {$G_{\bullet} = \mathrm{Diff}(\mathcal{C}) \ltimes C^{\infty}(\mathcal{C})_B$};

\node[right] (labelBox) at ($(boxR)+(0.35,0)$)
    {$G^+_{\square}\subset G_{\square} = C^{\infty}(\mathcal{C})_T$};

\node[right] (labelLine) at ($(lineRend)+(0.25,0.1)$)
    {$(X_{\diagup},\, \Omega_{\diagup})$};

% --- Curved arrows ---

% dot -> box (left side)
\draw[->, bend right=20]
    ([xshift=-1pt]labelDot.north west) to node[right] {$a^{\bullet,\square}$} ([xshift=-1pt]labelBox.south west);

% box -> line (center)
\draw[->, bend right=20]
    (labelBox.north) to node[right] {$a_+^{\square,\diagup}$} (labelLine.south);

% dot -> line (right side)
\draw[->, bend right=35]
    ([xshift=1pt]labelDot.north east) to node[right] {$a^{\bullet,\diagup}$} ([xshift=1pt]labelLine.south east);

\end{tikzpicture}
\caption{The classical data of a single hypersurface exhibit a nesting structure in which each piece of data acts on all of the data above it.}\label{Fig12}
\end{figure}

From this point of view, the kinematical data of the hypersurface consist of the presymplectic data $(X_{\diagup},\Omega_{\diagup})$ together with the admissible corner symmetry semigroup $G^+_{\sqrbullet}$, acting on $X_{\diagup}$ through
\beq
	a_+^{\sqrbullet}: G^+_{\sqrbullet} \times X_{\diagup} \rightarrow X_{\diagup}\,. 
\eeq
This action is symplectomorphic only for the subgroup $G_\bullet\subset G^+_{\sqrbullet}$. Infinitesimally, the corresponding vector fields are of the form
\beq
\un\xi_{(f_T,f_B,Y)}=(f_T+f_B)\un\ell+\un{Y}\,,
\eeq
with $f_T$ future-directed, $f_B$ vanishing at the initial corner, and $\un{Y}$ tangent to $\cC$. That is, 
\beq
	f_T(x) \geq 0\,, \qquad f_B\rvert_{\mathcal{C}} = 0, \qquad \underbrace{\cL_{\un{\ell}} \cdots\cL_{\un{\ell}}}_{{n \ \text{times}}} f_B \rvert_{\mathcal{C}} = 0\quad \forall \  n > 1\,, 
\eeq
where $f_T$ is understood as a corner function extended constantly along the generators.
We will denote the associated Noether currents by $j_{(f_T,f_B,Y)}$, the associated Hamiltonian functions by $H_{(f_T,f_B,Y)}$ and the Noether charges by $Q_{(f_T,f_B,Y)}$.

\section{Building Blocks of the Quantum CIVP} \label{sec: QKin}

Having described the classical structure of the CIVP -- including its equations of motion, constraints, and kinematical phase space -- we now turn to its quantization. Our approach proceeds in two steps. We first construct the quantum theory of the unconstrained kinematical data and only then impose the constraints as operator equations. This section carries out the first step. We quantize the data associated with a null hypersurface sector by sector: first the kinematical data $\diagup$, and then the corner data. The latter naturally decompose into the symplectomorphic sector $\bullet$ and the more intricate supertranslation sector $\square$. The section ends with a summary of the resulting kinematical quantum algebra and prepares the ground for the quantum implementation of the constraints.

\subsection{Quantizing the Hypersurface}

Classically, the bulk data on a null hypersurface define the phase space $(X_{\diagup},\Omega_{\diagup})$. A completely general quantization of this phase space is beyond the scope of the present paper.\footnote{
%although we comment on the associated subtleties in Appendix~\ref{app: SDQ}.
This being said, it seems possible that much of the analysis in this paper can be generalized by appealing to ideas from (strict) deformation quantization \cite{Fedosov1994,Kontsevich:1997vb,Cattaneo_2000,CATTANEO_2001,Landsman_1999,Collini:2016tqe}. We leave a more thorough exploration of this to future work.} The aspects of the quantization that will play a central role below -- most notably the quantum imposition of the CIVP constraints -- can already be captured without resolving these issues in full generality. Accordingly, in the main text we restrict to a standard linear Weyl quantization of the bulk sector. This should be viewed as the perturbative,
leading-order quantization of the local phase space around a chosen classical background, viewed as a formal power series in $\hbar$.

To perform our quantization we choose a point $x \in X_{\diagup}$ and then consider the symplectic vector space $S_{\diagup} =  T_x X_{\diagup}$ induced by the symplectic form $\Omega_{\diagup}$. One may interpret $x$ as a background choice so that $S$ corresponds to the (phase space) local fluctuations around this phase space point. To each $V \in S_{\diagup}$\footnote{Here $V$ is a generic tangent vector to field space, not necessarily the
Hamiltonian vector field associated with a symmetry.} we associate a Weyl symbol $\op{W}(V)$, the collection of which can be endowed with the structure of an abstract $*$-algebra with product and involution given by
\beq \label{Weyl Relations}
	\op{W}(V_1)\op{W}(V_2) = e^{-\frac{i}{2} I_{V_2} I_{V_1} \Omega_{\diagup}} \op{W}(V_1 + V_2), \qquad \op{W}(V)^* = \op{W}(-V). 
\eeq  
We will denote the algebra generated by these symbols by $\mathscr{A}_{\diagup}$. Strictly speaking, if $\Omega_{\diagup}$ has residual degeneracies, the Weyl
algebra is constructed after quotienting by its null directions.\\

The algebra $\mathscr{A}_{\diagup}$ can be realized as a concrete algebra of operators by implementing the GNS construction \cite{gelfand1994imbedding,segal1947irreducible}. The space of states on $\mathscr{A}_{\diagup}$ can be identified with the space of characteristic functions $C: S_{\diagup} \rightarrow \mathbb{C}$ such that $C(0) = 1$ and $S_{\diagup}^{\times 2}\ni (V_1,V_2)   \mapsto e^{i I_{V_2} I_{V_1} \Omega_{\diagup}} C(V_2 - V_1)$ is positive definite. Given a state, that is, a map $\omega: \mathscr{A}_{\diagup}\to \mathbb{C}$, its associated characteristic function is given by $C_{\omega}(V) = \omega(\op{W}(V))$. A state $\omega$ is called regular if $C_{\omega}(tV)$ is continuous in $t$ for each $V \in S_{\diagup}$.  

Let $\omega$ be a regular state on $\mathscr{A}_{\diagup}$.\footnote{Regularity is not necessary for the GNS construction, but we restrict our attention to this class of states for our discussion.} Its null space, $k_{\omega}$, consists of all elements $a \in \mathscr{A}_{\diagup}$ such that $\omega(a^* a) = 0$. The quotient space $\mathscr{A}_{\diagup}/k_{\omega}$ is a pre-closed inner product space with pre-inner product\footnote{``Pre'' is a topological designation indicating that the vector space may not be complete with respect to the topology induced by its inner product. One can always obtain a Hilbert space from a pre-closed inner product space completion.}
\beq \label{GNS inner}
	\langle \eta_{\omega}(a_1), \eta_{\omega}(a_2) \rangle_{\omega} = \omega(a_1^* a_2). 
\eeq
Here, $\eta_{\omega}: \mathscr{A}_{\diagup} \rightarrow \mathscr{A}_{\diagup}/k_{\omega}$ is the quotient map. The completion of $\mathscr{A}_{\diagup}/k_{\omega}$ with respect to \eqref{GNS inner} is a Hilbert space which we denote by $L^2(\mathscr{A}_{\diagup};\omega)$.\footnote{We emphasize that, although the algebra $\mathscr{A}_{\diagup}$ knows only about the right null hypersurface degrees of freedom, its standard Hilbert space acts as a purification. This is related to the observation that the GNS Hilbert space associated with a mixed state realizes an algebraic double in the sense that it admits commuting left and right representations of the algebra. \label{fn: GNSDouble}} The algebra $\mathscr{A}_{\diagup}$ is represented on $L^2(\mathscr{A}_{\diagup},\omega)$ by the GNS representation $\pi_{\omega}: \mathscr{A}_{\diagup} \to \mathfrak{B} (L^2(\mathscr{A}_{\diagup},\omega))$, defined by
\beq
	\pi_{\omega}(a_1) \eta_{\omega}(a_2) = \eta_{\omega}(a_1 a_2).
\eeq 
The Hilbert space $L^2(\mathscr{A}_{\diagup},\omega)$ contains a distinguished vector $\xi_{\omega}$ such that
\beq
	\omega(a) = \langle \xi_{\omega}, \pi_{\omega}(a) \xi_{\omega} \rangle_{\omega},
\eeq
which is by construction cyclic for the Hilbert space with respect to $\mathscr{A}_{\diagup}$ with $\eta_{\omega}(a) = \pi_{\omega}(a) \xi_{\omega}$ defining a dense set of vectors. \\

The preceding discussion allows us to make direct contact with familiar results, such as the connection between the GNS representation of the CCR algebra and the standard Fock Hilbert space. This is discussed in Appendix \ref{app: Fock}. Here, we stress that the algebraic construction of this note will only require the Weyl algebra and do not depend on any choice of Hilbert space representation. To this end, in what follows it will sometimes be useful to appeal to what is known as the universal representation of a $C^*$ algebra $\mathscr{A}$ \cite{kadison1983fundamentals1,kadison1986fundamentals2}. Heuristically, this is obtained as a direct sum over all GNS representations for all states on $\mathscr{A}$. We will denote this space by $L^2(\mathscr{A})$, distinguished from $L^2(\mathscr{A},\omega)$ which is dependent on a particular chosen state. From this point onward we will denote by $\pi: \mathscr{A} \rightarrow \mathfrak{B}(L^2(\mathscr{A}))$ the representation of $\mathscr{A}$ on its universal Hilbert space.\footnote{The reader may be concerned about the non-separability of the universal representation. Ultimately, we will only use the universal representation as an intermediary tool for constructing new, separable $C^*$ algebras. In fact, it is possible to carry out the analyses of this paper without fixing a Hilbert space representation at all, though this level of abstraction obscures some of the analysis.}

With the algebra $\mathscr{A}_{\diagup}$ in hand, we can now proceed to quantize the action of the corner degrees of freedom. Recall that the group $G_{\bullet}$ acts on $X_{\diagup}$ symplectomorphically via the action $a^{\bullet}: G_{\bullet} \times X_{\diagup} \rightarrow X_{\diagup}$. It is straightforward to see that this induces an action of $G_{\bullet}$ on $\mathscr{A}_{\diagup}$ via invertible, algebra preserving maps.
%\footnote{We are directly working with the von Neumann algebra instead of $\mathscr{A}_\diagup$ because our state $\omega$ is assumed to be invariant under $\alpha^\bullet$. {\color{red} This is necessary, but is this obvious/true?} \textcolor{blue}{In fact this is somewhat subtle when it comes to e.g. the superrotations. So I think it would be wiser to reformulate the analysis in the language of $C^*$ algebras.}} 
In other words, the action of $G_{\bullet}$ on $\mathscr{A}_{\diagup}$ is an automorphism. To be precise, given $g\in G_\bullet$, we define
%\footnote{Strictly speaking, we should always refer to elements of $\cM_\diagup$ as $\pi(W(V))$. We omit the $\pi$ in the rest of this section to ease the notation.}
\beq
	\alpha^{\bullet}: G_{\bullet} \rightarrow \text{Aut}(\mathscr{A}_{\diagup}), \qquad \alpha^{\bullet}_{g}(\op{W}(V)) = \op{W}(a^{\bullet}_{g}{}_* V).
\eeq
Clearly $\alpha^{\bullet}_{g} \circ \alpha^{\bullet}_{g^{-1}} = \text{id}$, and we can also write
\beq
	\alpha^{\bullet}_{g}\bigg(\op{W}(V_1) \op{W}(V_2)\bigg) = \alpha^{\bullet}_{g}(\op{W}(V_1)) \alpha^{\bullet}_{g}(\op{W}(V_2)),
\eeq
which simply follows from \eqref{Weyl Relations} and the fact that $a^{\bullet}_{g}{}^*\Omega_{\diagup} = \Omega_{\diagup}$. \\

Quantizing the action $a^{\square}: G_{\square} \times X_{\diagup} \rightarrow X_{\diagup}$ is more subtle, since it does not act as a symplectomorphism. This is consistent and a consequence of the fact that positive supertranslations are non-invertible and do not preserve the region, as already remarked. To promote these observations into the quantum setup, we make use of the following result \cite{Demoen1977,DEMOEN197927}. Abstractly, given a symplectic vector space $(S,\Omega)$ and a linear map $T: S \rightarrow S$ we can define $\Omega_T(V_1,V_2) = \Omega(V_1,V_2) - \Omega(T(V_1),T(V_2))$ which is generally a presymplectic form. We denote by $(S,\Omega_T)$ the resulting presymplectic vector space and by $\mathscr{A}_T$ its associated algebra endowed with canonical commutation relations (CCR). Then, given any state $\omega_T$ on $\mathscr{A}_T$, the map
\beq
    \alpha_T(\op{W}(V)) = \omega_T(\op{W}(V)) \op{W}(T(V))
\eeq
is completely positive with respect to $\mathscr{A}$, the CCR algebra of the original pair $(S,\Omega)$. Given any regular state $\omega$ on $\mathscr{A}$ such that $\omega(\op{W}(T(V))) \neq 0$, it is demonstrated in \cite{EvansLewis1977} that $\omega_T(\op{W}(V)) = \frac{\omega(\op{W}(V))}{\omega(\op{W}(T(V)))}$ is a state on $\mathscr{A}_T$. Thus, we may take
\beq
    \alpha_{T}^{\omega}(\op{W}(V)) = \frac{\omega(\op{W}(V))}{\omega(\op{W}(T(V)))} \op{W}(T(V)),
\eeq
where we have distinguished that this action depends implicitly upon the choice of state. By construction, the state $\omega$ can be seen to be invariant under $\alpha_{T}^{\omega}$:
\beq
    \omega\big(\alpha_{T}^{\omega}(\op{W}(V))\big) = \omega(\op{W}(V)), \qquad \forall \ V \in S. 
\eeq\\

Returning to our specific set-up, the abstract map $T:S\to S$ is supplied by the positive supertranslation action. For each $g\in G^+_\square$, the classical map
\beq
a^\square_g:X_{\diagup}\to X_{\diagup}
\eeq
induces, after linearizing around the chosen point $x\in X_{\diagup}$, a linear map on $S_{\diagup}=T_xX_{\diagup}$. By a slight abuse of notation, we denote this induced map by
\beq
a^\square_{g*}:S_{\diagup}\to S_{\diagup}.
\eeq
Thus, in the notation of the abstract construction above, we take (for each $g \in G_{\square}^+$)
\beq
T=a^\square_{g*}.
\eeq
Given a state\footnote{For example, the quasifree, Hadamard vacuum state constructed by Kay and Wald \cite{kay1991theorems}, trivially satisfies the property \eqref{nonzero}, as it is actually invariant under a supertranslation.} $\omega$ on $\mathscr{A}_{\diagup}$ such that
\beq\label{nonzero}
\omega\left(\op{W}(a^\square_{g*}V)\right)\neq 0,
\qquad
\forall \ V\in S_{\diagup},
\eeq
we obtain the completely positive map
\beq \label{Def of ST-CP Action}
\alpha^{\square,\omega}_g(\op{W}(V))
=
\frac{\omega(\op{W}(V))}
{\omega(\op{W}(a^\square_{g*}V))}
\,\op{W}(a^\square_{g*}V).
\eeq
Physically, we can think of $\alpha_{T}^{\omega}$ as a quantum channel: it is not an automorphism of the algebra, but rather a completely positive map. This is crucial in understanding how null time evolution is implemented at the quantum level.

It warrants noting that, by construction, $\alpha^{\square,\omega}$ is a semigroup homomorphism,
\beq
    \alpha^{\square,\omega}_{g_1} \circ \alpha^{\square,\omega}_{g_2} = \alpha^{\square,\omega}_{g_1 \circ g_2}, \qquad \forall g_1,g_2 \in G_{\square}^+. 
\eeq
In what follows we will denote this action simply by $\alpha^{\square}$, though it should be noted that its definition depends upon a choice of a state. The fact that $\alpha^{\square}$ defines a homomorphism of the supertranslation semigroup will be crucial in Section~\ref{sec: STCrossed}. \\

Before continuing, we briefly summarize the main takeaway messages. In this subsection we have quantized the phase space $(X_{\diagup},\Omega_{\diagup})$ of a single null hypersurface to obtain the operator algebra $\mathscr{A}_{\diagup}$. The classical phase space admitted symmetry actions $a^{\bullet}$ and $a^{\square}$ by $G_{\bullet}$ and $G_{\square}^+$, respectively. The former action by superboosts and superrotations is symplectomorphic and naturally quantizes to an automorphic algebra preserving action, $\alpha^{\bullet}$, on $\mathscr{A}_{\diagup}$. The latter action by supertranslations is not symplectomorphic and consequently quantizes to a non-algebra preserving action, $\alpha^{\square}$, on $\mathscr{A}_{\diagup}$ by quantum channels. The collection of data $(\mathscr{A}_{\diagup}, \alpha^{\square}, \alpha^{\bullet})$ constitutes a quantization of the hypersurface degrees of freedom and their symmetries. We next move on to quantizing the corner data, $\bullet$ and $\square$, which constitute independent degrees of freedom in the CIVP. 

\subsection{Quantizing the Corner Symmetry Algebras}

As we have emphasized, in addition to their action on the hypersurface degrees of freedom, the corner degrees of freedom are also independently quantized in the CIVP. In the next two subsections we will show how these degrees of freedom form a pair of operator algebras $\mathscr{A}_{\bullet}$ and $\mathscr{A}_{\square}$ which are, respectively, the group algebra of $G_{\bullet}$ and the semigroup algebra of $G^+_{\square}$. Physically, we interpret these as the operator algebras generated by the Noether charges $Q_{(f_B,\un{Y})}$ and $Q_{f_T}$ whose charge aspects are explicitly the corner supported initial data of the CIVP. 

\subsubsection{Superboosts and Superrotations}

Since $G_\bullet$ is non-locally compact, there is no Haar measure available. In the representation-theoretic framework used here, we therefore choose a quasi-invariant measure $\rd\nu_{\bullet}$; for diffeomorphism groups see \cite{SHIMOMURA2001406}.
In the gravitational crossed-product setting, this strategy was used in
\cite{Klinger:2026tws,Klinger:2025tvg} to demonstrate that such a group can be quantized. In particular, in \cite{Klinger:2026tws} a quasi-invariant measure for the group of superboosts was constructed explicitly. This can be completed to a quasi-invariant measure on the full group $G_{\bullet}$ by pairing it with a quasi-invariant measure on the group of superrotations, $\text{Diff}(\mathcal{C})$. In \cite{SHIMOMURA2001406} such a measure is exhibited for general diffeomorphism groups on compact manifolds, and thus we are assured the existence of a quasi-invariant measure on all of $G_{\bullet}$.\footnote{Alternatively, the group $G_{\bullet}$ can be quantized by identifying it with a groupoid and implementing a strict deformation quantization by means of a Haar system \cite{Williams2019Toolkit}. } 

Given the measure $\rd\nu_{\bullet}$ we can define a Hilbert space $L^2(G_{\bullet},\rd\nu_{\bullet})$ which consists of square integrable functions on $G_{\bullet}$ such that
\beq
	\langle \psi, \psi \rangle_{\bullet} = \int_{G_{\bullet}} \rd\nu_{\bullet}(g) \overline{\psi(g)} \psi(g) < \infty. 
\eeq
This Hilbert space admits a unitary representation of $G_{\bullet}$ given by
\beq
	\ell_{\bullet}: G_{\bullet} \rightarrow U(L^2(G_{\bullet},d\nu_{\bullet})), \qquad \bigg(\ell_{\bullet}(g) \psi\bigg)(g') = J_{\bullet}(g,g^{-1} \circ g')^{-1/2} \psi(g^{-1} \circ g'),
\eeq
where $J_{\bullet}(g,g')$ is the Jacobian of the quasi-invariant measure under left translations
\beq
	\rd\nu_{\bullet}(g \circ g') = J_{\bullet}(g,g') \rd\nu_{\bullet}(g'). 
\eeq
The group algebra associated with $G_{\bullet}$ is the $C^*$ algebra generated by $\ell_{\bullet}(g)$, which we will denote by $\mathscr{A}_{\bullet}$. We can identify
\beq
	\ell_{\bullet}(\text{exp}(f_B \un{\ell} + \un{Y})) = e^{i \op{Q}_{(f_B,Y)}},
\eeq
where $\op{Q}_{(f_B,\un{Y})}$ is the quantum operator associated to the corner charge representing the group $G_{\bullet}$. Thus, the operators $\ell_\bullet(g)$ are exponentiated quantum representatives of the corner charges whose charge aspects $Q_{(f_B,Y)}$ are the area element and the Hájiček one-form.

\subsubsection{Supertranslations} \label{sec: ST Alg}

As in the case of deducing its action, the algebra associated with the extended corner data is somewhat more involved. In the previous section, we saw that the relevant action of $G_{\square}$ implicates a sub-semigroup $G_{\square}^+$. Consequently, the algebra that quantizes the intrinsic degrees of freedom associated with $\square$ in the CIVP must respect this semigroup structure. Fortunately, there is a canonical way to quantize a semigroup when it is embedded inside of a group \cite{Nica1992,LACA1996415}. 

Let $\rd\nu_{\square}$ be a quasi-invariant measure on the full group $G_{\square}$, and denote by $L^2(G_{\square},\rd\nu_{\square})$ the Hilbert space of square integrable functions on $G_{\square}$ with respect to this measure.\footnote{We emphasize that such a quasi-invariant measure can be obtained by following the same infinite-dimensional Gaussian measure construction described in \cite{Klinger:2026tws} for the group of superboosts.} The full group $G_{\square}$ admits a unitary representation on this Hilbert space, which we denote by the unitary map $\ell_{\square}: G_{\square} \rightarrow \mathfrak{B}(L^2(G_{\square},\rd\nu_{\square}))$. Next, we can introduce a Hilbert subspace $L^2(G_{\square}^+,\rd\nu_{\square})$ which consists of all square integrable functions restricted to the domain $G_{\square}^+ \subset G_{\square}$. Let $P: L^2(G_{\square},\rd\nu_{\square}) \rightarrow L^2(G_{\square}^+,\rd\nu_{\square})$ be an isometric restriction to $L^2(G_{\square}^+,\rd\nu_{\square})$. Then, the unitary representation $\ell_{\square}$ descends to a representation by isometries of $G_{\square}^+$ on $L^2(G_{\square}^+,\rd\nu_{\square})$. In particular,
\beq
	w_{\square}: G_{\square}^+ \rightarrow {\mathfrak B}(L^2(G_{\square}^+,\rd\nu_{\square})), \qquad w_{\square}(g) = P \ell_{\square}(g) P^{\dagger}\qquad g\in G_\square. 
\eeq
This is called the Wiener-Hopf representation of $G_{\square}^+$ induced by the unitary representation $\ell_{\square}$ of $G_{\square}$.  

The algebra associated with $G_{\square}^+$ is the $C^*$ algebra generated by $w_{\square}$, which we denote by $\mathscr{A}_{\square}$. Formally, we can write
\beq
	w_{\square}(\text{exp}(f_T \un{\ell})) = e^{i \op{Q}_{f_T}}, 
\eeq	
where $\op{Q}_{f_T}$ should be understood as the generator of the positive supertranslation semigroup, not necessarily as a self-adjoint generator of a unitary group.
We note that the charge aspect of $Q_{f_T}$ is the expansion, and thus $\mathscr{A}_{\square}$ can be read as quantizing the $\square$ initial data in the CIVP. \\

As we have described around eqn. \eqref{Dot action on Square}, the group $G_{\bullet}$ acts on $G_{\square}^+$ in a way that preserves its semigroup structure. From the operator algebraic point of view, this action can be quantized to an automorphic action
\beq
    \alpha^{\bullet,\square}: G_{\bullet} \rightarrow \text{Aut}(\mathscr{A}_{\square}). 
\eeq
In particular, the action $a^{\bullet,\square}: G_{\bullet} \times G_{\square}^+ \rightarrow G_{\square}^+$ is
\beq\label{ab}
    a^{\bullet,\square}_{\text{exp}(f_B \un{\ell} + \un{Y})}(\text{exp}(f_T \un{\ell})) = \text{exp}(e^{b} \ \text{exp}_{Y}{}_*(f_T \un{\ell})),
\eeq
where $\text{exp}_{Y}{}_*$ denotes the push-forward by the finite diffeomorphism generated by $Y$,  while the factor $e^b$ accounts for the boost weight. The quantum version of \eqref{ab} is simply defined as
\beq \alpha^{\bullet,\square}_{g_{\bullet}}(w_{\square}(g_{\square})) = w_{\square} \circ \alpha^{\bullet,\square}_{g_{\bullet}}(g_{\square}), \qquad g_{\bullet} \in G_{\bullet},\qquad  g_{\square} \in G_{\square}^+. 
\eeq
This concludes the quantization  of the corner symmetry algebras, and thus of all the kinematic data involved in a single null hypersurface plus corner. We recollect its salient features in the next subsection.
    
\subsection{Summary of Kinematical Quantization} 

\begin{figure}[ht]
\centering
\begin{tikzpicture}[
    scale=2.0,
    line/.style={thick},
    box/.style={draw, rectangle, minimum size=6pt, inner sep=0pt},
    dot/.style={circle, fill=black, inner sep=1.5pt},
    every node/.style={align=center, font=\small}
]

% --- Spacing parameters ---
\def\gapdotbox{0.5}
\def\gapboxline{0.5}
\def\linelength{2}

% --- RIGHT RAY (45°) ---
\coordinate (dotR) at (0,0);                     
\coordinate (boxR) at ($(dotR)+(45:\gapdotbox)$);
\coordinate (lineRstart) at ($(boxR)+(45:\gapboxline)$);
\coordinate (lineRend) at ($(lineRstart)+(45:\linelength)$);

\node[dot] at (dotR) {};
\node[box] at (boxR) {};
\draw[line] (lineRstart) -- (lineRend);

% Labels (named nodes)
\node[right] (labelDot) at ($(dotR)+(0.35,-0.1)$)
    {$\mathscr{A}_{\bullet}$};

\node[right] (labelBox) at ($(boxR)+(0.35,0)$)
    {$\mathscr{A}_{\square}$};

\node[right] (labelLine) at ($(lineRend)+(0.25,0.1)$)
    {$\mathscr{A}_{\diagup}$};

% --- Curved arrows ---

% dot -> box (left side)
\draw[->, bend right=20]
    ([xshift=-1pt]labelDot.north west) to node[right] {$\alpha^{\bullet,\square}$} ([xshift=-1pt]labelBox.south west);

% box -> line (center)
\draw[->, bend right=20]
    (labelBox.north) to node[right] {$\alpha^{\square,\diagup}$} (labelLine.south);

% dot -> line (right side)
\draw[->, bend right=35]
    ([xshift=1pt]labelDot.north east) to node[right] {$\alpha^{\bullet,\diagup}$} ([xshift=1pt]labelLine.south east);

\end{tikzpicture}
\caption{Quantum data of a single hypersurface.}\label{fig13}
\end{figure}
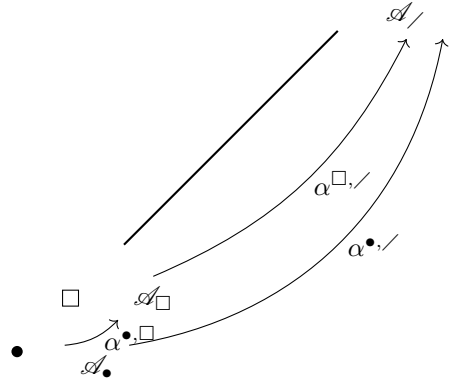

We have therefore arrived at the perturbative kinematical quantization of any single hypersurface theory appearing in the CIVP. Of course, this implies a quantization of the full set of kinematical data in the CIVP by doubling. Although not necessary, it can be helpful to interpret the operators appearing in the algebras $\mathscr{A}_{\diagup}$, $\mathscr{A}_{\bullet}$, and $\mathscr{A}_{\square}$ more directly in terms of the CIVP data.

Working in the GNS Hilbert space of a regular state on $\mathscr{A}_{\diagup}$ we can read the Weyl operators as exponentiated, smeared field observables. Recalling the symplectic analysis in section \ref{2.2}, a vector $V \in T_x X_{\diagup}$ can be written in the form
\beq
	V = V^{\mu} \frac{\delta}{\delta \mu} + V^{\Omega} \frac{\delta}{\delta \Omega} + V^{\ell} \frac{\delta}{\delta \ell} + V^{\Omega \pi} \frac{\delta}{\delta(\Omega \pi)} + V^{\bar{q}} \frac{\delta}{\delta \bar{q}} + V^{\Omega \sigma} \frac{\delta}{\delta(\Omega \sigma)} = V^{\Phi} \frac{\delta}{\delta \Phi} + V^{\Pi} \frac{\delta}{\delta \Pi},
\eeq
where $\Phi = (\mu,\ell, \bar{q})$ and $\Pi = (\Omega, \Omega \pi, \Omega \sigma)$ are local Darboux coordinates for $X_{\diagup}$ in the neighborhood of $x$. Then, under the representation $\pi_{\omega}$, the Weyl symbols become
\beq \label{smeared fields}
	\pi_{\omega}(\op{W}(V)) = e^{i(\op{\Phi}[V] + \op{\Pi}[V])} = e^{i \op{\phi}(V)} \sim e^{i \int_{\mathcal{N}_{\diagup}} \varepsilon_{\mathcal{N}_{\diagup}}^{(0)} \Big(V^{\mu} \op{\Omega} + V^{\Omega} \op{\mu} + V^{\ell} \op{\pi} + V^{\Omega \pi} \op{\Omega \pi} + V^{\bar{q}} \op{\bar{q}} + V^{\Omega \sigma} \op{\Omega \sigma}\Big)},
\eeq
with
\beq
	\op{\Phi}[V] = \int_{\mathcal{N}_{\diagup}} \varepsilon_{\mathcal{N}_{\diagup}}^{(0)} V^{\Pi} \op{\Phi}, \qquad \op{\Pi}[V] = \int_{\mathcal{N}_{\diagup}} \varepsilon_{\mathcal{N}_{\diagup}}^{(0)} V^{\Phi} \op{\Pi}.
\eeq	
Likewise, the corner charges $Q_{(f_B,Y)}$ and $Q_{f_T}$ are quantized to the generators of the group algebra $\mathscr{A}_{\bullet}$ and the semigroup algebra $\mathscr{A}_{\square}$, respectively. That is, we can read
\begin{flalign} \label{noether ops}
    &\ell_{\bullet}(\text{exp}(f_B \un{\ell})) = e^{i\op{Q}_{(f_B,0)}} \sim e^{-\frac{i}{8\pi G} \int_{\mathcal{C}} \ve_{\mathcal{C}}^{(0)} \mathcal{L}_{\un{\ell}} f_B \op{\Omega}}, \nonumber \\
    &\ell_{\bullet}(\text{exp}(\un{Y})) = e^{i \op{Q}_{(0,Y)}} \sim e^{-\frac{i}{8\pi G} \int_{\mathcal{C}} \ve_{\mathcal{C}}^{(0)} Y^a \op{\Omega} \op{\pi}_a}, \nonumber \\
    &w_{\square}(\text{exp}(f_T \un{\ell})) = e^{i\op{Q}_{f_T}} \sim e^{\frac{i}{8\pi G} \int_{\mathcal{C}} \ve_{\mathcal{C}}^{(0)} f_T \op{\Omega} \op{\theta}}. 
\end{flalign}
We emphasize that $(\Omega,\pi_a,\theta)$ are evaluated on the corner $\mathcal{C}$ and thus must be treated as independent of the quantities appearing in the pure hypersurface quantization.\footnote{We note that the operators $\op{W}(V)$, $\ell_{\bullet}(g_{\bullet})$ and $w_{\square}(g_{\square})$ for $V \in T_x X_{\diagup}$, $g_{\bullet} \in G_{\bullet}$ and $g_{\square} \in G_{\square}^+$ are the rigorously well defined objects which we formulate our analysis around. The parts of eqns. \eqref{smeared fields} and \eqref{noether ops} after the $\sim$ are nevertheless useful heuristics that relate these operators to more standard classical data.}  

Finally, the trio of algebras we have quantized exhibit the same nesting structure observed in the classical case. This is readily seen comparing Figure \ref{fig13} with Figure \ref{Fig12}. The algebra $\mathscr{A}_{\bullet}$ acts on both $\mathscr{A}_{\square}$ and $\mathscr{A}_{\diagup}$. The algebra $\mathscr{A}_{\square}$ acts on $\mathscr{A}_{\diagup}$. These symmetry actions will serve as the focal points by which we can reassemble the hypersurface -- a task which we turn our attention to now. This nesting of actions dictated the way we deconstructed the single null hypersurface data, and it is crucial to respect in order to achieve its proper quantization.

\section{Implementing Hypersurface Constraints} \label{sec: QCon}

With all of the kinematical degrees of freedom at our disposal, we now embark upon the task of sewing the hypersurface back together. The central ingredient which makes this possible is the hierarchy of symmetry actions described in the previous section. Using these relations, we can merge together $\mathscr{A}_{\diagup}$, $\mathscr{A}_{\bullet}$ and $\mathscr{A}_{\square}$ into amalgamated product algebras. 

Following the sequence of symmetries, we first glue together $\mathscr{A}_{\diagup}$ and $\mathscr{A}_{\square}$ to form the extended hypersurface algebra $\mathscr{A}_{\sqrdiagup}$. The formation of this algebra resembles closely that of a crossed product, but is complicated by the fact that $G_{\square}^+$ only acts on $\mathscr{A}_{\diagup}$ by completely positive maps. This difficulty is circumvented by appealing to Stinespring's theorem, which allows the completely positive maps representing supertranslations to be dilated to algebra preserving endomorphisms \cite{Longo:2017aet}. 

To complete the hypersurface algebra, we must further include the generators of $\mathscr{A}_{\bullet}$. These operators are appended to the algebra via a standard group-like crossed product construction using the automorphic action of $G_{\bullet}$ on $\mathscr{A}_{\diagup}$ and $\mathscr{A}_{\square}$ which induces an automorphic action on $\mathscr{A}_{\sqrdiagup}$. The resulting algebra, which we denote by $\mathscr{A}_{\sqrbulletdiagup}$, consists of all operators satisfying the quantum version of the gravitational constraints. 

\subsection{The Extended Hypersurface Algebra} \label{sec: STCrossed}

The first step is to form the extended hypersurface algebra
$\mathscr{A}_{\sqrdiagup}$, which incorporates the supertranslation sector, and can be regarded as the quantum analog of the extended phase space \cite{Ciambelli:2021nmv,Klinger:2023qna,Klinger:2023tgi}. Recall that $G_{\square}^{+}$ acts on $\mathscr{A}_{\diagup}$ by completely positive maps, as opposed to $G_{\bullet}$, which acts by automorphisms. This difference is crucial in what follows. We would like to construct an algebra, $\mathscr{A}_{\sqrdiagup}$, in which the action of  $G_{\square}^{+}$ is inner implemented. 

To achieve this goal, it is useful to first appeal to a generalized form of Stinespring dilation introduced by Longo in \cite{Longo:2017aet}. Given any unital, completely positive map on a $C^*$ algebra\footnote{We note that technically the completely positive map should be interpreted as mapping into a von Neumann closure of $\mathscr{A}$. See theorem 2.10 and corollary 2.11 in \cite{Longo:2017aet} for further discussion.}, $\alpha \in \text{CP}(\mathscr{A})$, there exist dilation pairs $(v,\theta)$ with $v \in \mathscr{A}$ ($v^* v = \op{1}$) and $\theta \in \text{End}(\mathscr{A})$ such that, given $a\in \mathscr{A}$,
\beq
	\alpha(a) = v^* \theta(a) v.  
\eeq	
In other words, any completely positive map $\alpha$ is inner isometric to an endomorphism $\theta$. 

Given the action $\alpha^{\square}: G_{\square}^+ \rightarrow \text{CP}(\mathscr{A}_{\diagup})$ we may therefore identify a family of dilation pairs $\{v^{\square}_{g}, \theta^{\square}_{g}\}_{g \in G_{\square}^+}$, such that
\beq \label{Stinespring Endo}
	\alpha^{\square}_g(a) = v_g^{\square *} \theta^{\square}_g(a) v^{\square}_g \qquad a\in \mathscr{A}_{\diagup}.
\eeq
We will further assume that these pairs can be chosen such that
\beq
	\theta^{\square}: G_{\square}^{+} \rightarrow \text{End}(\mathscr{A}_{\diagup})
\eeq
is a homomorphism, that is, $\theta_g\circ\theta_h=\theta_{g\circ h}$, and recall that $\theta^{\square}_{g}(a) \theta^{\square}_{g}(b) = \theta^{\square}_{g}(ab)$, for all $a,b\in \mathscr{A}_{\diagup}$.\\

The triple $(\mathscr{A}_{\diagup}, G_{\square}^+, \theta^{\square})$ defines a \emph{semigroup covariant system} \cite{LACA1996415}. Following \cite{LACA1996415}, we can associate to this covariant system a unique $C^*$ algebra embedding both $\mathscr{A}_{\diagup}$ and $\mathscr{A}_{\square}$, and implementing the action $\theta^{\square}$. This is achieved introducing the triple $(\mathscr{A}_{\sqrdiagup},i_{\diagup}, i_{\square})$, where $\mathscr{A}_{\sqrdiagup}$ is a $C^*$ algebra, $i_{\diagup}: \mathscr{A}_{\diagup} \rightarrow \mathscr{A}_{\sqrdiagup}$ is a unital homomorphism and $i_{\square}: G_{\square}^+ \rightarrow \textrm{Isom}(\mathscr{A}_{\diagup})$ is a semigroup homomorphism\footnote{Here, $\textrm{Isom}(\mathscr{A}_{\sqrdiagup})$ is the semigroup of isometries in $\mathscr{A}_{\sqrdiagup}$.} such that
\beq \label{ST CP cond}
    i_{\diagup} \circ \theta^{\square}_g(a) = i_{\square}(g) i_{\diagup}(a) i_{\square}(g)^{\dagger}, \qquad a \in \mathscr{A}_{\diagup}, g \in G_{\square}^+. 
\eeq
Provided $\mathscr{A}_{\diagup}$ is equal to the $C^*$ algebra generated by $i_{\diagup}(\mathscr{A}_{\diagup})$ and $i_{\square}(G_{\square}^+)$ satisfies the universality condition described in \cite{LACA1996415}, $\mathscr{A}_{\sqrdiagup}$ is referred to as the \emph{semigroup crossed product} of the covariant system $(\mathscr{A}_{\diagup},G_{\square}^+, \theta^{\square})$. In our context, we have used the notation $\mathscr{A}_{\sqrdiagup}$ to emphasize that this algebra combines together the physical degrees of freedom of $\diagup$ and $\square$. 

Combining eqn. \eqref{ST CP cond} and eqn. \eqref{Stinespring Endo} we find that
\beq
    \big(i_{\square}(g)^{\dagger} i_{\diagup}(v_g^{\square})\big)^{\dagger} i_{\diagup}(a) i_{\square}(g)^{\dagger} i_{\diagup}(v_g^{\square}) = i_{\diagup} \circ \alpha^{\square}_g(a),
\eeq
that is, the original action of $G_{\square}^+$ by completely positive maps is implemented by $i_{\square}(g)^{\dagger} i_{\diagup}(v_g^{\square}) \in \mathscr{A}_{\sqrdiagup}$. In the following, we will sometimes use the notation $\mathscr{A}_{\sqrdiagup} = \mathscr{A}_{\diagup} \times_{\alpha^{\square}} G_{\square}^+$ to emphasize that this algebra also implements the CP action $\alpha^{\square}$ innerly. In this sense, the algebra $\mathscr{A}_{\sqrdiagup}$ accomplishes our desired gluing of $\diagup$ and $\square$. \\

The abstract algebraic presentation of the algebra $\mathscr{A}_{\sqrdiagup}$ avoids the need to appeal to any explicit Hilbert space representation. With this being said, it will be useful later on to introduce a Hilbert space representation compatible with this algebra. A \emph{covariant representation} of the system $(\mathscr{A}_{\diagup}, G_{\square}^+, \theta^{\square})$ is a Hilbert space $\mathscr{H}_{\sqrdiagup}$ admitting an ordinary representation $\pi_{\theta^{\square}}: \mathscr{A}_{\diagup} \rightarrow \mathfrak{B}(\mathscr{H}_{\sqrdiagup})$ and an isometric representation\footnote{For the trivial covariant system $(\mathbb{C},G_{\square}^+,\text{id})$, the representation $w_{\square}$ here coincides with the Wiener-Hopf representation described in Section~\ref{sec: ST Alg}. In general it may differ, but the $C^*$ algebra it generates will coincide with the semigroup $C^*$ algebra, justifying the common notation.} $w_{\square}: G_{\square}^+ \rightarrow \mathfrak{B}(\mathscr{H}_{\sqrdiagup})$ satisfying the covariance relation
\beq \label{ST Covariance}
    w_{\square}(g) \pi_{\theta^{\square}}(a) = \pi_{\theta^{\square}} \circ \theta^{\square}_g(a) w_{\square}(g), \qquad a \in \mathscr{A}_{\diagup}, g \in G_{\square}^+. 
\eeq
The $C^*$ algebra generated by $\pi_{\theta^{\square}}(\mathscr{A}_{\diagup})$ and $w_{\square}(G_{\square}^+)$ -- in a suitable universal topology -- coincides with the semigroup crossed product introduced above. We note that a useful covariant representation is isomorphic to $L^2(\mathscr{A}_{\diagup}) \otimes L^2(G_{\square}^+,\rd\nu_{\square})$. We can regard this Hilbert space as densely spanned by states of the form $\ket{a,g}$ with $a \in \mathscr{A}_{\diagup}$ and $g \in G_{\square}^+$. Then,
\beq \label{square dressing}
    \pi_{\theta^{\square}}(a) \ket{b,h} = \ket{ab,h}, \qquad w_{\square}(g) \ket{b,h} = \ket{\theta^{\square}_g(b), gh}. 
\eeq
In what follows, when we write $\mathscr{H}_{\sqrdiagup}$ we refer to this representation.

We note that our construction realizes a version of the crossed product which is applicable for algebras acted upon by semigroups of completely positive maps, which we achieved thanks to the generalized Stinespring dilation theorem.\footnote{For semigroup crossed products and dilation constructions see
\cite{Demoen1977, EvansLewis1977, Nica1992, LACA1996415}.
For related applications of generalized crossed-product ideas in quantum gravity,
see \cite{AliAhmad:2025oli,Klinger:2026kqj}.} The result of this section is the algebra $\mathscr{A}_{\sqrdiagup}$, in which the supertranslation action is represented internally, after dilation to an endomorphic action. Although, as we will employ in the following section, the crossed product of an algebra by the automorphic action of a group can be interpreted as implementing a constraint, the algebra $\mathscr{A}_{\sqrdiagup}$ should be interpreted as essentially kinematical. In some sense, it reflects the dual role of the extended corner degrees of freedom $\square$ as an intermediary between the hypersurface and the corner. The algebra $\mathscr{A}_{\sqrdiagup}$ quantizes both the hypersurface degrees of freedom and the expansion, since the latter is the charge aspect associated to supertranslation, see eqn. \eqref{noether ops}. 

\subsection{Solving Raychaudhuri and Damour as Operator Equations} \label{sec: Ray+Dam}

By virtue of the discussion in section \ref{subsub}, and in particular eqns. \eqref{ham} and \eqref{Noe}, imposing the classical constraints of the theory is akin to the requirement
\beq \label{Constraint}
	H_{(f_B,Y)} \os Q_{(f_B,Y)}. 
\eeq
Given our kinematical quantization, we can now provide a quantum interpretation of this constraint. First, let us define by $\mathscr{A}_{\textrm{kin.}} = \mathscr{A}_{\sqrdiagup} \otimes \mathscr{A}_{\bullet}$ the full kinematical algebra. The Hamiltonian $H_{(f_B,Y)}$ should be interpreted as the infinitesimal generator of the action of the group $G_{\bullet}$ on the full kinematical algebra, $\beta^{\bullet}: G_{\bullet} \rightarrow \text{Aut}(\mathscr{A}_{\textrm{kin.}})$:
\beq
	\beta^{\bullet}_{g}(\pi_{\theta^{\square}}(a)) = \pi_{\theta^{\square}} \circ \alpha^{\bullet}_{g}(a), \;\; \beta^{\bullet}_{g}(w_{\square}(g_{\square})) = w_{\square} \circ \alpha^{\bullet,\square}_{g}(g_{\square}), \;\; \beta^{\bullet}_{g}(\ell_{\bullet}(g')) = \ell_{\bullet}(g \circ g' \circ g^{-1}). 
\eeq
On the other hand, the Noether charge $Q_{(f_B,Y)}$ should be interpreted as the infinitesimal generator of the action of the group $G_{\bullet}$ on itself via $\text{Ad}_{\ell_{\bullet}(g)}$. From this point of view, the constraint \eqref{Constraint} tells us that the physical class of operators are precisely those $\mathcal{O} \in \mathscr{A}_{\textrm{kin.}}$ such that\footnote{The form of eqn. \eqref{CP Constraint} makes it clear that the semigroup crossed product for actions by completely positive maps is not implementing such a constraint. As we have discussed, the resulting CP action, while inner implemented, is not implemented by the generators of the algebra of the symmetry in question.}
\beq \label{CP Constraint}
	\beta^{\bullet}_{g}(\mathcal{O}) = \text{Ad}_{\ell_{\bullet}(g)}(\mathcal{O}). 
\eeq
This is nothing but the quantization of \eqref{Constraint} applied as a conjugation to operators in the kinematical algebra:
\beq
    e^{i \op{H}_{(f_B,Y)}} \mathcal{O} e^{-i \op{H}_{(f_B,Y)}} = e^{i \op{Q}_{(f_B,Y)}} \mathcal{O} e^{-i \op{Q}_{(f_B,Y)}}. 
\eeq

We can rearrange eqn. \eqref{CP Constraint} into the form
\beq \label{CP Constraint 2}
    \text{Ad}_{\ell_{\bullet}(g^{-1})} \circ \beta^{\bullet}_{g}(\mathcal{O}) = \mathcal{O}, \qquad \forall g \in G_{\bullet}. 
\eeq
Infinitesimally, eqn. \eqref{CP Constraint 2} identifies the set of operators $\mathcal{O} \in \mathscr{A}_{\textrm{kin.}}$ which commute with $\op{C}_{(f_B,Y)} = \op{H}_{(f_B,Y)} - \op{Q}_{(f_B,Y)}$, the generator of $\text{Ad}_{\ell_{\bullet}(g^{-1})} \circ \beta^{\bullet}_{g}$. This is precisely the quantum constraint operator. As shown in \cite{Ciambelli:2024swv}, the quantum constraint algebra may acquire an anomaly,
obstructing the naive representation of the Lie algebra of $G_\bullet$. One possible path to implement the constraint in this case is to perform a mode splitting, as described, e.g., in \cite{Green:1987sp,DiFrancesco:1997nk,Green:2012pqa}. Alternatively, and along the lines of the recent relevant result \cite{Freidel:2026stu}, the constraint can be implemented even in the presence of the anomaly by centrally extending the group $G_{\bullet}$. To avoid overloading notation, we will continue to denote the $\bullet$ data simply by $G_{\bullet}$. However, we emphasize that this may be centrally extended if an anomaly is present. See Appendix \ref{app: TCP} for more discussion. \\

The algebra generated by operators satisfying the covariance condition Eqn. \eqref{CP Constraint} is
naturally represented by a crossed product \cite{Klinger:2023auu,AliAhmad:2024wja,AliAhmad:2024vdw,Daele_1978}. To be precise, let $\gamma^{\bullet}: G_{\bullet} \rightarrow \text{Aut}(\mathscr{A}_{\sqrdiagup})$ denote the automorphism action of $G_{\bullet}$ on the extended hypersurface algebra. That is
\beq
	\gamma^{\bullet}_{g}(\pi_{\theta^{\square}}(a)) = \pi_{\theta^{\square}} \circ \alpha^{\bullet}_{g}(a), \qquad \gamma^{\bullet}_{g}(w_{\square}(g_{\square})) = w_{\square} \circ \alpha^{\bullet,\square}_{g}(g_{\square}). 
\eeq
Then, we claim that the physical subalgebra of $\mathscr{A}_{\textrm{kin.}}$ is given by $\mathscr{A}_{\sqrbulletdiagup} = \mathscr{A}_{\sqrdiagup} \times_{\gamma^{\bullet}} G_{\bullet}$, following an analogous procedure as described in the previous subsection.\footnote{Since any semigroup is also a group, the definition of the semigroup crossed product envelops that of the standard group crossed product.} Concretely, this can be understood as the algebra on $\mathscr{H}_{\sqrbulletdiagup} = \mathscr{H}_{\sqrdiagup} \otimes L^2(G_{\bullet},\rd\nu_{\bullet})$ generated by $\ell_{\bullet}(G_{\bullet})$ and the representation (for all $\mathfrak{X}\in \mathscr{A}_{\sqrdiagup}$)
\beq \label{bullet dressing}
	\pi_{\gamma^{\bullet}}: \mathscr{A}_{\sqrdiagup} \rightarrow \mathfrak{B}(\mathscr{H}_{\sqrbulletdiagup}), \qquad \bigg(\pi_{\gamma^{\bullet}}(\mathfrak{X}) \psi\bigg)(g) = \gamma^{\bullet}_{g^{-1}}(\mathfrak{X})\bigg(\psi(g)\bigg), 
\eeq
where $\psi$ is a function from $G_\bullet$ into $\mathscr{H}_{\sqrdiagup}$, since any vector in $\mathscr{H}_{\sqrbulletdiagup}$ can be thought of as such a map.
It is easy to compute
\beq
	\text{Ad}_{\ell_{\bullet}(g)}\bigg(\pi_{\gamma^{\bullet}}(\mathfrak{X})\bigg) = \beta^{\bullet}_{g}\bigg(\pi_{\gamma^{\bullet}}(\mathfrak{X})\bigg).
\eeq	
At the same time, the constraint \eqref{CP Constraint} is trivially satisfied for the generators $\ell_{\bullet}(g)$. These facts together imply that $\mathscr{A}_{\sqrbulletdiagup}$ is contained inside the physical subalgebra of $\mathscr{A}_{\textrm{kin.}}$. Under the hypotheses of the commutation theorem \cite{Daele_1978}, this exhausts the full physical
subalgebra, and we therefore identify
$\mathscr{A}_{\sqrbulletdiagup}$ as the sought-after on-shell algebra of a single null hypersurface. \\

In summary, the full physical algebra of a single hypersurface is given by
\beq
    \mathscr{A}_{\sqrbulletdiagup} = \bigg(\mathscr{A}_{\diagup} \times_{\alpha^{\square}} G_{\square}^+\bigg) \times_{\gamma^{\bullet}} G_{\bullet}. 
\eeq
This may be interpreted as the $C^*$ algebra of operators on the Hilbert space\footnote{Given a reference vacuum state $\omega$ on $\mathscr{A}_{\diagup}$, the algebra $\mathscr{A}_{\sqrbulletdiagup}$ will only act on
\beq
    \int_{G_{\sqrbullet}}^{\oplus} \rd\nu_{\bullet}(g_{\bullet}) \; \rd\nu_{\square}(g_{\square}) \; L^2(\mathscr{A}_{\diagup}, \omega \circ \alpha^{\square}_{g_{\square}} \circ \alpha^{\bullet}_{g_{\bullet}}) \subset \mathscr{H}_{\sqrbulletdiagup}.
\eeq
This is a separable Hilbert space, resounding the claim that the algebra $\mathscr{A}_{\sqrbulletdiagup}$ remains separable despite the appearance of the non-separable Hilbert space $L^2(\mathscr{A})$ in its construction (see, e.g., \cite{Klinger:2025tvg,Klinger:2026tws}).}
\beq
    \mathscr{H}_{\sqrbulletdiagup} \simeq L^2(\mathscr{A}_{\diagup}) \otimes L^2(G_{\square}^+,\rd\nu_{\square}) \otimes L^2(G_{\bullet},\rd\nu_{\bullet}),
\eeq
generated by
\beq
    \{\pi_{\gamma^{\bullet}} \circ \pi_{\theta^{\square}}(\op{W}(V)), \pi_{\gamma^{\bullet}}(e^{i \op{Q}_{f_T}}), e^{i \op{Q}_{(f_B,Y)}} \; | \; V \in T_x X_{\diagup}, f_T \un{\ell} \in \mathfrak{g}_{\square}^+, f_B \un{\ell} + \un{Y} \in \mathfrak{g}_{\bullet}\},
\eeq
where $\pi_{\gamma^{\bullet}}: \mathscr{A}_{\sqrdiagup} \rightarrow \mathfrak{B}(\mathscr{H}_{\sqrbulletdiagup})$ (eqn. \eqref{bullet dressing}) and $\pi_{\theta^{\square}}: \mathscr{A}_{\diagup} \rightarrow \mathfrak{B}(\mathscr{H}_{\sqrdiagup})$ (eqn. \eqref{square dressing}) are dressing representations for $G_{\bullet}$ and $G_{\square}^+$, respectively. These operators represent the physical degrees of freedom appearing in the CIVP for either null hypersurface:

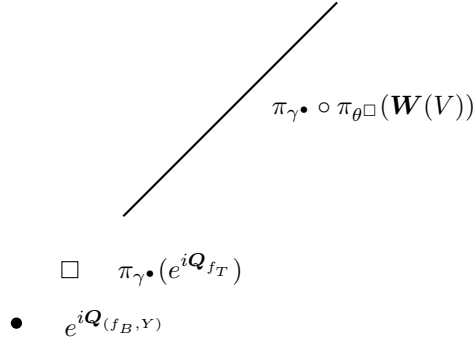
\begin{figure}[H]
\centering
\begin{tikzpicture}[
    scale=2.0,
    line/.style={thick},
    box/.style={draw, rectangle, minimum size=6pt, inner sep=0pt},
    dot/.style={circle, fill=black, inner sep=1.5pt},
    every node/.style={align=center, font=\small}
]

% --- Spacing parameters ---
\def\gapdotbox{0.5}
\def\gapboxline{0.5}
\def\linelength{2}

% --- RIGHT RAY (45°) ---
\coordinate (dotR) at (0,0);                     
\coordinate (boxR) at ($(dotR)+(45:\gapdotbox)$);
\coordinate (lineRstart) at ($(boxR)+(45:\gapboxline)$);
\coordinate (lineRend) at ($(lineRstart)+(45:\linelength)$);

\node[dot] at (dotR) {};
\node[box] at (boxR) {};
\draw[line] (lineRstart) -- (lineRend);

% Labels (named nodes)
\node[right] (labelDot) at ($(dotR)+(0.25,0)$)
    {$e^{i\op{Q}_{(f_B,Y)}}$};

\node[right] (labelBox) at ($(boxR)+(0.25,0)$)
    {$\pi_{\gamma^{\bullet}}(e^{i\op{Q}_{f_T}})$};

\node[right] (labelLine) at ($(lineRend)+(-0.5,-0.7)$)
    {$\pi_{\gamma^{\bullet}} \circ \pi_{\theta^{\square}}(\op{W}(V))$};

\end{tikzpicture}
\caption{Constraint quantization of a single hypersurface. The hierarchy of actions is translated into a hierarchy of dressings; every piece of data is dressed to all data below it.}
\end{figure}

It is important to stress again that the derivation of gauge-invariant quantum operators is largely dictated by the hierarchy of symmetries: $\bullet$ acts on $\square$, and both act on $\diagup$. The action of $\square$ is particularly subtle, as emphasized above. Physically, boosts and spatial rotations implement the constraints through the crossed product, whereas supertranslations move the cut and therefore give rise, at the quantum level, only to completely positive maps. The construction of the final algebra $\mathscr{A}_{\sqrbulletdiagup}$ is a novel and central result of the manuscript.

\section{Corner Gluing} \label{sec: gluing}

In the previous sections, we have constructed bit by bit the algebra of on-shell observables on a single null hypersurface, $\mathscr{A}_{\sqrbulletdiagup}$. This was done by starting with the algebra in the bulk of the hypersurface and appending to it the corner data. Doing so properly accounts for the initial cut while simultaneously implementing the intrinsic Einstein constraints -- the Raychaudhuri and Damour equations -- as quantum operator relations. Therefore, $\mathscr{A}_{\sqrbulletdiagup}$ is the final on-shell algebra of a portion of a null hypersurface, from an initial cut to caustics.

As we have emphasized in Sec.~\ref{sec: Classical}, this algebra is not sufficient on its own to encode the full collection of diffeomorphism invariant observables for a spacetime region. This is because the CIVP requires two intersecting hypersurfaces to reconstruct a local spacetime region. Thus, $\mathscr{A}_{\sqrbulletdiagup}$ must be supplemented by an analogous algebra, $\mathscr{A}_{\sqrbulletdiagdown}$, corresponding to the complementary hypersurface along with a set of gluing conditions which specify how these two hypersurfaces fit together at their shared corner. \\

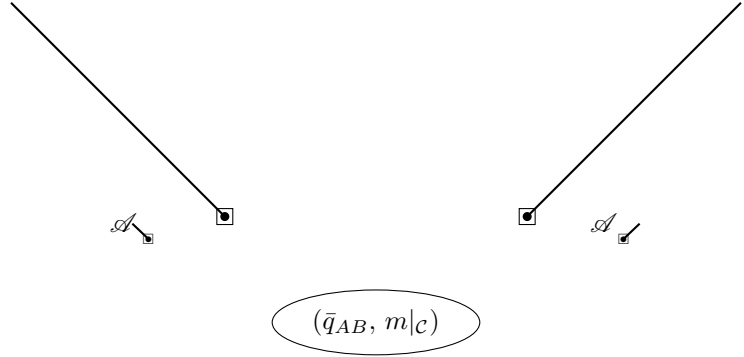
\begin{figure}[ht] \label{fig:CIVPDeconstructed2}
\centering
\begin{tikzpicture}[
    scale=2.0,
    line/.style={thick},
    sqdot/.style={draw, rectangle, minimum size=6pt, inner sep=0pt},
    dot/.style={circle, fill=black, inner sep=1.2pt},
    oval/.style={draw, ellipse, minimum width=1.2cm, minimum height=0.5cm},
    every node/.style={align=center, font=\small}
]

% --- Spacing parameters ---
\def\linelength{2}
\def\ovalgap{0.7}

% --- RIGHT RAY (45°) ---
\coordinate (lineRstart) at (1,0);
\coordinate (lineRend) at ($(lineRstart)+(45:\linelength)$);

\draw[line] (lineRstart) -- (lineRend);

% square-with-dot at bottom
\node[sqdot] (sqR) at (lineRstart) {};
\node[dot] at (sqR.center) {};

% Right labels
\node[right] at ($(lineRstart)+(0.35,-0.1)$) {$\mathscr{A}_{\sqrbulletdiagup}$};
\node[right] at ($(lineRend)+(0.25,0.1)$) {};

% --- LEFT RAY (135°) ---
\coordinate (lineLstart) at (-1,0);
\coordinate (lineLend) at ($(lineLstart)+(135:\linelength)$);

\draw[line] (lineLstart) -- (lineLend);

% square-with-dot at bottom
\node[sqdot] (sqL) at (lineLstart) {};
\node[dot] at (sqL.center) {};

% Left labels
\node[left] at ($(lineLstart)+(-0.35,-0.1)$) {$\mathscr{A}_{\sqrbulletdiagdown}$};
\node[left] at ($(lineLend)+(-0.25,0.1)$) {};

% --- CENTRAL OVAL ---
\node[oval] at (0,-\ovalgap) {$(\bar{q}_{AB},\, m\rvert_{\mathcal{C}})$};

\end{tikzpicture}
\caption{The CIVP algebra prior to gluing.}
\end{figure}

\subsection{The Glued Algebra}

The algebra $\mathscr{A}_{\sqrbulletdiagdown}$ is formed by following the same sequence of steps as was described in Sec.~\ref{sec: STCrossed}-\ref{sec: Ray+Dam}, only applied to the kinematical variables of the left hypersurface. Consequently, the algebras $\mathscr{A}_{\sqrbulletdiagdown}$ and $\mathscr{A}_{\sqrbulletdiagup}$ will be isomorphic. To prepare these algebras for gluing, it is useful to represent them simultaneously on a single common Hilbert space. A natural candidate for this Hilbert space is
\beq
	\mathscr{H}_{\sqrbulletV} = \mathscr{H}_{\diagup} \otimes \mathscr{H}_{\square_u} \otimes \mathscr{H}_{\square_v} \otimes \mathscr{H}_{\bullet},
\eeq
where\footnote{As in footnote \ref{fn: GNSDouble}, the Hilbert space $\mathscr{H}_{\diagup}$ knows about more than just the $\diagup$ degrees of freedom. This reflects the standard left-right representation of the hypersurface algebra: the two null branches are represented by mutually commuting left and right actions on the same
Hilbert space, rather than by two independent tensor factors. The fact that $\mathscr{H}_{\diagup}$ admits commuting left and right representations of $\mathscr{A}_{\diagup}$ is central to the following construction.} $\mathscr{H}_{\diagup} = L^2(\mathscr{A}_{\diagup})$, $\mathscr{H}_{\square_u} \simeq \mathscr{H}_{\square_v} = L^2(G_{\square}^+,\rd\nu_{\square})$, and $\mathscr{H}_{\bullet} = L^2(G_{\bullet},\rd\nu_{\bullet})$. We can regard this Hilbert space as densely spanned by states of the form $\ket{a, g_{\square}^u, g_{\square}^v, g_{\bullet}}$ where $a \in \mathscr{A}_{\diagup}, g_{\square}^u, g_{\square}^v \in G_{\square}^+$, and $g_{\bullet} \in G_{\bullet}$. Then, the algebras $\mathscr{A}_{\sqrbulletdiagup}$ and $\mathscr{A}_{\sqrbulletdiagdown}$ are embedded into $\mathfrak{B}(\mathscr{H}_{\sqrbulletV})$ via the following representations: $\pi_{\sqrbulletdiagup},\pi_{\sqrbulletdiagdown}: \mathscr{A}_{\sqrbulletdiagup} \rightarrow \mathfrak{B}(\mathscr{H}_{\sqrbulletV})$
\begin{flalign}
	&\pi_{\sqrbulletdiagup}(a^u) \ket{b,g_{\square}^u, g_{\square}^v, g_{\bullet}} = \ket{\gamma^{\bullet}_{g_{\bullet}^{-1}}(a^u) b, g_{\square}^u, g_{\square}^v, g_{\bullet}}, \nonumber \\
	&\pi_{\sqrbulletdiagup}(h_{\square}^u) \ket{b,g_{\square}^u, g_{\square}^v, g_{\bullet}} = \ket{\theta^{\square_u}_{\gamma^{\bullet}_{g_{\bullet}^{-1}}(h_{\square}^u)}(b),\gamma^{\bullet}_{g_{\bullet}^{-1}}(h_{\square}^u) \circ g_{\square}^u, g_{\square}^v, g_{\bullet}}, \nonumber \\
	&\pi_{\sqrbulletdiagup}(h_{\bullet}^u) \ket{b,g_{\square}^u, g_{\square}^v, g_{\bullet}} = \ket{b, g_{\square}^u, g_{\square}^v, h_{\bullet}^u \circ g_{\bullet}}, \nonumber \\
	&\pi_{\sqrbulletdiagdown}(a^v) \ket{b,g_{\square}^u, g_{\square}^v, g_{\bullet}} = \ket{b \gamma^{\bullet}_{g_{\bullet}^{-1}}(a^v)^*, g_{\square}^u, g_{\square}^v, g_{\bullet}}, \nonumber \\
	&\pi_{\sqrbulletdiagdown}(h_{\square}^v) \ket{b,g_{\square}^u, g_{\square}^v, g_{\bullet}} = \ket{\theta^{\square_v}_{\gamma^{\bullet}_{g_{\bullet}^{-1}}(h_{\square}^v)}(b), g_{\square}^u, \gamma^{\bullet}_{g_{\bullet}^{-1}}(h_{\square}^v) \circ g_{\square}^v, g_{\bullet}}, \nonumber \\
	&\pi_{\sqrbulletdiagdown}(h_{\bullet}^v) \ket{b,g_{\square}^u, g_{\square}^v, g_{\bullet}} = \ket{\gamma^{\bullet}_{h_{\bullet}^v}(b), g_{\square}^u, \gamma^{\bullet}_{h_{\bullet}^v}(g_{\square}^v), g_{\bullet} \circ h_{\bullet}^v{}^{-1}}. 
\end{flalign}
Here, $\theta^{\square_u}$ and $\theta^{\square_v}$ are the dilated endomorphism actions of the semigroups $G_{\square_u}^+$ and $G_{\square_v}^+$ which are treated as independent and commuting at this stage. In the first three equations, $\pi_{\sqrbulletdiagup}$ defines $\diagup$, $\square$, and $\bullet$ operators acting on the left in the Hilbert space $\mathscr{H}_{\sqrbulletV}$. We note that the representation of $\diagup$ is dressed by the actions of $\bullet$ and $\square$ such that intertwining with $\pi_{\sqrbulletdiagup}(h^u_{\square})$, and $\pi_{\sqrbulletdiagup}(h^u_{\bullet})$ implement supertranslations, and superboosts/superrotations, respectively. Likewise, the representation of $\square$ is dressed by the action of $\bullet$ such that intertwining with $\pi_{\sqrbulletdiagup}(h^u_{\bullet})$ implements superboosts/superrotations on $\pi_{\sqrbulletdiagup}(h^u_{\square})$. At the same time, in the latter three equations $\pi_{\sqrbulletdiagdown}$ defines an action of the $\diagdown$, $\square$ and $\bullet$ operators from the right on the same Hilbert space. These actions possess an analogous hierarchy of dressing ensuring the inner implementation of supertranslations, superboosts and superrotations on the right hypersurface degrees of freedom. The representation is slightly modified relative to $\pi_{\sqrbulletdiagup}$ to ensure that, by construction, the algebras
\beq
	\pi_{\sqrbulletdiagup}(\mathscr{A}_{\sqrbulletdiagup}) = \{\pi_{\sqrbulletdiagup}(a^u), \pi_{\sqrbulletdiagup}(h_{\square}^u), \pi_{\sqrbulletdiagup}(h_{\bullet}^u) \; | \; a^u \in \mathscr{A}_{\diagup}, h_{\square}^u \in G_{\square}^+, h^u_{\bullet} \in G_{\bullet}\},
\eeq
and 
\beq
	\pi_{\sqrbulletdiagdown}(\mathscr{A}_{\sqrbulletdiagup}) = \{\pi_{\sqrbulletdiagdown}(a^v), \pi_{\sqrbulletdiagdown}(h_{\square}^v), \pi_{\sqrbulletdiagdown}(h_{\bullet}^v) \; | \; a^v \in \mathscr{A}_{\diagup}, h_{\square}^v \in G_{\square}^+, h^v_{\bullet} \in G_{\bullet}\}
\eeq
define two, mutually commuting copies of the algebra $\mathscr{A}_{\sqrbulletdiagup}$ contained in $\mathfrak{B}(\mathscr{H}_{\sqrbulletV})$.\\

The final part of the quantization is to implement the appropriate gluing conditions indicated by the CIVP as drawn in Figure \ref{fig:CIVP = CIVPDecon}. That is, we will aim to define our complete CIVP algebra as
\beq \label{Gluing Schematic}
	\mathscr{A}_{\sqrbulletV} = \bigg(\pi_{\sqrbulletdiagdown}(\mathscr{A}_{\sqrbulletdiagup}) \otimes_{\bigcirc} \pi_{\sqrbulletdiagup}(\mathscr{A}_{\sqrbulletdiagup})\bigg)/ \sim_\bullet. 
\eeq
This algebra is constructed in two steps as indicated by $\sim_{\bullet}$ and $\otimes_{\bigcirc}$:
\begin{enumerate}
    \item Step one relates the $\bullet$ data of the two hypersurface theories, since the final algebra must refer to a single joint initial cut. This is realized by enforcing the equivalence relation $\sim_{\bullet}$.
    \item Step two relates the $\square$ data of the two hypersurface theories. This is a more subtle procedure: the two $\square$ operators are independent, but as we will show they may satisfy a non-trivial commutation relation determined by $\bigcirc$.
\end{enumerate}
To reiterate, the CIVP fixes the required structure of the gluing problem. We here provide a general algebraic framework for this gluing and then propose a natural realization motivated by the geometry of double-null transport.

\subsection{Charge Matching and Expansion Algebra}

We can realize these steps by constructing a final representation
\beq
	\pi_{\sqrbulletV}: \pi_{\sqrbulletdiagdown}(\mathscr{A}_{\sqrbulletdiagup}) \otimes \pi_{\sqrbulletdiagup}(\mathscr{A}_{\sqrbulletdiagup}) \rightarrow \mathfrak{B}(\mathscr{H}_{\sqrbulletV}). 
\eeq
The simplest aspect of the gluing is dealing with the (dressed) hypersurface data. The gluing only implicates data at the shared corners and so we take
\beq
	\pi_{\sqrbulletV} \circ \pi_{\sqrbulletdiagup}(a^u) = \pi_{\sqrbulletdiagup}(a^u), \qquad \pi_{\sqrbulletV} \circ \pi_{\sqrbulletdiagdown}(a^v) = \pi_{\sqrbulletdiagdown}(a^v). 
\eeq

Addressing step one of the gluing is also rather straightforward. The tensor product algebra above overcounts the $\bullet$ degrees of freedom. There should be only one copy of this data contained in the final, physical algebra. However, this identification must be implemented in such a way as to ensure the Raychaudhuri and Damour constraints are undisturbed. At the purely formal level, this may be represented by a diagonal projection on the two copies of the pure corner group. Denoting the two generators by $\pi_{\sqrbulletdiagdown}(g_{\bullet}^u)$ and $\pi_{\sqrbulletdiagup}(g_{\bullet}^v)$, we write
\beq
    \pi_{\sqrbulletV}\bigg(\pi_{\sqrbulletdiagdown}(g_{\bullet}^u) \otimes \pi_{\sqrbulletdiagup}(g_{\bullet}^v)\bigg)=\delta_\Delta(g^u_\bullet,g^v_\bullet)\pi_{\sqrbulletdiagdown}(g_{\bullet}^u{}^{-1}) \pi_{\sqrbulletdiagup}(g_{\bullet}^u)\,,
\eeq
where $\delta_\Delta(g^u_\bullet,g^v_\bullet)$ denotes the delta distribution supported on
the diagonal $g^u_\bullet=g^v_\bullet$. This formula is only a shorthand for the algebraic quotient procedure that
identifies the two pure-corner copies and retains the diagonal action. In a fully analytic treatment, the diagonal projection would have to be defined after choosing the appropriate representation and quasi-invariant measure on $G_\bullet$.\\

This representation retains the property as discussed in eqn. \eqref{CP Constraint} of implementing the gravitational constraints in the form of an operator equation as
\beq \label{Double CP Constraint}
	\text{Ad}_{\pi_{\sqrbulletV}(g_{\bullet})}(\mathfrak{X}) = \beta^{\bullet}_{g_{\bullet}}(\mathfrak{X}),
\eeq
where $\beta^{\bullet}$ is the action of $G_{\bullet}$ on all of the kinematical degrees of freedom now including \emph{both} null hypersurfaces. Eqn. \eqref{Double CP Constraint} holds for any operator $\mathfrak{X}$ of the form $\pi_{\sqrbulletdiagup}(a^u), \pi_{\sqrbulletdiagdown}(a^v), \pi_{\sqrbulletdiagup}(g_{\square}^u), \pi_{\sqrbulletdiagdown}(g_{\square}^v)$, and $\pi_{\sqrbulletV}(g_{\bullet})$, which follows from the fact that $\pi_{\sqrbulletdiagup}$ and $\pi_{\sqrbulletdiagdown}$ are commuting representations of $\mathscr{A}_{\sqrbulletdiagup}$. This concludes step one in the construction of $\mathscr{A}_{\sqrbulletV}$. \\

Step two tells us how to relate the two expansions down at the corner. It is not as easy as simply retaining the same representation, as for the pure hypersurface data, or restricting to a diagonal subalgebra, as for the pure corner data, since we know that the CIVP specifies two independent but interacting expansion scalars. This is the reason why we have included two independent copies of $\mathscr{H}_{\square}$ in $\mathscr{H}_{\sqrbulletV}$. In what follows, we present a general approach to constructing the relative tensor product $\otimes_{\bigcirc}$. We then propose a specific implementation of our general construction, motivated by a series of physical considerations. 

A productive way to think about the corner gluing for the $\square$ data is as specifying the action of a $u$-supertranslation on a $v$-supertranslation, and vice versa. In general, let us suppose that there are a pair of such actions
\beq
    \bigcirc^{ij}: G_{\square_i}^+ \times G_{\square_j}^+ \rightarrow G_{\square_j}^+, 
\eeq
where $\bigcirc^{uv}$ denotes the action of a $u$-supertranslation on a $v$-supertranslation and $\bigcirc^{vu}$ the action of a $v$-supertranslation on a $u$-supertranslation. Given these data, we can form the bi-crossed product\footnote{This is also known as the Zappa-Szép product, see e.g. \cite{Brin_2005,BROWNLOWE20143937} and the references therein for a further discussion.} semigroup $G_{\square_u}^+ \Join_{\bigcirc} G_{\square_v}^+$, which is based on the vector space $G_{\square_u}^+ \times G_{\square_v}^+$ endowed with algebraic relations
\beq \label{ST Twisting}
    (g_{\square}^u, g_{\square}^v) \circ (h_{\square}^u, h_{\square}^v) = \bigg(g_{\square}^u \circ \bigg(\bigcirc^{vu}_{g_{\square}^v}(h_{\square}^u)\bigg), \bigg(\bigcirc^{uv}_{h_{\square}^u}(g_{\square}^v)\bigg) \circ h_{\square}^v\bigg). 
\eeq\\

Assuming that the actions $\bigcirc^{ij}$ are unital and denoting by $e_{\square}^i$ the unit element of $G_{\square_i}^+$ we find that
\beq
    (g_{\square}^u, e_{\square}^v) \circ (e_{\square}^u, g_{\square}^v) = (g_{\square}^u, g_{\square}^u).
\eeq
Thus, we can regard the semigroup $G_{\square_u}^+ \Join_{\bigcirc} G_{\square_v}^+$ as being generated by $(g_{\square}^u,e_{\square}^v)$ and $(e_{\square}^u, g_{\square}^v)$ for $g_{\square}^u \in G_{\square_u}^+, g_{\square}^v \in G_{\square_v}^+$. With this observation in mind, we can construct the representation $\pi_{\sqrbulletV}$ to respect the bi-crossed product structure by identifying
\beq
    \ket{b,h_{\square}^u,h_{\square}^v, h_{\bullet}} = \ket{b, (h_{\square}^u, h_{\square}^v), h_{\bullet}}
\eeq
and defining
\begin{flalign}
    &\pi_{\sqrbulletV}(g_{\square}^u) \ket{b,(h_{\square}^u,h_{\square}^v), h_{\bullet}} = \ket{\theta^{\square_u}_{\gamma^{\bullet}_{h_{\bullet}^{-1}}(g_{\square}^u)}(b), (\gamma_{h_{\bullet}^{-1}}(g_{\square}^u),e_{\square}^v) \circ (h_{\square}^u, h_{\square}^v), h_{\bullet}} \nonumber \\
    &\pi_{\sqrbulletV}(g_{\square}^v) \ket{b,(h_{\square}^u, h_{\square}^v), h_{\bullet}} = \ket{\theta^{\square_v}_{\gamma^{\bullet}_{h_{\bullet}^{-1}}(g_{\square}^v)}(b), (h_{\square}^u, h_{\square}^v) \circ (e_{\square}^u, \gamma^{\bullet}_{h_{\bullet}^{-1}}(g_{\square}^{v})), h_{\bullet}}
\end{flalign}
Due to the dressing of the other operators in the algebra, if the representation $\pi_{\sqrbulletV}$ is chosen to respect the non-trivial action \eqref{ST Twisting} the $u$-supertranslations will inherit a non-trivial action on the operators in the $v$-hypersurface, and vice versa. This provides a general algebraic implementation of step two. Therefore, we have completely characterized the algebra $\mathscr{A}_{\sqrbulletV}$ in \eqref{Gluing Schematic}. As stressed, the formal construction up to here is fully general. \\

\paragraph{A specific construction} We can now provide a speculative construction of this gluing pattern, with far-reaching consequences. This is where the physics fun begins! Indeed, there is a natural (emergent) spacetime interpretation for the action \eqref{ST Twisting}. Namely, we regard the semigroup $G_{\square_u}^+ \Join_{\bigcirc} G_{\square_v}^+$ as furnishing a representation of the algebra of parallel transports in the null directions $u$ and $v$. In this sense, the information required to determine this action is encoded entirely in the data $\bigcirc$ from the point of view of the CIVP -- hence our notation. Recall, $\bigcirc$ includes the relative angle between the two null hypersurfaces and the unimodular part of the metric at the corner. The former informs us how the $u$-supertranslation projects onto the $v$-hypersurface and vice versa. The latter can be traded for a Riemann curvature \cite{Reisenberger:2007ku,Reisenberger:2018xkn,Reisenberger:2012zq} which informs us of the commutator between covariant derivatives along the two clock vector fields.\footnote{From the point of view of the present quantization, we are treating this data as essentially fixed. It is interesting to ask whether this data should too be quantized to dynamical degrees of freedom and included as operators in the CIVP algebra. We leave this question to future work.}

Writing
\beq
    \pi_{\sqrbulletV}(\text{exp}(f_T^u \un{\ell}_u)) = e^{i \op{Q}^u_{f_T^u}}, \qquad \pi_{\sqrbulletV}(\text{exp}(f_T^v \un{\ell}_v)) = e^{i \op{Q}^v_{f_T^v}},
\eeq
we can reformulate the bi-crossed product analysis at the infinitesimal level. From the representation theoretic point of view, tying together the two hypersurfaces entails gluing together two copies of the corner symmetry algebra $\mathfrak{g}_{\sqrbullet}$. The resulting, extended corner symmetry algebra will then be of the form
\beq
    \mathfrak{g}_{\square \bigcircbullet \square} = \mathfrak{g}_{\bullet} \ltimes (\mathfrak{g}_{\square} \Join_{\bigcirc} \mathfrak{g}_{\square}).
\eeq
One way to ensure that $\mathfrak{g}_{\square \bigcircbullet \square}$ obeys the requisite properties of a Lie algebra (e.g. the Jacobi identity) is to take the commutation relation between the supertranslation charges to be of the form
\beq
    [\op{Q}^u_{f_T^u}, \op{Q}^v_{f_T^v}] = i K_{\bigcirc}(f_T^u, f_T^v) \op{1}.
\eeq
Here $K_{\bigcirc}$ is assumed to define a bilinear antisymmetric field independent integral kernel on $C^{\infty}(\mathcal{C})$, compatible with the $G_\bullet$ action, that is, a central extension. This means that, in this specific construction, the $u$ and $v$ supertranslations cease to commute at the quantum level, in a minimal way. We will explore some consequences of this in the next section. This algebra has actually appeared previously in gravitational contexts, such as t' Hooft's analysis of gravitational shockwaves \cite{tHooft:1996rdg}. It can also be understood as an infinite dimensional generalization of the central extension to the corner symmetry algebra, as discussed in \cite{Ciambelli:2024qgi}. It remains an important question whether this central extension is the unique consistent form of the gluing condition; this is currently under investigation. 

\section{Discussion and Looking Forward} \label{sec: disc}

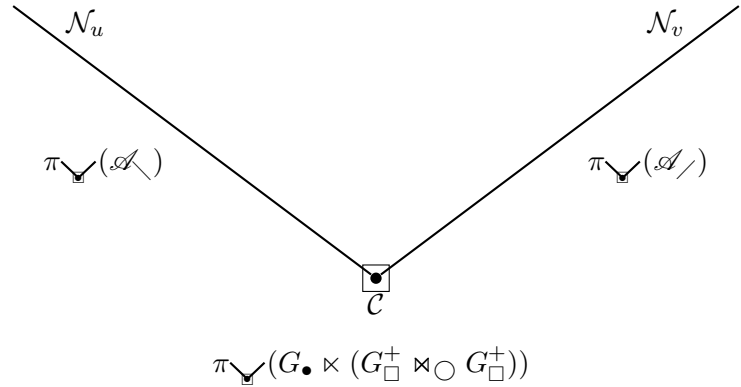
\begin{figure}[H] 
\centering
\begin{tikzpicture}[
    scale=1.2,
    every node/.style={align=center},
    line/.style={thick},
    point/.style={circle, fill=black, inner sep=1.5pt}
]

% Intersection point
\node[point, label=below:$\mathcal{C}$] (C) at (0,0) {};

% Left and right lines (forming a V)
\draw[line] (C) -- (-4,3);
\draw[line] (C) -- (4,3);

% Labels for the null hypersurfaces
\node at (-3.2,2.8) {$\mathcal{N}_u$};
\node at (3.2,2.8) {$\mathcal{N}_v$};
\node [draw, rectangle, minimum width=0.35cm, minimum height=0.35cm] at (C) {};

% Data on left line
\node at (-3,1.2) {$\pi_{\sqrbulletV}(\mathscr{A}_{\diagdown})$};

% Data on right line
\node at (3,1.2) {$\pi_{\sqrbulletV}(\mathscr{A}_{\diagup})$};

% Data at the intersection point
\node at (0,-1) {
$\pi_{\sqrbulletV}(G_{\bullet} \ltimes (G_{\square}^+ \Join_{\bigcirc} G_{\square}^+))$
};

\end{tikzpicture}
\caption{Operator algebra of the CIVP.}
\label{fig:OA_CIVP}
\end{figure}

The goal of this paper was to systematically construct the quantum physical (i.e., on shell) algebra associated with the CIVP which we found to be of the form\footnote{The notation $\vee$ stands for the $C^*$ algebraic union. Given a pair of $C^*$ algebras $A,B$ with representations $\pi_A: A \rightarrow \mathfrak{B}(H)$ and $\pi_B: B \rightarrow \mathfrak{B}(H)$, $\pi_A(A) \vee \pi_B(B)$ is the $C^*$ algebra of operators on $H$ obtained by sums and sequences of operators of the form $\pi_A(a) \pi_B(b)$ with respect to the $C^*$ operator topology of $\mathfrak{B}(H)$.}
\beq\label{finAlg}
    \mathscr{A}_{\sqrbulletV} &=& \pi_{\sqrbulletV}(\mathscr{A}_{\diagdown}) \vee \pi_{\sqrbulletV}(G_{\bullet}) \vee \pi_{\sqrbulletV}(G_{\square}^+ \Join_{\bigcirc} G_{\square}^+) \vee \pi_{\sqrbulletV}(\mathscr{A}_{\diagup})\\
    &=&\bigg(\pi_{\sqrbulletdiagdown}(\mathscr{A}_{\sqrbulletdiagup}) \otimes_{\bigcirc} \pi_{\sqrbulletdiagup}(\mathscr{A}_{\sqrbulletdiagup})\bigg)/ \sim_\bullet,
\eeq
as depicted in Figure \ref{fig:OA_CIVP}. Here, we recall that $G_{\square}^+ \Join_{\bigcirc} G_{\square}^+$ is the semigroup of positive supertranslations in the $u$ or $v$ directions equipped with an algebra determined by the gluing data, $\bigcirc$, in the CIVP. 

To construct this algebra, we first constructed the classical gravitational phase space on null hypersurfaces, and formulated the CIVP. Emphasis was put on the universal symmetry structure, and in particular the supertranslations
semi-group. This classical part also allowed us to set up the useful symbolic notation $\diagup$, $\square$, and $\bullet$, associated with degrees of freedom in the bulk of the hypersurface, supertranslation charge at the cut (expansion), and superboost and superrotation charges at the cut, respectively. 

We then quantized each block separately, and showed how they act on each other at the kinematical level. In particular, $\bullet$ acts on $\square$ and both of them act on $\diagup$. It is crucial to appreciate and respect this nesting in the quantization procedure. Moreover, it was important that the supertranslation semigroup is embedded inside a group in order to canonically quantize it. 

We then moved to the physical algebra by suitably gluing together the various blocks. The algebra $\mathscr{A}_\bullet$ acts automorphically on the rest and so we can use the crossed product construction to impose the gravitational constraints. Conversely, $\mathscr{A}_\square$ acts only by completely positive maps, and we resorted to Stinespring's theorem to dilate this action to algebra-preserving endomorphisms. The outcome of this procedure was the construction of the physical algebra of a single null hypersurface, $\mathscr{A}_{\sqrbulletdiagup}$.

Eventually, we glued together two such hypersurfaces in a two-step procedure. First, we needed to pass from two independent initial cuts $\bullet$ to a single shared one, and we did so by equating their data while simultaneously respecting the gravitational constraints. Secondly, the $\square$ data must be consistently joined together. This was more subtle, and we did so by constructing a relative tensor product dictated by the intrinsic data of the initial cut $\bigcirc$. This long and tortuous road led us to the final physical algebra $\mathscr{A}_{\sqrbulletV}$ reported in \eqref{finAlg}.\\

As we now discuss, this algebra encompasses and expands upon many important observations that have emerged in the study of subregions in quantum gravity in the last several years. It also lends itself to a number of fascinating questions and directions for future work. To see this, it is a useful exercise to decompose and subsequently reassemble this algebra in various ways.

\subsection{Subcases} 

To begin, we note that the algebras
\beq
	\mathscr{A}_{\bulletdiagdown} = \pi_{\sqrbulletV}(\mathscr{A}_{\diagdown}) \vee \pi_{\sqrbulletV}(G_{\bullet}), \qquad \mathscr{A}_{\bulletdiagup} = \pi_{\sqrbulletV}(\mathscr{A}_{\diagup}) \vee \pi_{\sqrbulletV}(G_{\bullet})
\eeq
are crossed-product algebras associated with imposing the
corner symmetry constraints on a single null branch. These algebras generalize the constructions described in \cite{Klinger:2023tgi,Chandrasekaran:2022eqq,Chandrasekaran:2022cip,Jensen:2023yxy,AliAhmad:2023etg,Kudler-Flam:2023qfl,Klinger:2026tws,Faulkner:2024gst} which consider the invariant algebras with respect to subgroups of $G_{\bullet}$. For instance, in \cite{Chandrasekaran:2022eqq,Chandrasekaran:2022cip,Kudler-Flam:2023qfl} only the global boost and isometry constraints are considered. In \cite{Klinger:2026tws}, the full group of boost supertranslations and global isometries were treated. Finally, in \cite{Klinger:2025tvg} it was discussed how to include also superrotations in addition to superboosts, leading to the full algebra $\mathscr{A}_{\bulletdiagup}$. \\

Subregions in quantum gravity are only well posed for analysis in the event that they can be specified in a diffeomorphism invariant fashion. To this point, the presence of the two algebras $\mathscr{A}_{\bulletdiagdown}$ and $\mathscr{A}_{\bulletdiagup}$ in the CIVP can be read as a consequence of the relational character of the initial value problem in gravity. The region of interest is defined invariantly by the intersection of two null hypersurfaces. Any observable in the $u$-theory is implicitly described \emph{relative} to the $v$-theory and vice versa. As recently advocated in \cite{AliAhmad:2024wja,AliAhmad:2024vdw}, this relational feature is realized with respect to intrinsically well defined gravitational degrees of freedom, without the need to add any adjunct ingredients. \\

The complementary subalgebra
\beq
	\mathscr{A}_{\square \bigcirc \square} = \pi_{\sqrbulletV}(G_{\square}^+ \Join_{\bigcirc} G_{\square}^+)
\eeq
can be interpreted as a generalization of `t Hooft's gravitational shockwave algebra \cite{tHooft:1996rdg}. Indeed, the representation $\pi_{\sqrbulletV}$ is designed specifically such that the expansion operators act on the dressed hypersurface algebras $\mathscr{A}_{\bulletdiagdown}$ and $\mathscr{A}_{\bulletdiagup}$ so as to implement the non-algebra preserving action \eqref{Def of ST-CP Action}. Supertranslations play a critical role in understanding the physics of gravitational subregions, and this was recently used in \cite{Chandrasekaran:2026pnc} to construct a von Neumann algebra for subregions starting from null hypersurfaces. In \cite{Chandrasekaran:2026pnc}, supertranslations have been implemented on the hypersurface as an algebra preserving map. Here, we differ from that analysis since, thanks to the symmetry structure of the CIVP as described in Sec.~\ref{sec: Classical}, supertranslations form a semigroup which acts by completely positive maps rather than automorphisms. It would be interesting to reformulate the results in \cite{Chandrasekaran:2026pnc} in light of this feature.

\subsection{Real Time Evolution}

The overall algebra $\mathscr{A}_{\sqrbulletV}$ can be read as
\beq
	\mathscr{A}_{\sqrbulletV} = \mathscr{A}_{\bulletdiagdown} \vee \mathscr{A}_{\square \bigcirc \square} \vee \mathscr{A}_{\bulletdiagup}.
\eeq
We note that the algebras $\mathscr{A}_{\bulletdiagdown}$ and $\mathscr{A}_{\bulletdiagup}$ share a copy of $\pi_{\sqrbulletV}(G_{\bullet})$ which is not treated as distinct across the joining operation $\vee$. Thus, the algebra $\mathscr{A}_{\sqrbulletV}$ includes only one copy of the superboost/superrotation algebra, as desired. The algebra $\mathscr{A}_{\square \bigcirc \square}$ crucially encodes the non-trivial commutation relation between the expansions of the two hypersurfaces. In this sense, we are allowed to consider general sums and products of these generators and their action on the hypersurface. Taking inspiration from the ambient spacetime picture, one argues that the operator
\beq
	\op{\mathcal{H}}_{t} = \op{Q}^u_t + \op{Q}^v_t
\eeq	
may be regarded as the generator of time-like supertranslations. In this way, the algebra $\mathscr{A}_{\sqrbulletV}$ acquires a fully relational, diffeomorphism invariant notion of real time evolution, realizing in concreteness the hypothesis formulated in \cite{Klinger:2025hjp}. In the CIVP algebra this should be understood not as an ordinary reversible Hamiltonian evolution, but as a relational semigroup evolution: it moves the shared corner upward to its timelike future and thereby produces a nested family of CIVP algebras.

\subsection{Algebraic Type}

One point that we have not addressed thus far is the type of the algebra $\mathscr{A}_{\sqrbulletV}$, or more precisely its von Neumann closure. The von Neumann algebra invariant under isometries \cite{Chandrasekaran:2022cip} and superboosts \cite{Klinger:2026tws} for the exterior of a black hole or a de Sitter static patch is of type II. The inclusion of superrotations and supertranslations introduces obstructions to retaining this conclusion relative to the algebra $\mathscr{A}_{\sqrbulletV}$. The proof that the algebra is semifinite in \cite{Klinger:2026tws} was based on \cite{AliAhmad:2024eun} which observed that the crossed product of a type III von Neumann algebra $\mathscr{A}$ with a group $G$ is semifinite provided there exists a faithful, semifinite, normal weight $\varphi$ on $\mathscr{A}$ whose modular automorphism group is contained in $G$ and which is left invariant under the full action of $G$. This result holds in the cases described in \cite{Klinger:2026tws} because the Unruh/Hartle-Hawking states have geometric modular flows and are left invariant under all isometries and superboosts. However, these states are not preserved by superrotations which means their inclusion into the algebra may obstruct the type change. For more general regions it is unclear whether or not there even exists a state with geometric modular flow \cite{Jensen:2023yxy}.\footnote{See \cite{Caminiti:2025hjq} for a study of geometric modular flows in CFTs, which already shows in field theories that it is in general hard to construct states with geometric flows, making therefore the working assumption in \cite{Jensen:2023yxy} particularly delicate.} Finally, even if one could find a state with geometric modular flow which is preserved under the action of superboosts and superrotations, the algebra $\mathscr{A}_{\sqrbulletV}$ now includes also supertranslations which are not even algebra preserving. As such, one would require a generalization of the proof described in \cite{AliAhmad:2024eun} to address this case. \\

Having made these comments, there exists an alternative argument that the von Neumann closure of $\mathscr{A}_{\sqrbulletV}$ may be of type II. When forming the algebra $\mathscr{A}_{\sqrbulletV}$ we took an algebraic union of $\mathscr{A}_{\diagup}$ and $\mathscr{A}_{\diagdown}$, which were regarded as commutant algebras. This may imply that the total hypersurface contributes an algebra of type I. The hypersurface degrees of freedom are typically the obstruction to semifiniteness, contributing a factor of type III in standard QFT constructions. The remaining part of the algebra $\mathscr{A}_{\sqrbulletV}$ is associated with the (semi)groups $G_{\bullet}$ and $G_{\square}^+$ which can be shown to be semifinite under suitable assumptions on their quasi-invariant measures. This argument is appealing, but requires more careful analysis. For instance, this would hold only if we can extend the two null sheets to infinity, which in general fails due to caustics. We leave the question of determining the type of $\mathscr{A}_{\sqrbulletV}$ to future work. We plan to explore the possibility of having a trace on the quantum CIVP, and thus being able to construct entropy for subregions, in the spirit of \cite{Chandrasekaran:2022cip}. Similarly, we plan to address the properties of the nested subalgebras obtained by time-like supertranslations, and whether they realize a half-sided modular inclusion of semifinite algebras like the one proposed in \cite{Chandrasekaran:2026pnc}.

\subsection{Vacuum Transitions and Observables}

Another interesting way to organize the algebra of the CIVP is as
\beq
	\mathscr{A}_{\sqrbulletV} = \mathscr{A}_{\diagdown} \vee \mathscr{A}_{\square \bigcircbullet \square} \vee \mathscr{A}_{\diagup},
\eeq
where
\beq
	\mathscr{A}_{\square \bigcircbullet \square} = \pi_{\sqrbulletV}(G_{\bullet} \ltimes (G_{\square}^+ \Join_{\bigcirc} G_{\square}^+))
\eeq
is the full operator algebra of the corner including both non-extended and extended operators and we have identified $\pi_{\sqrbulletV}(\mathscr{A}_{\diagdown}) \simeq \mathscr{A}_{\diagdown},\pi_{\sqrbulletV}(\mathscr{A}_{\diagup}) \simeq \mathscr{A}_{\diagup}$. From this point of view, we can regard $\mathscr{A}_{\square \bigcircbullet \square}$ as encoding the vacuum sector relative to the standard hypersurface theory. In particular, the algebra $\mathscr{A}_{\sqrbulletV} \subset \mathfrak{B}(\mathscr{H}_{\sqrbulletV})$ admits a vacuum sector
\beq
	\ket{\omega, g_{\square}^u, g_{\square}^v, g_{\bullet}}. 
\eeq
Here, $\ket{\omega} \in \mathscr{H}_{\diagup}$ is a cyclic vector for the naive hypersurface algebra. More to the point, we might interpret $\ket{\omega,g_{\square}^u, g_{\square}^v, g_{\bullet}}$ as describing the real time evolution of the corner degrees of freedom. In this sense, the overlaps
\beq
	\bra{\omega, g_{\square}^u, g_{\square}^v, g_{\bullet}} \ket{\omega, h_{\square}^u, h_{\square}^v, h_{\bullet}}
\eeq
should be intimately related to the vacuum transitions described in \cite{Verlinde:2019xfb,Verlinde:2022hhs,Klinger:2025tvg}. In future work we plan to explore how these vacuum transition amplitudes can be related to proposed experiments for observing signatures of quantum gravity that are amplified above the Planck scale. 

\pagebreak

\appendix
\renewcommand{\theequation}{\thesection.\arabic{equation}}
\setcounter{equation}{0}

\section*{Acknowledgments}

We would like to thank Shadi Ali Ahmad, Rodrigo Andrade e Silva, Ven Chandrasekaran, Eanna Flanagan, Laurent Freidel, Ted Jacobson, Josh Kirklin, Rob Leigh, Temple He, Juan Maldacena, Rob Myers, Giulio Neri, Geoff Penington, Javi Peraza, Michael Reisenberger, Gautam Satishchandran, Jon Sorce and Kathryn Zurek for interesting discussions. The work of M.S.K. was supported by the Heising-Simons foundation ``Observable Signatures of Quantum Gravity" collaboration and the Walter Burke Institute for Theoretical Physics. This material is also based upon work supported by the U.S. Department of Energy, Office of Science, Office of High Energy Physics, under Award Number DE-SC0011632. Research at Perimeter Institute is supported in part by the Government of Canada through the Department of Innovation, Science and Economic Development Canada and by the Province of Ontario through the Ministry of Colleges and Universities. L.C. is supported by the Celestial Holography Simons collaboration.

\section{Fock Space Representations of the CCR Algebra} \label{app: Fock}

As in the main text, let $\mathscr{A}$ be the CCR algebra generated by the Weyl symbols associated with a symplectic vector space $(X,\Omega)$. Let us fix a regular state $\omega$ to describe how this quantization can be related to more familiar notions from quantum field theory. In this appendix we will write $\mathscr{H} =L^2(\mathscr{A},\omega)$, $\pi=\pi_\omega$, $C=C_\omega$, $\xi=\xi_\omega$, $\langle \ \rangle=\langle \ \rangle_\omega$, to refer to the GNS Hilbert space of $\omega$, its associated representation, characteristic function, vector representative and inner product, respectively.  

By the regularity of $\omega$, the map
\beq
	t \in \mathbb{R} \mapsto \pi(\op{W}(tV))
\eeq
defines a strongly continuous one parameter family of unitary operators on $\mathscr{H}$. By the Stone theorem we therefore obtain an assignment of each $V \in S$ to a self adjoint operator
\beq
	\op{\phi}(V) = -i\frac{\rd \pi(\op{W}(tV))}{\rd t}\big\rvert_{t = 0}, \qquad \pi(\op{W}(V)) = e^{i \op{\phi}(V)}. 
\eeq
The operator $\op{\phi}(V)$ should be interpreted as a smeared field operator and satisfies the canonical commutation relations
\beq	
	[\op{\phi}(V_1), \op{\phi}(V_2)] = i I_{V_2} I_{V_1} \Omega . 
\eeq\\

A regular state $\omega$ is called quasi-free if its characteristic function is Gaussian, namely if there exist a real linear functional $\mu: S \to \RR$ and a real symmetric positive bilinear form $c: S \times S \to \RR$ such that
\beq
	C(V) = e^{i \mu(V) - \frac{1}{2} c(V,V)}.
\eeq
Equivalently, all higher correlation functions are determined by the one- and two-point functions through Wick contractions. The one-point function is
\beq
	\mu(V) = \langle \xi, \op{\phi}(V) \xi \rangle,
\eeq
while $c$ encodes the symmetrized covariance,
\beq
	c(V_1,V_2) = \frac12\langle \xi, \{\op{\phi}(V_1) - \mu(V_1),\op{\phi}(V_2) - \mu(V_2)\} \xi\rangle. 
\eeq\\

With our Weyl convention, the ordered two-point function is therefore
\beq
\langle \xi,\op{\phi}(V_1)\op{\phi}(V_2)\xi\rangle=\mu(V_1)\mu(V_2)+c(V_1,V_2)+\frac{i}{2}\,I_{V_2}I_{V_1}\Omega.
\eeq
Thus $c$ is the symmetric covariance, while the antisymmetric part of the ordered two-point function is fixed by the canonical commutation relations.\\ 

The quasi-free state determines a positive Hermitian form on the complexification of $S$,
\beq
h(V_1,V_2)=c(V_1,V_2)+\frac{i}{2}\,I_{V_2}I_{V_1}\Omega.
\eeq
The corresponding one-particle Hilbert space $\mathfrak h$ is obtained by quotienting $S^{\mathbb C}$ by the null space of $h$ and completing with respect to this inner product. In general, the GNS representation of a quasi-free state is the quasi-free representation determined by $h$. When the state is pure, this representation is the usual bosonic Fock representation over $\mathfrak h$.\\

To introduce creation and annihilation operators one must choose a polarization of the real symplectic space $S$. We encode this polarization by a complex structure
\beq
J:S \to S,\qquad J^2=-1,
\eeq
compatible with the symplectic form in the sense that
\beq
I_{JV_2}I_{JV_1}\Omega =I_{V_2}I_{V_1}\Omega,
\eeq
which infers that $J$ preserves the canonical commutation relations. We furthermore require $J$ to be positive, namely that the symmetric bilinear form $I_{JV_2}I_{V_1}\Omega$ is positive definite. This turns $S$ into a complex pre-Hilbert space, with multiplication by $i$ implemented by $J$.

For a pure quasi-free state, the symmetric covariance is determined by this positive complex structure:
\beq
c(V_1,V_2)=\frac12 I_{JV_2}I_{V_1}\Omega.
\eeq
Conversely, a positive compatible complex structure $J$ defines a pure quasi-free state with this covariance, and thus a choice of Fock polarization.

In this case the creation and annihilation operators can be written as
\beq
\op{b}(V)=\frac12\left(\op{\phi}(V)+i\op{\phi}(J V)\right),\qquad \op{b}^\dagger(V)=\frac12\left(\op{\phi}(V)-i\op{\phi}(J V)\right),
\eeq
so that
\beq
\op{\phi}(V)=\op{b}(V)+\op{b}^\dagger(V).
\eeq
They satisfy
\beq
[\op{b}(V_1),\op{b}^\dagger(V_2)]=h(V_1,V_2),
\qquad
[\op{b}(V_1),\op{b}(V_2)]=0.
\eeq
For a quasi-free state with non-vanishing one-point function $\mu$, this Fock representation is understood as the corresponding coherent displacement of the centered quasi-free representation.\\

\section{Twisted Crossed Products and Central Extensions} \label{app: TCP}

In the main text we were faced with the problem of implementing grouplike constraints in a scenario in which the symmetry algebra of the constraint group may acquire a central extension under quantization. In this appendix we describe a convenient mathematical formalism which accomplishes this task. The set up is as follows, we denote by $(A,G,\alpha,\omega)$ a system consisting of a $C^*$ algebra $A \subset \mathfrak{B}(\mathscr{H})$, a group $G$ acting on $A$ by automorphisms as $\alpha: G \rightarrow \text{Aut}(A)$, and a group two-cocycle $\omega$ taking values in the center of the algebra $A$.\footnote{In the case of a factor algebra this cocycle is valued in the complex numbers.} 

We can cast the problem of constraint quantization in an algebraic language as follows. Physical operators $\cO$ belong to the subalgebra $A_{\textrm{phys.}}$ of $A \otimes \mathfrak{B}(L^2(G))$ which satisfy the relation
\beq \label{twisted CP constraint}
	\text{Ad}_{\lambda_{\omega}(g^{-1})} \circ \beta_g(\mathcal{O}) = \mathcal{O}, \qquad \forall g \in G. 
\eeq
Here, $\beta: G \rightarrow \text{Aut}(A \otimes \mathfrak{B}(L^2(G))$ is the automorphism action of $G$ on the full kinematical algebra
\beq
	\beta_g(a \otimes \op{1}) = \alpha_g(a) \otimes \op{1}, \qquad \beta_g(\op{1} \otimes \mathcal{U}) = \op{1} \otimes \text{Ad}_{\lambda_{\omega}(g)}(\mathcal{U}), \qquad a \in A, \ \ \mathcal{U} \in \mathfrak{B}(L^2(G)),
\eeq		
and $\lambda_{\omega}: G \rightarrow \mathfrak{B}(L^2(G))$ is a representation of the group $G$. We interpret $\beta_g$ as the off-shell symmetry action generated by the full Hamiltonians $H_{\un{\xi}}$ while $\text{Ad}_{\lambda_{\omega}(g)}$ is the symmetry action generated purely by the corner charges, $Q_{\un{\xi}}$. The automorphism $\text{Ad}_{\lambda_{\omega}(g^{-1})} \circ \beta_g$ can be interpreted as the symmetry action of $G$ implemented infinitesimally by the constraints $c_{\un{\xi}} = H_{\un{\xi}} - Q_{\un{\xi}}$. Consequently, the condition \eqref{twisted CP constraint} is tantamount to the requirement that physical operators commute with the constraint, as we explained in section \ref{sec: Ray+Dam}.\\

The discussion up to this point is completely analogous to the interpretation of the standard crossed product. Where things change is in how the physical subalgebra $A_{\textrm{phys.}}$ solving \eqref{twisted CP constraint} is constructed. In particular, $A_{\textrm{phys.}}$ corresponds to the subalgebra of $A \otimes \mathfrak{B}(L^2(G))$ generated by the representations
\begin{flalign}
	&\pi_{\alpha}: A \rightarrow \mathfrak{B}(\mathscr{H} \otimes L^2(G)), \qquad \bigg(\pi_{\alpha}(a) \psi\bigg)(g) = \alpha_{g^{-1}}(a) \bigg(\psi(g)\bigg), \nonumber \\
	&\lambda_{\omega}: G \rightarrow \mathfrak{B}(\mathscr{H} \otimes L^2(G)), \qquad \bigg(\lambda_{\omega}(h) \psi\bigg)(g) = \omega(g^{-1},h) \psi(h^{-1} g). 
\end{flalign} 
The crucial difference to the standard crossed product construction is that the representation of the group, $\lambda_{\omega}$, obtains a twisting from the cocycle $\omega$. As a result, the algebra generated by this representation takes the form
\beq \label{Twisting}
	\lambda_{\omega}(g) \lambda_{\omega}(h) = \pi_{\alpha}(\omega(g,h)) \lambda_{\omega}(gh), \qquad g,h \in G. 
\eeq
In this respect, $\lambda_{\omega}$ is a \emph{projective} rather than a unitary representation. Despite this change, it is straightforward to check by direct computation that
\beq
	\text{Ad}_{\lambda_{\omega}(g)}(\pi_{\alpha}(a)) = \pi_{\alpha} \circ \alpha_g(a), \qquad \forall a \in A, \ \  g \in G. 
\eeq	 
This is equivalent to the observation of \cite{Ciambelli:2024swv} that the appearance of a central charge anomaly does not inhibit the standard interpretation of invariant \emph{operators} at the algebraic level. The physical algebra in this instance is therefore $A \times_{\alpha,\omega} G = \pi_{\alpha}(A) \vee \lambda_{\omega}(G)$, which is called the \emph{twisted} crossed product associated to the twisted covariant system $(A,G,\alpha,\omega)$ \cite{Sutherland1980CohomologyI}. The relations \eqref{twisted CP constraint} and \eqref{Twisting}
\beq
	\text{Ad}_{\lambda_{\omega}(g^{-1})} \circ \beta_g(\mathcal{O}) = \mathcal{O}, \qquad \lambda_{\omega}(g) \lambda_{\omega}(h) = \pi_{\alpha}(\omega(g,h)) \lambda_{\omega}(gh)
\eeq
reproduce the conditions described in \cite{Ciambelli:2024swv} to realize a physical algebra in the presence of a central charge anomaly. \\

As a final note, in the case\footnote{A more general analysis of twisted crossed products and their relationship with anomalies will appear in \cite{aliahmad2026}.} that $\omega$ takes values in $U(1)$ one can always pass from a projective representation of a group to a unitary representation of its central extension. Consequently, we can construct an isomorphism 
\beq \label{Twisted to Central Extension}
	A \times_{\alpha,\omega} G \simeq A \times_{\alpha^{\omega}} G^{\omega},
\eeq
where here $G^{\omega}$ is the central extension of $G$, $\alpha^{\omega}$ is the induced automorphism action of $G^{\omega}$ on $A$, and $A \times_{\alpha^{\omega}} G^{\omega}$ is the standard crossed product associated with the standard covariant system $(A,G^{\omega}, \alpha^{\omega})$. To be precise, we recall that as a space $G^{\omega} = G \times U(1)$. It is given the structure of a group via the product
\beq
	(g_1, z_1) (g_2, z_2) = (g_1 g_2, \omega(g_1,g_2) z_1 z_2).
\eeq
This group acts on $A$ via the automorphism
\beq
	\alpha^{\omega}_{(g,z)}(a) = \alpha_g(a),
\eeq	
and realizes a unitary representation
\beq
	\lambda(g,z) = z \lambda_{\omega}(g). 
\eeq
By construction, $(\mathscr{H} \otimes L^2(G), \pi_{\alpha}, \lambda)$ is a covariant representation of the covariant system $(A, \alpha^{\omega}, G^{\omega})$ and thus the crossed product $A \times_{\alpha^{\omega}} G^{\omega}$ is the algebra generated by $\pi_{\alpha}(A) \vee \lambda(G^{\omega})$. Since the extension is assumed to take values in the center of $A$, we can realize an isomorphism
\beq
	\pi_{\alpha}(a) \mapsto \pi_{\alpha}(a), \qquad \lambda_{\omega}(g) \mapsto \lambda(g,1), \qquad \pi_{\alpha}(z) \mapsto \lambda(e,z), \qquad a \in A, g \in G, z \in Z(A). 
\eeq
The relation \eqref{Twisted to Central Extension} seems closely related to the observation made in \cite{Freidel:2026stu} that the invariant algebra of a null ray with respect to null time translations is a Virasoro crossed product, which could otherwise have been identified as a twisted crossed product with $\textrm{Diff}(\mathbb{R})$.\footnote{We note that, strictly speaking, one should regard $\textrm{Diff}(\mathbb{R})$ as a loop group. This leads to important subtleties as compared to $\textrm{Diff}(S^1)$ and the Virasoro algebra. We expect the algebraic results described here should carry over to the loop group case, but leave a detailed analysis to future work.} 

\pagebreak

\bibliographystyle{uiuchept}
\bibliography{LCMSKarXivV1.bib}

\end{document}